\documentclass[usenatbib,iop,numberedappendix]{aeb_emulateapj_2015}
\usepackage{amsmath}
\usepackage{amssymb}
\usepackage{xspace}
\usepackage[normalem]{ulem}
\usepackage{mathrsfs}
\usepackage{longtable}
\usepackage{comment}

\usepackage{natbib}

\usepackage[usenames,dvips]{color}

\definecolor{epcol}{rgb}{0.398, 0.0, 0.797}

\def\del#1{{}}

\newcommand\noentry[1]{--}

\def\s{{\rm s}} 
\def\yr{{\rm yr}} 

\def\m{{\rm m}} 
\def\cm{{\rm c}\m} 
\def\pc{{\rm pc}} 
\def\kpc{{\rm k}\pc} 
\def\Mpc{{\rm M}\pc} 

\def\eV{{\rm eV}} 
\def\meV{{\rm m}\eV} 
\def\GeV{{\rm G}\eV} 
\def\TeV{{\rm T}\eV} 

\def\G{{\rm G}} 

\newcommand\bmath[1] {\mbox{\boldmath$\rm #1$}}

\def\Fermi{{\em Fermi}\xspace}

\def\Dpp{D_{\rm pp}}
\providecommand{\e}[1]{\ensuremath{\times 10^{#1}}}

\def\bx{\bmath{x}}

\def\bp{\bmath{p}}

\def\Dpp{D_{\rm pp}}

\def\bmu{\bmath{\mu}}
\def\rmn{\mathrm}

\def\ph{{\rm ph}}
\def\deg{{\rm deg}}

\def\P{{\mathcal P}}

\newcommand\circp[2]{#1^{\circ}\!\!\!.#2}

\begin{document}

\title{
  Bow Ties in the Sky II:\\
  Searching for Gamma-ray Halos in the \Fermi Sky Using Anisotropy
}

\author{
Paul Tiede\altaffilmark{1},
Avery E.~Broderick\altaffilmark{1,2},
Mohamad Shalaby\altaffilmark{1,2,3},
Christoph Pfrommer\altaffilmark{4},\\
Ewald Puchwein\altaffilmark{5},
Philip Chang\altaffilmark{6},
and
Astrid Lamberts\altaffilmark{7}
}
\altaffiltext{1}{Department of Physics and Astronomy, University of Waterloo, 200 University Avenue West, Waterloo, ON, N2L 3G1, Canada}
\altaffiltext{2}{Perimeter Institute for Theoretical Physics, 31 Caroline Street North, Waterloo, ON, N2L 2Y5, Canada}
\altaffiltext{3}{Department of Physics, Faculty of Science, Cairo University, Giza 12613, Egypt}
\altaffiltext{4}{Heidelberg Institute for Theoretical Studies, Schloss-Wolfsbrunnenweg 35, D-69118 Heidelberg, Germany}
\altaffiltext{5}{Institute of Astronomy and Kavli Institute for Cosmology, University of Cambridge, Madingley Road, Cambridge, CB3 0HA, UK}
\altaffiltext{6}{Department of Physics, University of  Wisconsin-Milwaukee, 3135 North Maryland Avenue, Milwaukee, WI 53211, USA}
\altaffiltext{7}{TAPIR, Mailcode 350-17, California Institute of Technology, Pasadena, CA 91125, USA}

\shorttitle{Detecting Anisotropic Gamma-ray Halos}
\shortauthors{Tiede et al.}

\begin{abstract}
  Many-degree-scale gamma-ray halos are expected to surround extragalactic high-energy gamma ray sources.  These arise from the inverse Compton emission of an intergalactic population of relativistic electron/positron pairs generated by the annihilation of $\gtrsim100~\GeV$ gamma rays on the extragalactic background light.  These are typically anisotropic due to the jetted structure from which they originate or the presence of intergalactic magnetic fields.  Here we propose a novel method for detecting these inverse-Compton gamma-ray halos based upon this anisotropic structure.  Specifically, we show that by stacking suitably defined angular power spectra instead of images it is possible to robustly detect gamma-ray halos with existing \Fermi Large Area Telescope (LAT) observations for a broad class of intergalactic magnetic fields.  Importantly, these are largely insensitive to systematic uncertainties within the LAT instrumental response or associated with contaminating astronomical sources.
\end{abstract}


\maketitle

\section{Introduction} \label{sec:I}

The extragalactic gamma-ray sky at TeV energies is dominated by blazars, a subclass of active galactic nuclei (AGNs) with powerful relativistic outflows directed at us. The relativistic jets are powered by accretion onto a central nucleus, presumably a supermassive black hole. In the unified picture of AGNS, their emission properties depend on the orientation of the AGN relative to the line of sight \citep{Urry:1995}. There exist two categories of AGNs that differ in their accretion mode and in the physical processes that dominate their emission.
\begin{itemize}
\item {\em Thermal/disk-dominated AGNs.} Infalling matter assembles in a thin
  disk and radiates thermal emission with a range of temperatures. This emission
  is then Comptonized by a hot corona above the disk to produce power-law X-ray
  emission, defining the class of quasars or Seyfert galaxies.

\item {\em Non-thermal/jet-dominated AGNs.}  Highly energetic electrons that
  have been accelerated in the relativistic jet interact with the jet magnetic
  field and emit synchrotron radiation from the radio to X-ray regime. In
  addition, the same population of electrons can Compton up-scatter seed photons
  that are either provided by the synchrotron radiation itself or by an external
  photon field into the gamma-ray regime. Hence, the broadband spectral energy
  distribution of these objects is characterized by two peaks. This defines the
  class of radio-loud AGNs which can furthermore be subdivided into blazars
  (with the line of sight intersecting the jet opening angle) and non-aligned
  non-thermal dominated AGNs.

\end{itemize}

As a result, the gamma-ray emission of blazars benefits from the relativistic Doppler boosting, shifting the upper end of the gamma-ray emission into the GeV/TeV energy regime.  Blazars exhibit a continuous sequence: their luminosity anti-correlates with the peak energy of their synchrotron spectrum, i.e., the objects emitting very high-energy gamma rays (VHEGRs) at TeV energies have the lowest intrinsic luminosity \citep[e.g.,][]{Fossati:1998,Ghisellini:1998}.

The main observational representatives of both AGN classes, quasars and radio galaxies, exhibit a strong redshift evolution with a steeply rising comoving luminosity density up to a redshift $z\approx2$ and a decline thereafter \citep{Hopkins+07}. In contrast, we can only observe nearby TeV blazars that typically reach out to redshifts of $z\lesssim0.3$ (\citealt{TeVCat:2008}\footnote{See {\tt http://tevcat.uchicago.edu}, Catalog Version 3.4}). The reason for this apparent contradiction lies in the low luminosity of TeV blazars and the finite mean free path of TeV photons as they propagate through space \citep{Fermi_EBL2012,Domi_etal:13}, precluding the detection of high-redshift blazars (if they exist). The opacity of the Universe to TeV photons is due to the annihilation and pair production of TeV photons of energy $E_\gamma$ on the extragalactic background light \citep{Goul-Schr:67,Sala-Stec:98}.  The mean free path of these VHEGRs is
\begin{equation}
\Dpp(E_\gamma,z) \approx
5.8
\left(\frac{1+z}{2}\right)^{-\zeta}
\left(\frac{E_\gamma}{6~{\rm TeV}}\right)^{-1}
~{\rm Mpc}\,,
\label{eq:Dpp}
\end{equation}
where $\zeta=4.5$ for $z<1$ and $\zeta=0$ for $z\ge1$ \citep{Knei_etal:04,Nero-Semi:09}.  Momentum conservation ensures that the pairs propagate essentially in the same direction of the parent TeV photon and energy conservation implies a pair energy of $E_{e^\pm} \approx E_\gamma/2$.

The resulting ultra-relativistic pairs of electrons and positrons are commonly assumed to lose energy primarily through inverse Compton (IC) scattering with photons of the cosmic microwave background (CMB), cascading the original TeV emission down to \mbox{(multi-)}GeV energies on a mean free path for the scattering process of
\begin{equation}
  D_{\rmn{IC}}
  = \frac{3  m_e^2 c^4}{4 \sigma_{\rmn{T}} u_{\rmn{CMB}} E_{e^\pm}}
  \approx  0.1 \,\left(\frac{E_{e^\pm}}{3\,\rmn{TeV}}\right)^{-1}\,(1+z)^{-4}\,\rmn{Mpc},
  \label{eq:D_IC}
\end{equation}
where $m_e c^2$ is the electron rest mass energy, $\sigma_{\rmn{T}}$ is the Thompson cross section, and $u_{\rmn{CMB}}$ is the CMB energy density.

However, the inverse Compton cascaded (ICC) multi-GeV emission has not been observed by the \Fermi Large Area Telescope (LAT), indicating that some additional physics needs to be considered \citep[see, e.g.,][]{Nero-Vovk:10,Tave_etal:10a,Tave_etal:10b,Derm_etal:10,Tayl-Vovk-Nero:11,Taka_etal:11,Dola_etal:11,HESS:2014,Prokhorov:2016}. The presence of intergalactic magnetic fields (IGMFs) would deflect the beam of $e^+/e^-$ pairs with a Larmor radius of
\begin{equation}
  r_{\rmn{L}}=\frac{E_{e^\pm}}{e B} \approx 
  30\,\left(\frac{E_{e^\pm}}{3\,\rmn{TeV}}\right)\,
  \left(\frac{ B}{10^{-16}\,\rmn{G}}\right)^{-1}\,\rmn{Mpc}
  \label{eq:rL}
\end{equation}
out of our line of sight. This reduces the ICC flux and thus provides a lower limit on the strength $B$ of the IGMF. For the associated ICC photons with energies of
\begin{equation}
  E_{\rm IC} \approx 2 \gamma^2 E_{\rm CMB} \approx
  70 \left(\frac{E_{e^\pm}}{3~\TeV}\right)^2 \frac{E_{\rm CMB}}{1~\meV}~\GeV\,
  \label{eq:ICE}
\end{equation}
the \Fermi angular resolution is $\theta \approx\circp{0}{2}$ or $3\times10^{-3}$~rad \citep[using the 1--$\sigma$ containment angle of combined events, see Fig. 57 of][]{LAT_perf}. Hence, a deflection of the pairs by an angle $D_{\rmn{IC}}/r_{\rmn{L}} > \theta$ implies a lower limit on the IGMF of $B\gtrsim 10^{-16}\,\rmn{G}$ \citep[see e.g.,][]{Nero-Vovk:10} with important implications for primordial magnetogenesis \citep[for reviews, see][]{Kandus:2011, Durrer:2013}.

Alternatively, the process of ultra-relativistic pairs propagating through the intergalactic medium can be viewed as two counter-propagating beams that are subject to plasma instabilities, and, in particular, to the oblique instability \citep{PaperI}. The linear growth rate of the oblique instability is larger than the cooling rate due to IC scattering of the pairs \citep{PaperI, Schl_etal:12}. If this dominance of the instability growth rate carries over to the regime of non-linear saturation (\citealt{Chang:2014,Schlickeiser:2013}, but see also \citealt{Sironi:2014,Mini-Elyv:12}), this causes the kinetic energy of the pairs to be transferred to the unstable electromagnetic modes in the background plasma. This energy should eventually be dissipated, heating the intergalactic medium \citep{PaperII} with potentially far-reaching implications for the Lyman-$\alpha$ forest \citep{PaperIV, Lamberts:2015}, non-linear structure formation \citep{PaperIII}, and for the blazar luminosity function and the extragalactic gamma-ray background \citep{PaperV,PaperVa}. Estimates suggest that $\sim300\,\yr$ after the onset of TeV emission, the pair beam density has grown sufficiently for plasma beam instabilities to dominate its evolution. This would randomize the beam, and potentially suppress the ICC emission on which the IGMF limits are based \citep[rendering these limits dubious,][]{PaperI}.

How can we determine the ultimate fate of these pairs? Clearly an unambiguous detection of the deflected pair halo emission would immediately prove the deflection hypothesis. So far, all work has concentrated on measuring excess halo power at large angular scales through stacking analyses of blazar images. However, those resulted in null results \citep[e.g.,][]{FLAT-stack:2013} despite some earlier claims \citep{Ando:2010} that were subsequently disproven \citep{Nero-Semi-Tiny-Tkac:11} because of the uncertainty of the exact shape and side lobes of \Fermi's point-spread function \citep{LAT_perf}. A more recent attempt \citep{Chen:2015} utilizes the most recent PSF; nevertheless, it exhibits similar sensitivies to the uncertain instrument response.

Here we report a novel method for extracting signatures of the ICC component that exploits the large degree of anticipated anisotropy \citep[see, e.g.,][]{2010ApJ...719L.130N,2015JCAP...09..065L,BowTiesI}. This is caused either by the structure of the initial VHEGR jet (if its opening angle is not intersecting the line of sight) or by the fact that both electrons and positrons are produced by the VHEGR annihilation on the EBL and deflected in opposite directions by an IGMF.  The details of the structure depends on both the mechanism and properties of the IGM.  However, in both cases they can produce halos that are dominated by bi-lobed features.  Such features are generic for the gamma-ray bright blazars observed by \Fermi \citep{BowTiesI}.

In principle, this angular asymmetry can improve our ability to detect ICC halos in two different ways. First, the anisotropy implies a larger surface brightness for the ICC emission, and therefore {\em if we can properly orient the images} any excesses can be detected with a higher significance over the putative symmetric backgrounds.  In practice, we do this not by rotating and stacking images, which proves to be unfeasible due to the inability of determining the orientation of the deflecting magnetic field, but rather by constructing and stacking orientation-independent measures of the angular anisotropy in the gamma-ray sky about individual sources. In particular, we propose to exploit the pair halo anisotropy by computing suitably defined angular power spectra and stacking those. This approach is routinely applied in cosmological data analyses to determine cosmological parameters (e.g., from the CMB anisotropies) -- the key idea consists of averaging over the (unknown) orientation of the phases and accumulating signal in the multipoles of the dominant angular structures.\footnote{A similar strategy is described in \citet{2017arXiv170101501D}, where the utility of the $Q$ statistic is explored, a quantity that is closely related to what we descibe as the quadrupolar power.  A key distinction between the approaches presented here and in \citet{2017arXiv170101501D} is that we also construct and utilize the power at the other multipoles, providing an independent characterization of potential systematic errors.}

Second, averaged over the long live-time of the \Fermi LAT the resulting PSF is very nearly isotropic.  Small residual angular structure in the PSF arising from the cubical geometry of the LAT enters at the hexadecapole ($m=4$) order.  As a result, unlike searches that focus on radial excesses, schemes that exploit the nearly quadrupolar nature of the halos are easily distinguished from systematic effects due to the PSF. While this paper presents the detailed methodology and addresses the (known) systematics, we apply our blind experiment to \Fermi LAT data in a companion letter \citep{BowTiesIII}.

Of particular practical importance is the fact that the \Fermi data set is not easily repeatable.  As a result we have expended considerable effort to predict the anticipated anisotropy signal from various ICC halo models, design and optimize the construction of ICC halo diagnostics prior to analyzing the \Fermi data.  Here we report the results of this optimization investigation, i.e., the manner in which mock realizations were produced (Section \ref{sec:mocks}), a cursory description of the characteristics of the \Fermi hard gamma-ray blazar sample (Section \ref{sec:sample}), investigation of potential confounding features (Section \ref{sec:signatures}), and the confidence levels with which the various ICC models could be excluded given a null result (Section \ref{sec:MCCLs}).  It is in this last step that the dividends of having done so become manifest -- by training the analysis on simulated data (only weakly informed by the gross properties of the \Fermi sample) we ensure that the resulting conclusions are governed by {\em a priori} statistics, and are therefore well understood.

\section{Method Overview} \label{sec:cartoon}
\begin{figure*}
  \begin{center}
    \includegraphics[width=0.45\textwidth]{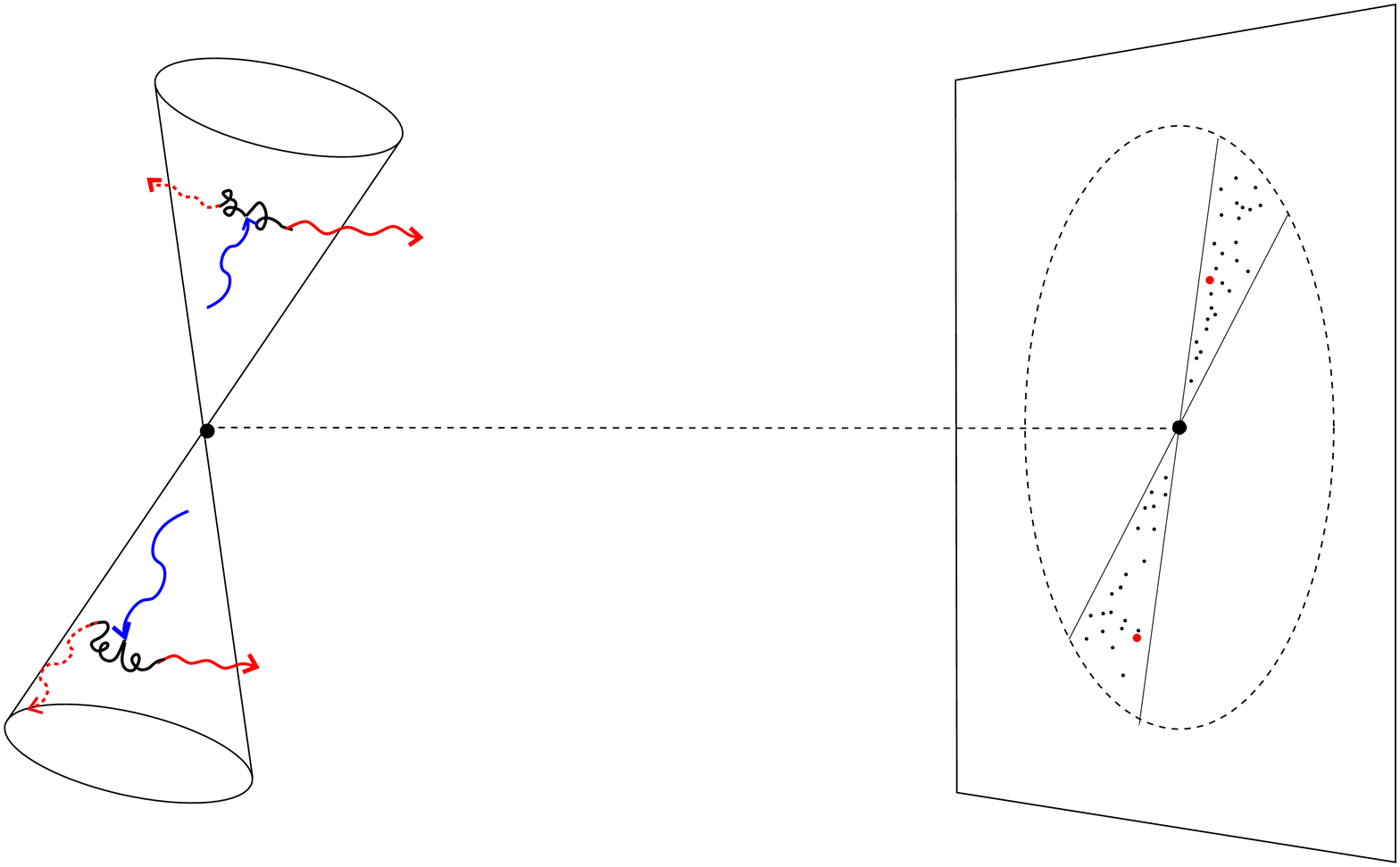}
    \includegraphics[width=0.45\textwidth]{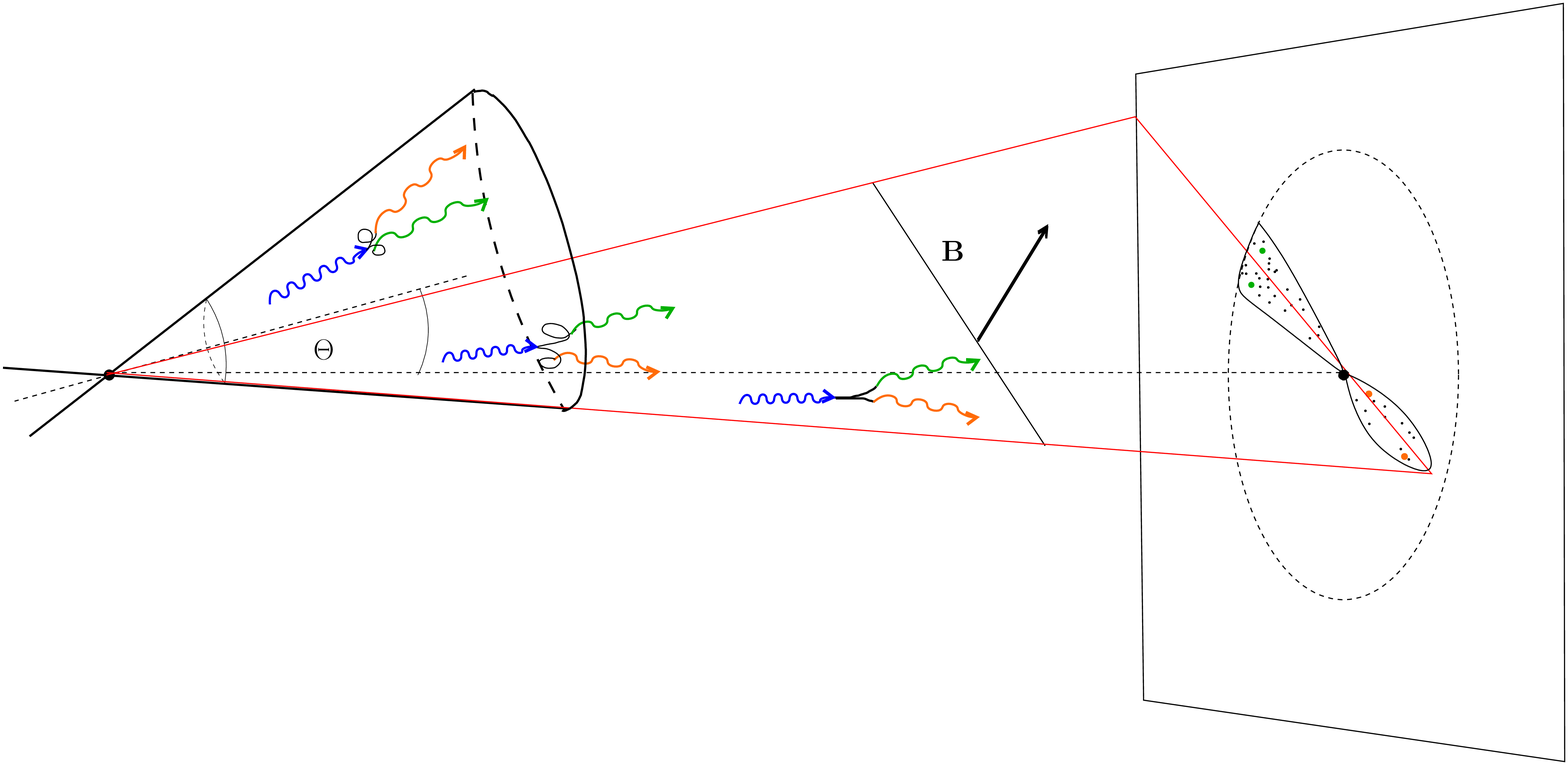}
  \end{center}
  \caption{Cartoons of the mechanisms by which anisotropy in the ICC halos is generated, distinguished by the structure of the underlying IGMF.  Left: For an IGMF tangled on small scales ($\lambda_B\ll3~\Mpc$) the anisotropy is due to the structure of the gamma-ray jet.  Right: For an IGMF that is uniform across the gamma-ray jet (IGMF coherence length $\lambda_B\gg100~\Mpc$) the anisotropy is due to the geometry of the gyrating, relativistic pairs.}\label{fig:cartoons}
\end{figure*}

Before describing the creation of physical realistic ICC halos, statistical measures of the anisotropy, and estimates of its detectability, we will begin with a summary of the key ideas underlying the signal we hope to find.  This is predicated on the standard picture of the ICC halo formation described in Section \ref{sec:I}: VHEGRs emitted from an AGN travel cosmological distances prior to generating energetic electron-positron pairs on the EBL which then inverse Compton up-scatter the CMB to GeV energies.  However, for two independent reasons, these ICC halos are anisotropic.

First, the VHEGRs are originally beamed along the jet axis.  This is evidenced by the overwhelming dominance of blazars in the extragalactic gamma-ray AGN sample \citep{2LAC,3LAC}.  Because the VHEGR mean free path is long in comparison to the distance traveled during the inverse-Compton cooling time of the resulting pairs this implies that the emission is essentially local, and therefore arises from a pair of conical regions indicated by the radio jet of the source AGN.  If the inverse-Compton gamma rays are isotropically emitted, arising, e.g., from a highly tangled IGMF, the spatial structure in the gamma rays generates a resultant structure in the GeV image.  This is shown explicitly in the left-hand panel of Figure~\ref{fig:cartoons}, along with the associated gamma-ray image.

\begin{figure*}
  \begin{center}
    \includegraphics[width=\columnwidth]{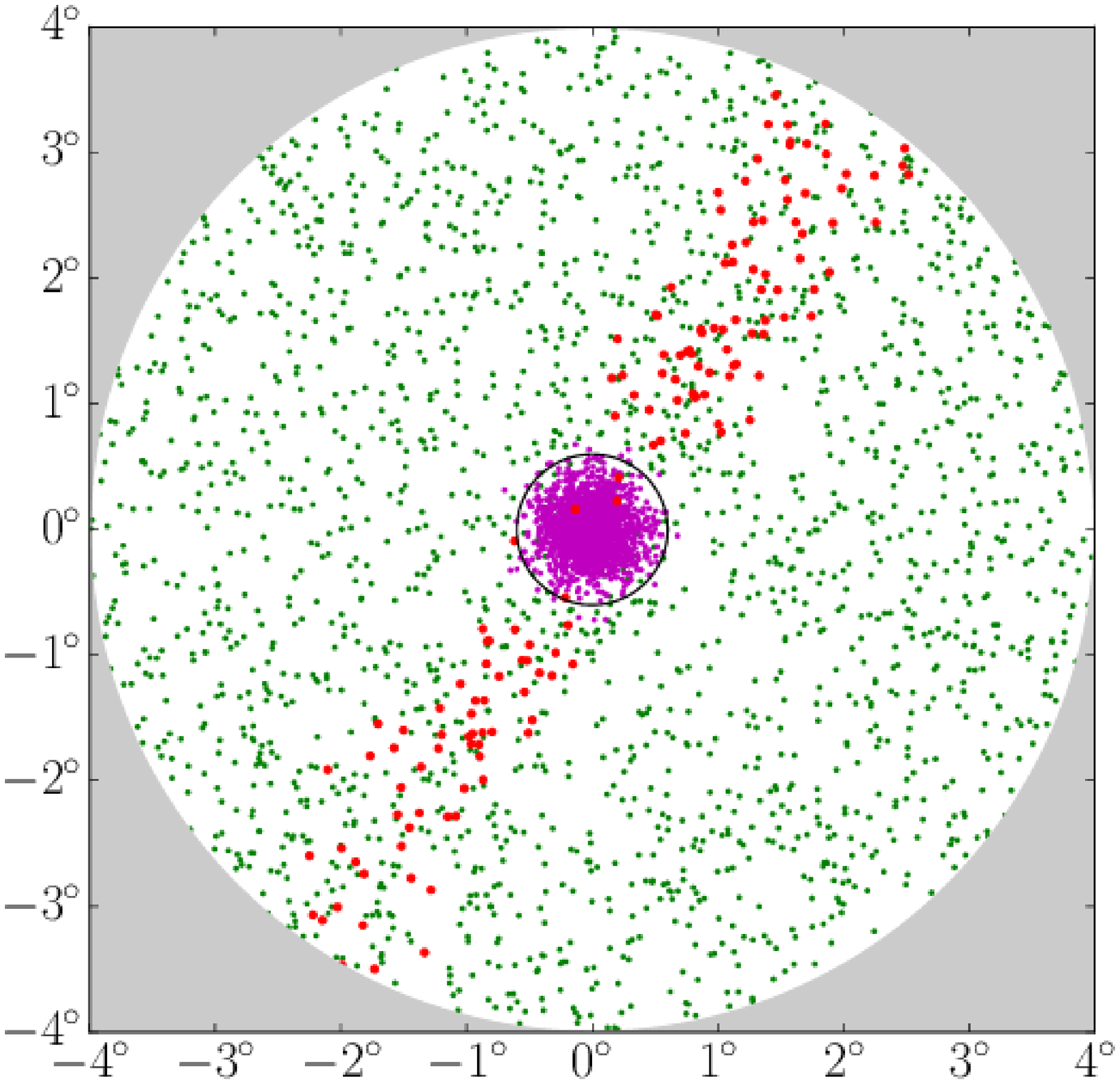}
    \includegraphics[width=\columnwidth]{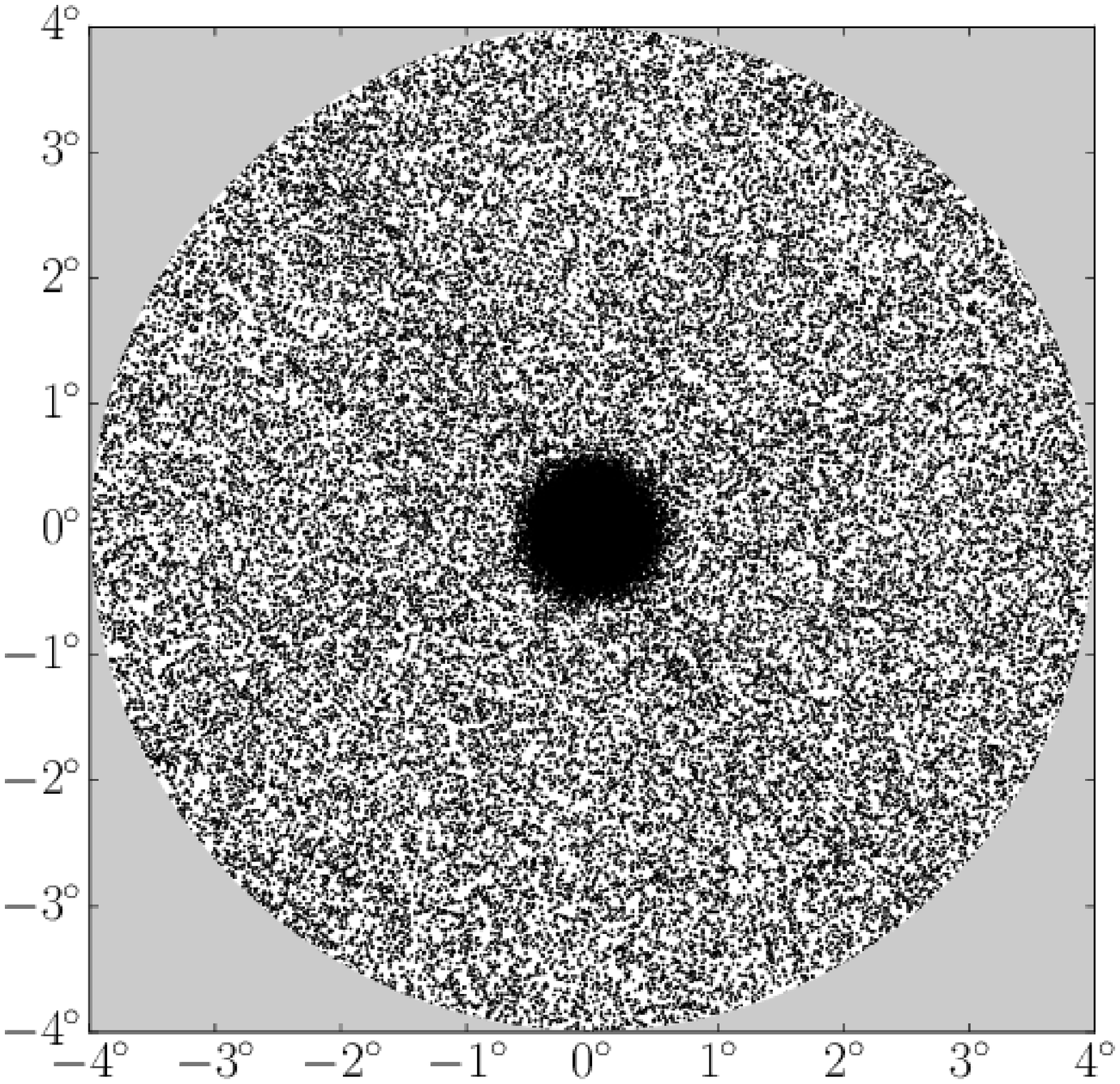}
  \end{center}
  \caption{Left: Realization of $4000$ photons from a toy model that exhibits the anisotropy arising from the mechanisms described in Figure~\ref{fig:cartoons}.  The contribution from the central source, diffuse background, and ICC halo component are shown in purple, green, and red, respectively.  The excluded central region is shown in black.  Note that despite the clear structure within the ICC halo component it is strongly subdominant to the otherwise isotropic source and background components (contributing roughly 5\% of the total photons).  Right: Stacked toy model images of 18 sources with arbitrary ICC halo orientations.  At this point the structure is effectively erased.}\label{fig:chreal}\label{fig:chstack}
\end{figure*}

Alternatively, the process of gyration in the IGMF also can impart structure on the image.  In the presence of an IGMF that is homogeneous on scales comparable to $\Dpp$ electrons and positrons will gyrate on fixed trajectories that emit towards an observer only for a subset of initial injection positions.  This is still superimposed on the jet structure, resulting in a potentially asymmetric image structure, shown in the right-hand panel of Figure~\ref{fig:cartoons}.  Gamma rays on opposite sides of the original AGN are produced predominantly by different lepton species, i.e., positrons on one side and electrons on the other\footnote{Strictly speaking, the identification of lobe sides with particular leptons does assume that the gyration timescale is long in comparison to the inverse-Compton cooling timescale.  Should this be the case today, it will continue to be true at higher redshift.  Note that even if the inverse-Compton cooling time is longer than the gyration period, the halo will still exhibit the bimodal structure.}.

A toy example that provides many of the key features of the ICC halos we will describe in detail in Section \ref{sec:mocks}, is shown in Figure~\ref{fig:chreal}.  This includes equal contributions from a uniform background and from a central source, totaling 4000 photons and comparable to a typical bright \Fermi AGN.   In addition, there is an anisotropic halo component (red) containing 10\% of the photons in the source drawn from the ad hoc flux distribution indicated by the contours.  All components have been convolved with a Gaussian PSF with standard deviation $\circp{0}{2}$, comparable to the scale of the Pass 8R2\_V6 PSF for front-converted events in the LAT.

\begin{figure}
  \begin{center}
    \includegraphics[width=\columnwidth]{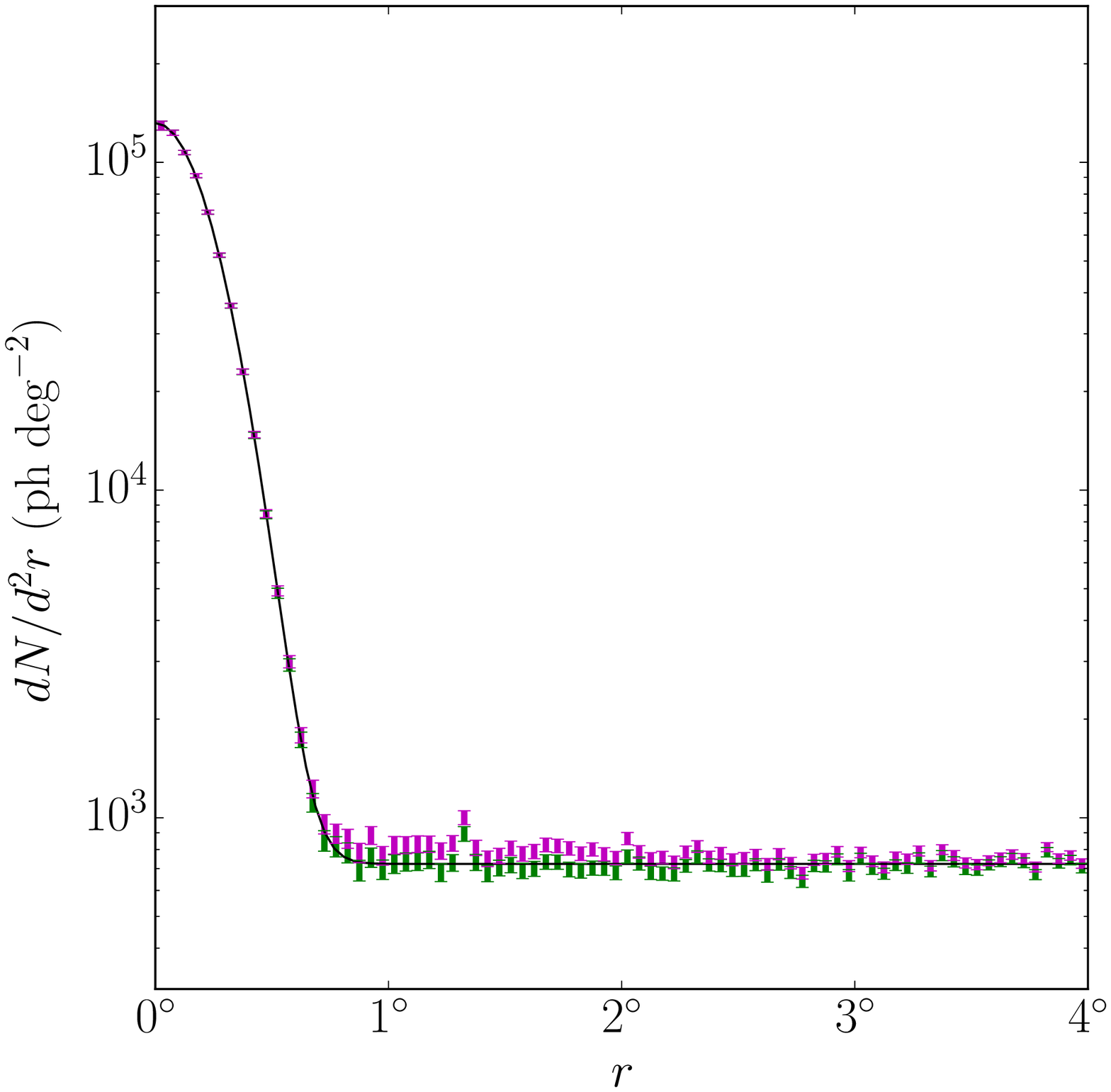}
  \end{center}
  \caption{Radial distribution of gamma-ray surface brightness from the stacked toy model images in Figure~\ref{fig:chstack} with (purple) and without (green) the ICC halo included.  This is compared with a model comprised of a central point source and uniform background, convolved with a toy Gaussian PSF with standard deviation $\circp{0}{2}$.  While the normalizations for the two components differ slightly from their input values, they are capable of providing a good fit even in the presence of the ICC halos.  In any case any disparity is insufficient to convincingly rule out modifications of the large-angle wings of the PSF.}\label{fig:chbin}
\end{figure}

\begin{figure}
  \begin{center}
    \includegraphics[width=\columnwidth]{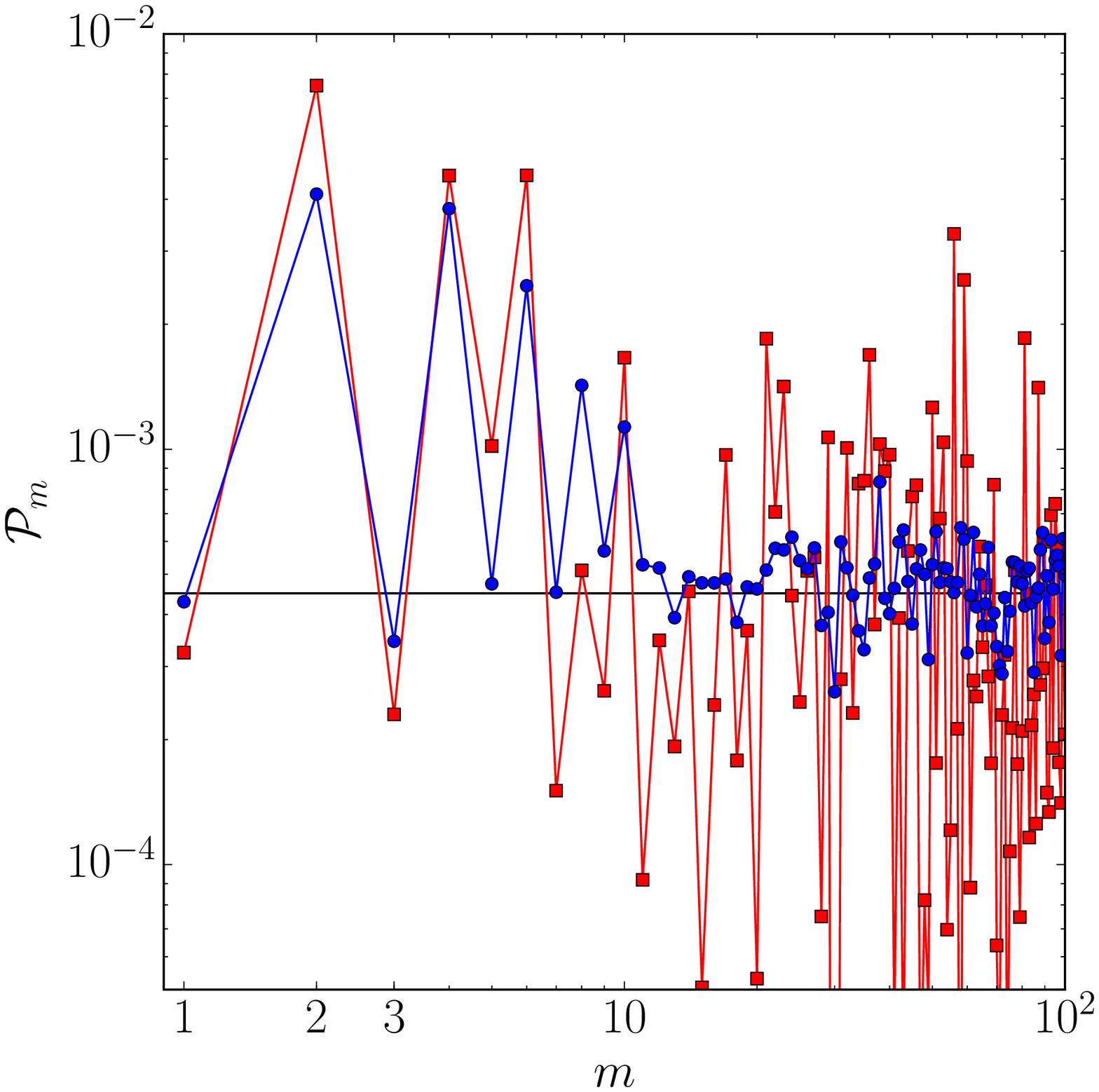}
  \end{center}
  \caption{Azimuthal power spectra of a single realization (red squares) and stacked azimuthal power spectra of the 18 realizations (blue circles) of the toy ICC halo model.  In these the anisotropic halo component generates a clear signal in even multipoles, beginning with the quadrupole, due to the underlying symmetry of the ICC halo images.  For reference, the Poisson noise limit is show by the black line, to which all models asymptote at large $m$.}\label{fig:chps}
\end{figure}

In the absence of the component color coding the sub-dominance of the ICC excess makes it difficult to identify directly from a single source.  Typically, this is dealt with via stacking multiple images, increasing the statistical significance with which the halo component can be isolated.  We therefore show a stacked image of 18 realizations of the same cartoon halo in Figure~\ref{fig:chstack}.  Because the orientation of the putative ICC halo feature is also randomly varied (corresponding to different source and IGM orientations), in contrast to the single image, the stacked image exhibits nearly no angular structure.  Nevertheless, there is a small gamma-ray excess at large angular scales.  In Figure~\ref{fig:chbin} this is shown explicitly in comparison to the case when the halo is absent, beginning near angular scales of $1^\circ$ (where the central source ceases to dominate over the background).  The interpretation of the excess is complicated, however, by uncertainty in structure of the large-scale tails of the PSF or the background flux: even marginal modifications of either can absorb the halo signal in its entirety.

Instead we focus on the anisotropic structure of the ICC halo, which presents a unique signature that is difficult to confuse with instrumental response.
Explicitly, we construct an angular power spectrum of the surrounding photon positions about the source, defined by:
\begin{equation}
  \P_m \equiv \frac{1}{N^2} \left| \sum_j e^{i m \theta_j} \right|^2
  \label{eq:PS}
\end{equation}
where $\theta_j$ is the polar angle of the $j$th gamma ray about the image center relative to a fiducial direction and $N$ is the total number of gamma rays.
To remove the bulk of the source contribution and eliminate the unresolved structure near the origin we mask the inner regions prior to constructing $\P_m$; for illustrative purposes here we simply exclude the inner $\circp{0}{6}$, though in practice we implement an energy-dependent mask (see Section \ref{sec:Mask}).  The ICC halos generate a characteristic power spectrum due to their bimodal structure, shown by the red points in Figure~\ref{fig:chps}, that is dominated by $m=2$ and the even multipoles that follow and qualitatively distinct from most potential image contaminants.  In contrast, Poisson noise from a cylindrically symmetric source (e.g., the PSF-convolved central AGN or background) is flat, shown by the black line in Figure~\ref{fig:chps}, and therefore easily distinguished from the anisotropic ICC halos.

\begin{deluxetable*}{cccccccccccc}
  \tablecaption{Optimized Source List for a Large-scale, Uniform IGMF \label{tab:sample}} 
  
  \tablehead{
    3FGL Name & Common Names & $E_p$\tablenotemark{a}  & $\Gamma_l$\tablenotemark{b} & $\Gamma_h$\tablenotemark{c}  & $z$  & $R_{\rm{img}}$\tablenotemark{d}  & $N_{\rm{ph,f}}$\tablenotemark{e} & $\mathcal{B}_{\rm{f}}$\tablenotemark{f} & $N_{\rm{ph,b}}$\tablenotemark{g} & $\mathcal{B}_{\rm{b}}$\tablenotemark{h} & $B_0$\tablenotemark{i}\\
    & & (GeV) &  &  &  & (deg) &(ph) & ($\rm {ph}/deg^2$) &(ph) & ($\rm {ph}/deg^2$) & (G)
  }
  \startdata
  3FGL J1104.4+3812& Mkn 421                &	95.38	&	1.77&		$2.14^{+0.11}_{-0.10}$	&	0.03	&	2.0	&	4999	&	43.42	&	4746	&	75.37   & $>10^{-17}$ \\
  3FGL J2347.0+5142&	\noentry{1ES 2344+514} &	1.81	&	1.69&		$1.90^{+0.05}_{-0.04}$	&	0.044	&	2.5	&	3176	&	144.57	&	2900	&	133.47  & \noentry{}\\
  3FGL J1653.9+3945&	Mkn 501               &	236.30	&	1.72&		$2.72^{+0.51}_{-0.37}$	&	0.034	&	1.8	&	2028	&	56.06	&	1990	&	74.14   & \noentry{}\\
  3FGL J2000.0+6509&	\noentry{1ES 1959+650} &   658.00	&	1.87&		$4.25^{+1.58}_{-1.60}$	&   0.047	&	3.5	&	6610	&	142.40	&	6257	&	135.76  & \noentry{}\\
\tableline\\
  3FGL J1015.0+4925&	\noentry{1ES 1011+496} & 	2.71	&	1.75&		$1.93^{+0.03}_{-0.03}$	&	0.212	&	2.5	&	1797	&	34.97	&	1756	&	38.88   & $>10^{-16}$ \\
  3FGL J1444.0-3907&  \noentry{PKS 1440-389}  &   11.9    &   1.67&       $2.14^{+0.13}_{-0.09}$  &   0.065   &   2.5 &   2908    &   124.25  &   2781    &   118.81  & \noentry{}\\
  3FGL J0650.7+2503&	\noentry{1ES 0647+250}  &	281.60	&	1.67&		$2.29^{+1.64}_{-1.35}$	&	0.203	&	2.5	&	2142	&	92.84	&	2069	&	90.54   & \noentry{}\\
  3FGL J1120.8+4212&	\noentry{RBS 0970}      &	28.65	&	1.56&		$2.15^{+0.29}_{-0.19}$	&	0.124	&	3.0	&	1123	&	34.57	&	1166	&	35.49   & \noentry{}\\
  3FGL J1442.8+1200&	\noentry{1ES 1440+122}  &	0.49    &	2.69&		$1.81^{+0.08}_{-0.07}$	&	0.163	&	2.5	&	921	    &	41.95	&	998	    &	47.33   & \noentry{}\\
  3FGL J0508.0+6736&	\noentry{1ES 0502+675}  &	2.32	&   1.81&		$1.43^{+0.03}_{-0.03}$	&	0.34	&	2.0	&	1927	&	129.53	&	1817	&	129.53  & \noentry{}\\
\tableline\\
  3FGL J0303.4-2407&	\noentry{PKS 0301-243}&    	1.02	&	1.78&		$1.97^{+0.03}_{-0.03}$	&	0.26	&	2.0	&	984	    &	30.18	&	1000	&	41.48   & $>10^{-15}$ \\
  3FGL J0543.9-5531&	\noentry{1RXS J...}   &  	0.84    &	0.46&		$1.77^{+0.05}_{-0.04}$	&	0.273	&	2.0	&	1051	&	73.31	&	1039	&	70.55   & \noentry{}\\
  3FGL J1436.8+5639&	\noentry{RBS 1409}    &   	1.05	&	2.54&		$1.79^{+0.10}_{-0.09}$	&	0.15	&	2.0	&	599	    &	39.55	&	559	    &	40.65   & \noentry{}\\
  3FGL J2329.2+3754&	\noentry{NVSS J...}   &   	2.19	&	2.19&		$1.81^{+0.11}_{-0.08}$	&	0.264	&	2.0	&	1194	&	83.78	&	1217	&	92.32   & \noentry{}\\
  3FGL J0958.6+6534&	\noentry{S4 0954+65}  &       46.64	&   2.35&		$1.35^{+0.68}_{-0.34}$	&	0.367	&	3.0	&	2317	&	62.36	&	2118	&	60.70   & \noentry{}\\
\tableline\\
  3FGL J0449.4-4350&  \noentry{PKS 0447-439}&     23.03   &   1.81&       $2.30^{+0.11}_{-0.10}$  &   0.205   &   2.0 &   1653    &   38.17   &   1532    &   41.48   & $>10^{-14}$ \\  
  3FGL J0757.0+0956&  \noentry{PKS 0754+100}&      59.98   &   2.25&       $1.62^{+0.93}_{-0.61}$  &   0.27    &   2.0 &   775     &   46.71   &   740     &   47.40  & \noentry{} \\
  3FGL J0622.4-2606&  \noentry{PMN J0622..} &      7.75    &   2.12&       $1.63^{+0.14}_{-0.11}$  &   0.414   &   2.0 &   976     &   57.74   &   1001    &   62.28  & \noentry{}\\
\enddata
\tablenotetext{a}{Energy of power spectrum break in \GeV.}
\tablenotetext{b}{Low-Energy photon spectral index.}
\tablenotetext{c}{High-Energy photon spectral index with $1\sigma$ error.}
\tablenotetext{d}{Radius of image selected in degrees.}
\tablenotetext{e}{Number of front converted photons within radius selected.}
\tablenotetext{f}{Estimate of number of front converted photons per degree squared from background sources.}
\tablenotetext{g}{Number of back converted photons within radius selected.}
\tablenotetext{h}{Estimate of number of back converted photons per degree squared from background sources.} 
\tablenotetext{i}{Values of present-day magnetic field for which the source is in the optimized source list for a large-scale, uniform IGMF (see Section \ref{sec:opt}).}
\end{deluxetable*}

More importantly, the $\P_m$ are independent of the orientation of the halo structure.  Thus it is possible to stack the $\P_m$ from many sources directly, improving their estimation and thereby improving the significance with which halos may be detected.  For example, the $\P_m$ arising from stacking the angular power spectra of the same 18 realizations used to generate Figure~\ref{fig:chstack}  is shown explicitly by the blue line in Figure~\ref{fig:chps}.  In this the halo structure signal at $m<10$ is clearly evident in comparison to the Poisson fluctuations that dominate at $m>15$.

Executing this in practice requires physically realistic halo flux distributions that connect the energy-dependent flux distributions to the underlying physical properties of the VHEGR emission and IGMF and the construction of mock \Fermi images, to which we now turn our attention.

\newpage
\section{Generating Mock Images of Gamma-ray Sources} \label{sec:mocks}
Key to assessing any scheme to detect the ICC halos is the creation of credible theoretical realizations of gamma-ray images of potential sources.  While the Section \ref{sec:cartoon} introduced the qualitative reasons to expect anisotropic structure in the ICC halo structure, how to do this quantitatively was presented in \citet{BowTiesI}, which we summarize here.

The energies of the gamma rays that comprise the putative ICC halos lie typically below 100~GeV; much higher energy gamma rays are absorbed on the EBL.  Below 1~GeV the \Fermi LAT PSF typically broadens substantially, limiting efforts to find asymmetric features and extending the contaminating influence of bright sources.  Between 1~GeV and 100~GeV the \Fermi LAT response functions only modestly depends on energy.  Therefore, we restrict our attention to this energy range.

At the most granular level, \Fermi images consist of collections of individual photons, numbered in the thousands for a single bright source, each with a reported sky location and energy.  Thus, in principle this procedure consists of first, identifying the joint probability distribution of photons from various emission components with a given energy and location, $dF/dEd^2x$, and second, efficiently drawing a random realization from this, $\{E_j,\bmath{x}_j\}$.  In practice, this is further modified by the \Fermi LAT response, which primarily impacts the images via the PSF.  We consider a three component model comprised of a uniform background, an intrinsic point source, and a putative ICC halo.  The former two are well defined and have parameters fixed by \Fermi directly.  Less clear are the ICC halos.  Their brightness and morphology depend on the poorly constrained VHEGRs, and thus require some spectral and collimation model that extends the known \Fermi properties to TeV energies.  This introduces a variety of additional poorly known parameters.

\begin{figure}
  \begin{center}
    \includegraphics[width=\columnwidth]{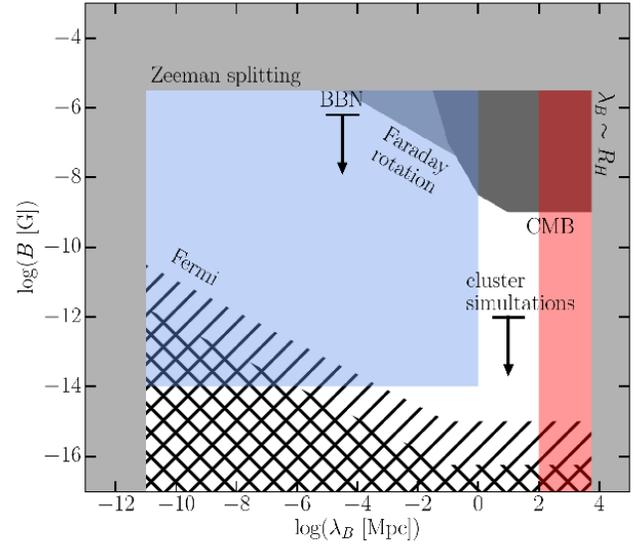}
  \end{center}
  \caption{Limits on the strength and correlation length of the IGMF, adapted from \citet{Nero-Vovk:10}.  The regions described by the small-scale, tangled IGMF and large-scale, uniform IGMF halo models are highlighted in blue and red, respectively.  The lower limit on the magnetic field strength in the small-scale, tangled IGMF halo models arises from the requirement that the pairs execute roughly a full gyration within a cooling time.  For comparison, limits arising from the SEDs of 1ES~0229+200 and 1ES~0347-121 are shown by the single- and cross-hatched regions, respectively.  Limits arising from direct atomic measurements, Faraday rotation, Big Bang nucleosynthesis, and the cosmic microwave background are also shown \citep[For details on the origins of the limits see][]{Nero-Semi:09}.}\label{fig:Blims}
\end{figure}

In addition to the intrinsic parameters of the source, the structure of the ICC halo depends critically upon the assumed geometry of the IGMF.  Here we consider the two limits described in Section \ref{sec:cartoon}: a small-scale, tangled field and a large-scale, uniform field.  In principle, these correspond to different assumptions about the IGMF power spectrum.  In practice, they imply distinct evolution models for the ultra-relativistic electron/positron pairs following their generation by VHEGR photons from the gamma-ray blazars; the pairs' momenta are rapidly isotropized in the former case, while pairs gyrate around the B-field in the latter case, emitting (toward the observer) only when their momentum is directed toward the observer.  We imagine that the general situation lies between these two limits, though for a large range of potential IGMF correlation lengths, $\lambda_B$, either will be applicable.  The regions where each limit applies in relation to the current constraints on the strength and correlation length of the IGMF are shown in Figure~\ref{fig:Blims}.

\subsection{Small-scale, Tangled IGMF}
The first limit is characterized by a rapid isotropization of the pair momenta.  A necessary, though not sufficient, condition for this limit is a strong IGMF, i.e., that the pairs gyrate through $\gtrsim 2\pi$ radians.  Hence, the local IGMF strength 
\begin{equation}
B\gtrsim 2\times10^{-14} \left(\frac{E_{\rm IC}}{10~\GeV}\right) (1+z)^4~\G\,.
\end{equation}
Additionally, the IGMF must be dominated by small-scale structures, varying over length scales that permit gyration around a number of axes.  Ostensibly, this implies that $\lambda_B$ must be small in comparison to the typical inverse-Compton cooling length, $\approx700~\kpc$ (see Equation~\ref{eq:D_IC}).   However, in practice it is sufficient to have isotropization in the statistical sense, i.e., multiple independent domains of locally ordered IGMF within the VHEGR jet.  This places a weaker constraint, requiring only that $\lambda_B$ is small in comparison to the width of the jet, typically of order a few Mpc.
These produce ICC halos that for the \Fermi blazars are characterized by only weak anisotropy.  The reason for this is the large foreshortening associated with the gamma-ray blazars that suppresses the angular structure that is dramatic at oblique angles.  Nevertheless we consider this case for completeness.

\begin{figure*}
  \begin{center}
    \includegraphics[width=0.24\textwidth]{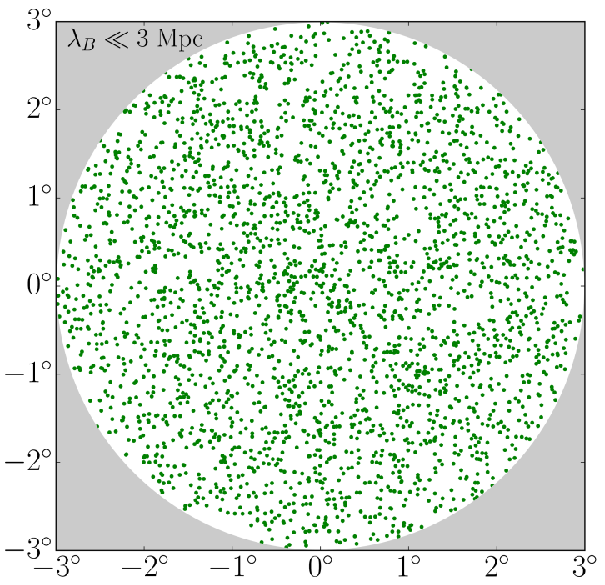}
    \includegraphics[width=0.24\textwidth]{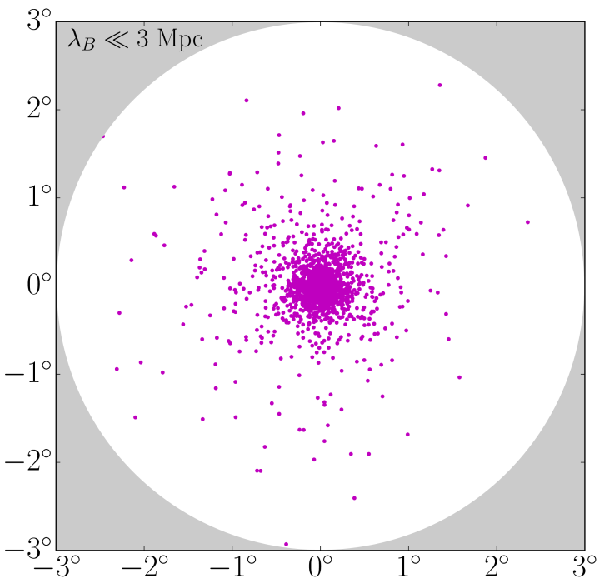}
    \includegraphics[width=0.24\textwidth]{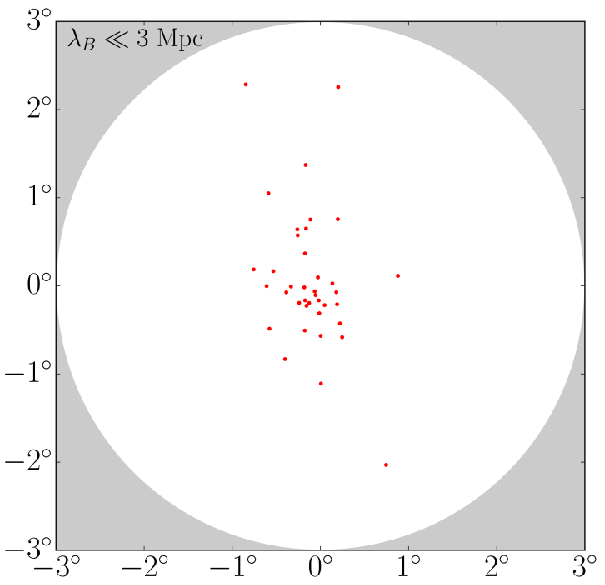}
    \includegraphics[width=0.24\textwidth]{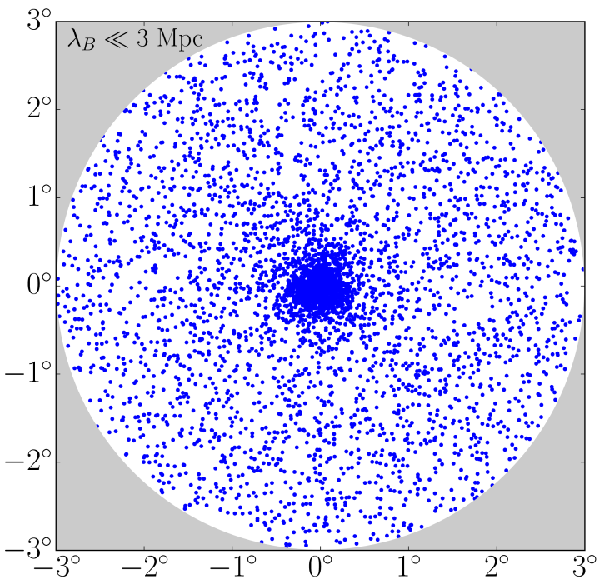}\\
    \includegraphics[width=0.24\textwidth]{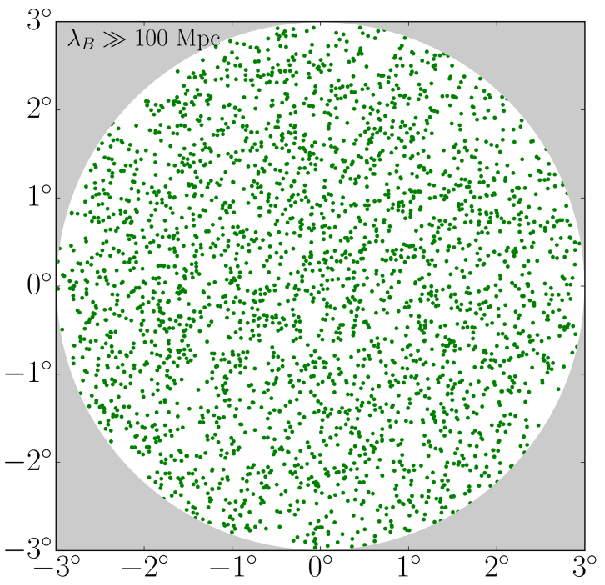}
    \includegraphics[width=0.24\textwidth]{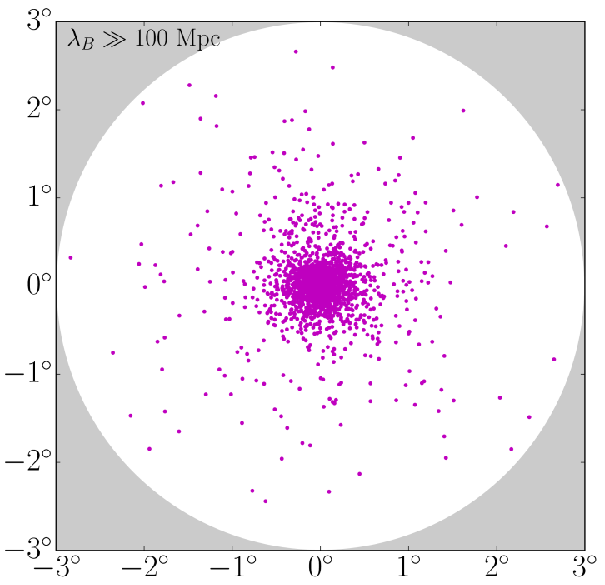}
    \includegraphics[width=0.24\textwidth]{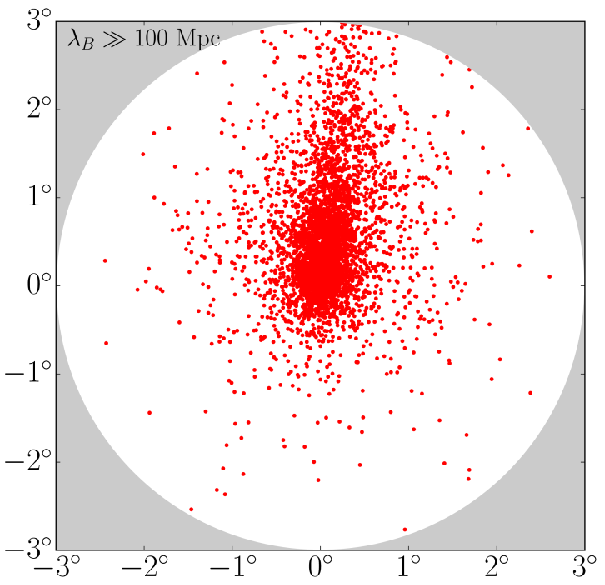}
    \includegraphics[width=0.24\textwidth]{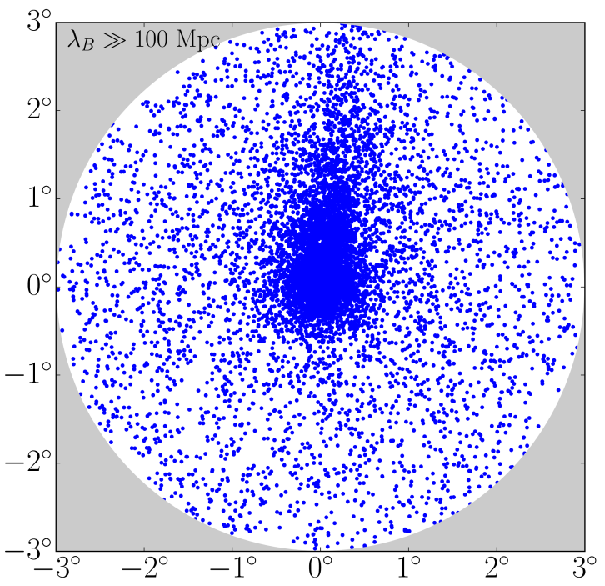}
  \end{center}
  \caption{Components of a typical realization of a gamma-ray map of a \Fermi AGN with an ICC halo associated with a small-scale, tangled IGMF (top) and a large-scale, uniform IGMF with strength $10^{-15}~\G$ (bottom).  Each has the background (left), source (center-left), halo (center-right), and combined (right) photon maps.  The assumed source parameters are $\Gamma_l = 1.7$, $\Gamma_h=2.5$, $E_p=1~\TeV$, $\theta_{\rm jet}=3^\circ$, $\Theta=5^\circ$, with on-axis fluence $5000~\ph$, a background photon density of $50~\ph/\deg^2$, and a magnetic field oriented $60^\circ$ to the line of sight and $170^\circ$ from the horizontal axis.} \label{fig:haloreals}
\end{figure*}

As described in \citet{BowTiesI} the flux of halo photons, shown for a typical realization in Figure~\ref{fig:haloreals}, is spatially and energy dependent.  Here we will ultimately be interested in constructing mock realizations of the \Fermi sky, and will therefore generate realizations that include an ICC halo component, a direct emission component, and a diffuse background component.  As in \citet{BowTiesI} we will assume that the source is a point source, and therefore broadened only via the \Fermi Pass 8R2\_V6 PSF.  Furthermore, we will usually assume that the background is locally homogeneous, though will explore departures from this in Section \ref{sec:signatures}.   The detailed gamma-ray distribution then depends on the total source luminosity, distance, spectral shape, orientation, jet geometry, and background characteristics.  We parameterize these in terms of eight quantities:
\begin{itemize}
\item The source redshift, $z$.
\item The 1~GeV--100~GeV fluence, $N_{35}$.
\item The low-energy photon spectral index, $\Gamma_l$.
\item The high-energy photon spectral index, $\Gamma_h$.
\item The energy of the spectral break, $E_p$.
\item The gamma-ray jet opening angle, $\theta_{\rm jet}$.
\item The gamma-ray jet viewing angle, $\Theta$.
\item The local background photon density, $N_B$.
\end{itemize}
The values of $z$, $N_{35}$, $\Gamma_l$, and $N_B$ are reported or may be estimated directly for the appropriate sub-sample of \Fermi AGN.  The values of $\Gamma_h$, $\theta_{\rm jet}$, and $\Theta$ typically are known only for a larger population, and must be constructed from the appropriate distribution for the sources of interest.  We defer a discussion of what these are and how the sampling is done until Section \ref{sec:sample}.

\subsection{Large-scale, Uniform IGMF}
The second limit is characterized by a uniform IGMF across the extent of the gamma-ray jet.  In principle, this requires uniformity on scales of $\Dpp$.  In practice this is reduced for nearby objects (e.g., Mkn 421) and at high energies; in combination these typically require $\lambda_B\gg100~\Mpc$.  Unlike the small-scale, tangled IGMF, there is no condition on the magnetic field strength a priori.  However, weak fields necessarily produce more compact image features pursuant to their smaller deflection angles.  Typically, to produce an observable ICC halo feature that extends beyond the source mask.  While this condition depends on gamma-ray energy, in practice this requires halos larger than roughly $\circp{0}{6}$, and thus we require
\begin{equation}
B\gtrsim4\times10^{-17} \left(\frac{E_{\rm IC}}{10~\GeV}\right) (1+z)^4~\G\,.
\end{equation}
For smaller magnetic fields the ICC halo structure will typically be overwhelmed by the direct emission component and its anisotropy substantially degraded by the \Fermi PSF.  These produce ICC halos that for the \Fermi blazars are characterized by strong anisotropy with an extent dictated by the magnetic field strength, an example of which is shown in Figure~\ref{fig:haloreals}.  This differs from the previous scenario primarily in the origin of the image structure -- here not due to the anisotropy of the gamma-ray emission but rather the anisotropy in the pair distribution function and the strong beaming of the inverse-Compton emission.

As with the previous case we are ultimately interested in producing mock realizations of the \Fermi sky, which we assume is comprised of a halo, direct emission, and background components.  Thus, for the large $\lambda_B$ case we will require all seven of the parameters in the small $\lambda_B$ case, as well as parameters describing the IGMF strength and orientation.  That is the mock \Fermi images in the presence of an large-scale IGMF are characterized by ten quantities:
\begin{itemize}
\item The source redshift, $z$.
\item The 1~GeV--100~GeV fluence, $N_{35}$.
\item The low-energy photon spectral index, $\Gamma_l$.
\item The high-energy photon spectral index, $\Gamma_h$.
\item The energy of the spectral break, $E_p$.
\item The gamma-ray jet opening angle, $\theta_{\rm jet}$.
\item The gamma-ray jet viewing angle, $\Theta$.
\item The local background photon density, $N_B$.
\item The IGMF, $\bmath{B}$.
\end{itemize}
As before, some of these are obtained from values reported for a sub-sample of \Fermi AGN while others must be sampled from the appropriate distributions;  these are discussed in detail in Section \ref{sec:sample}.  In addition we must define $\bmath{B}$.  While we will review this in Section \ref{sec:BSO}, here we note that we do this by specifying independently an orientation and magnitude with the latter set via the current IGMF strength, $B_0$.

\subsection{\Fermi Point Spread Function} \label{sec:PSF}
We assume the same PSF as described in \citet{BowTiesI}, to which we direct the interested reader for details on implementation, and only summarize salient points here.

Because ICC halos have yet to be unambiguously detected, we consider the Pass 8R2\_V6 ULTRACLEANVETO photon sample; these are the photons that are confidently associated with an astronomical origin and not necessarily nearby bright sources.  The form of the Pass 8R2\_V6 PSF is described in \citet{BowTiesI} and for the events of interest here substantially simplified by the weak PSF dependence on energy above 1~GeV and the fact that the collection of events within the Pass 8R2\_V6 ULTRACLEANVETO sample is distributed among a large number of potential bore angles.  As a result the collective PSF for the Front and Back detectors are well approximated for each by that at a single bore angle bin, corresponding to $\circp{36}{9}$--$\circp{45}{6}$ in both cases \citep[for details see][]{BowTiesI}.

In principle, the square geometry of the LAT imposes a strong dependence on the azimuthal angle of the photon \citep{LAT_perf}.  However, in practice the long duration of the \Fermi observations (8 years) combined with the solar tracking and eight-fold symmetry result in a nearly cylindrical symmetry \citep{LAT_perf}.  This may be broken for short duration or bursty events, and thus if the gamma-ray AGN of interest underwent periods of substantial variability a small residual angular structure may appear.  However, as discussed in Appendix \ref{app:PSFPS}, such structure will enter first at the hexadecapole, i.e., $m=4$, mode, and therefore is easily distinguishable from that due to ICC halo structure.

\subsection{Source-Halo-Background Confusion} \label{sec:confusion}
The direct emission from the source, background, and ICC halo are not spatially distinct.  For large-scale uniform IGMF geometries in particular the halos are strongly centrally concentrated, and therefore will suffer from confusion with the source photons.  Therefore, applying the observational constraints provided by the known source and background source counts requires a method to partition the ICC halo component between the source and background in a self-consistent manner.

The LAT PSF provides a natural definition of those events that would be identified as ``source'' photons.  Any substantial emission component beyond the 68\% containment radius of the Pass 8R2\_V6 ULTRACLEANVETO PSF would be identified as extended and therefore not included in the point-source flux estimates.  This is also consistent with the energy- and detector-dependent mask that we apply to the images to reduce Poisson noise (see Section \ref{sec:Mask}).

Therefore, we implicitly set the normalization of the ICC halo component by generating image realizations with an appropriate number of ``source'' photons inside the appropriate 68\% containment radius, including those from all components.  For strong sources with weak halos this makes little difference.  For weak sources with strong halos (e.g., those with very hard VHEGR SEDs) this curtails the halo emission appropriately.

Extending the ``source'' region farther begins to rapidly increase the angular size of the region as a result of the broad power-law tails on the Pass 8R2\_V6 ULTRACLEANVETO PSFs.  However, we have verified that extending this to the 95\% containment radius makes little difference to our ability to detect ICC halos.

\subsection{Near-Source Halo Suppression}
A small subset of \Fermi AGN are closer than $\Dpp$, and therefore the assumption that the sources were sufficiently far for the full ICC halo is violated.  It is possible to generate ICC halo models in this case for which the region contributing to the ICC halos is restricted to that between the Earth and the source.  However, for the large-scale, uniform IGMF models the small-angle contributions to the ICC halos are nearly uniformly distributed along the line of sight, enabling a simpler optical depth correction.  That is, we reduce the anticipated halo flux by the energy-dependent factor $1-e^{-D_P/\Dpp}$, where $D_P$ is the proper distance to the VHEGR source.  For the small-scale, tangled IGMF models this over-reduces the contribution for blazar sources arising from the counter-jet; nevertheless, we conservatively adopt the same optical depth correction factor.

\subsection{Time Delays and Duty Cycles}
Generally, the contributions to the ICC halos at different positions on the sky are not contemporaneous -- there is a delay between ICC halo gamma rays produced along the line of sight and those off.  The typical delay times are geometric in nature and therefore correlated with the angular diameter distance $D_A$ from the central gamma-ray source:
\begin{equation}
  \delta t \approx \frac{D_A^2}{2\Dpp c} \theta^2\,.
\end{equation}
where $\theta$ is the angular size of observed halo.
Therefore, the magnitude of the delay anticipated is limited by size of the ICC halo.  For ICC halos from gamma-ray blazars the ICC halo is limited by both the size of the magnetic field deflections and the width of the gamma-ray jet.  For the latter this gives $\theta\lesssim \theta_{\rm jet} \Dpp/D_A$, hence conservatively
\begin{equation}
  \delta t \lesssim \frac{\Dpp}{2c} \theta_{\rm jet}^2\,.
\end{equation}
For a typical $\theta_{\rm jet}\approx3^\circ$ and $\Dpp\approx300~\Mpc$ at $z=0.3$ for a TeV VHEGR this gives $\delta t \lesssim 10^6~\yr$.  While considerably larger than the present observing time, this is comfortably short in comparison to the typical radio duty cycles of a few times $10^7~\yr$ to a few times $10^8~\yr$ \citep{1987MNRAS.225....1A,2005ApJ...628..629N,2005Natur.433...45M,2008MNRAS.388..625S}, suggesting that the current gamma-ray flux is indicative of that responsible for a putative ICC halo.

\section{The \Fermi Gamma-ray Blazar Targets} \label{sec:sample}
Here we summarize a variety of essential properties of the observed \Fermi sample of bright, nearby gamma-ray blazars, i.e., objects likely to have detectable ICC halos.  Among these are the intrinsic source parameters, e.g., observed flux, redshift, etc., as well as extrinsic source context, e.g., local gamma-ray background, PSF, and the putative IGMF.  Some of these can be specified for each source explicitly, others must be determined probabilistically.  The list of targets that met all requirements within the \Fermi Pass 8R2\_V6 ULTRACLEANVETO class is presented in Table~\ref{tab:sample}.

Because ICC halos are essentially reprocessed VHEGR emission, they exist solely around VHEGR-bright objects.  Therefore, the goal of initial source class identification is to estimate VHEGR brightness of individual gamma-ray AGN.  To do this we exploit the 2FHL, which is optimized for the detection of objects above 50~GeV \citep{2FHL}.  We further restrict our attention to objects that also appear in the 3LAC \citep{3LAC} and have a measured redshift, yielding 122 objects.  These are dominated by BL Lac-like (BLL) objects, as opposed to the flat-spectrum radio quasars that compose the vast majority of the remainder of the \Fermi AGN population.  This is consistent with with the strong correlation between AGN type and spectral hardness, for which reason BLLs typically dominate at high energies.

The \Fermi PSF provides a direct lower-limit on the size of any detectable ICC halo, both by smearing the anisotropic structure and through contamination from the much brighter direct emission from the source.  Beyond $z=1$ the angular size of the region over which pairs are efficiently produced, i.e., the angular size of $\Dpp$, is near the \Fermi PSF.  A more stringent limit comes from the typical deflections in a large-scale IGMF:
\begin{equation}
\Delta\alpha_{\rm def}
  \approx
  \frac{\omega_B t_{\rm IC}}{\gamma^2}
  \approx
  \circp{1}{5} \left(\frac{B}{10^{-16}~\G}\right) \left(\frac{E_{\rm IC}}{10~\GeV}\right)^{-1} \left(1+z\right)^{-4}
  \label{eq:alphadef}
\end{equation}
where $\omega_B\equiv eB/m_{\rm e} c$ and $t_{\rm IC}\equiv3m_{\rm e} c/4\sigma_T u_s\approx2.4\times10^{12}(1+z)^{-4}~\yr$, which sets the size of the ICC halos generated by emission beamed towards us.  Noting that the IGMF strength today is reshifted relative to that at high $z$, for a fixed current IGMF magnitude $\Delta\alpha_{\rm def}\propto(1+z)^{-2}$ (see Section \ref{sec:BSO}).  Thus for a current $10^{-16}~\G$ IGMF, when $z>0.6$ the halo size is typically smaller than $\circp{0}{6}$, and hence likely to be confused with the central source.  This is a moderately strong function of the IGMF strength and ICC halo gamma-ray energy, growing to $z\lesssim 4$ for a present-day $10^{-15}~\G$ IGMF or 1~GeV ICC halo photons.  However, in those cases the halo size is limited by the jet width (Section \ref{sec:angles}).  For this reason we impose a limit of $z<0.5$, yielding a set of 84 sources.

The above comprise the intrinsic source selection criteria: appearance in the 2FHL and 3LAC with source identification and known redshifts.  

\subsection{Intrinsic SEDs} \label{sec:SEDs}
The tabulated SEDs in neither the 3FGL \citep[the parent catalog of the 3LAC][]{3FGL} nor 2FHL are good estimators of the {\em intrinsic} TeV brightness for at least two reasons.  First, curvature in the SED within the energy bands for which the photon spectral indexes are reported in the 3FGL (100~MeV-100~GeV) and 2FHL (50~GeV-2~TeV) makes any extrapolation to the VHEGR band of interest, 1~TeV-10~TeV, highly uncertain.  Second, absorption on the EBL for sources with $z>0.1$ can have a substantial impact on the SED above 100~GeV, rendering the {\em observed} VHEGR SEDs poor estimators of the {\em intrinsic} VHEGR SEDs.

For a handful of sources there exist reported deabsorbed SEDs from air Cherenkov telescopes (e.g., MAGIC, VERITAS, H.E.S.S.).  However, these suffer from a number of additional limitations.  Typically they provide measurements over a very limited temporal window of the highly-variable emission from gamma-ray bright blazars, and hence are often poor estimators of the time-averaged fluence over long periods.  Additionally, the deabsoprtion prescriptions vary substantially among sources and thus they do not provide a homogeneous class of SED estimates.  Finally, for all but the brightest sources the reported Cherenkov telescope SEDs are limited to below the VHEGR energy band of interest, producing the same uncertainties that arise from the 2FHL and 3FGL.  They do, however, indicate the degree of variability we may expect over decadal timescales.

Therefore, we independently generate composite SEDs by collating the 3FGL and 2FHL band-specific energy flux measurements for each of the 84 common sources with redshifts below 0.5 as described in Appendix \ref{app:SEDs}.  That is, gamma-ray SEDs were produced by compiling the 0.1-0.3~GeV, 0.3-1~GeV, 1-3~GeV, 3-10~GeV, and 10-100~GeV flux measurements reported in the 3FGL and 50-171~GeV, 171-585~GeV, and 585-2000~GeV reported in the 2FHL. The former (3FGL) are averaged over 4 years (Aug 2008-Jul 2012) and the latter (2FHL) are averaged over 7 years (Aug 2008-Aug 2015), and thus present nearly contemporaneous ranges extending over many years.

\begin{figure}
  \begin{center}
    \includegraphics[width=\columnwidth]{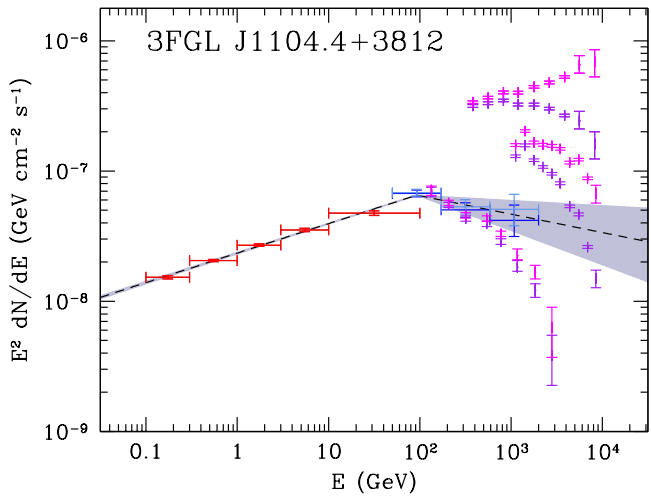}
  \end{center}
  \caption{Composite SED for Mkn 421 taken from the 3FGL (red) and 2FHL (dark blue) catalogs and their deabsorbed counterparts (orange and light blue, respectively, though the former are indistinguishable from the observed values).  A broken power-law fit is shown by the dashed line, the grey band denotes the uncertainty in the SED reconstruction.  For comparison VHEGR observations by ground-based air Cherenkov telescopes (Whipple~observations of the high state in 2000/2001, H.E.S.S.~observations in 2004, and MAGIC observations in 2004/2005, top to bottom, adapted from Figure 10 of \citealt{2007ApJ...663..125A}) are shown in purple (observed) and magenta (deabsorbed).} \label{fig:SEDex}
\end{figure}

These were deabsorbed using the pair-creation optical depth given by
\begin{equation}
  \tau_{\rm obs}(E,z)
  =
  \int_0^z dz' \frac{dD_P/dz}{\Dpp[E(1+z),z]} \,,
\label{eq:tauobs}
\end{equation}
where $D_P$ is the proper distance to the source, evaluated at the geometric center of the energy bin, i.e. we set 
\begin{equation}
  F_{E,\rm deabs} = F_{E,\rm obs} e^{\tau_{\rm obs}(E,z)}\,,
\label{eq:deabs}
\end{equation}
where $F_{E,\rm obs}$ is the specific energy flux.  Both the observed and deabsorbed SED for Mkn 421 are shown in Figure~\ref{fig:SEDex}, with the remainder of the sources used here (i.e., those listed in Table~\ref{tab:sample}) shown in Appendix \ref{app:SEDs}.

The deabsorbed SEDs are generally well fit by a broken power-law SED, defined by a normalization, a pivot energy $E_p$, and photon spectral indexes above and below $E_p$, $\Gamma_h$ and $\Gamma_l$, respectively:
\begin{equation}
  \frac{dN}{dE}
  =
  f_0
  \begin{cases}
    (E/E_p)^{-\Gamma_l} & E\le E_p\\
    (E/E_p)^{-\Gamma_h} & E>E_p\,.\\
  \end{cases}
\end{equation}
A maximum-likelihood fit of the broken-power-law SED model was performed to each candidate source.  The result is also shown by the dashed line in Figure~\ref{fig:SEDex} for Mkn 421.  We have visually verified that small variations in the assumed initial starting point results in negligible variations in the final fit parameter values (though large variations can result in erroneous fit results).

Two classes of qualitatively different SEDs were found from the deabsorbed spectral fits: sources with spectral breaks that are convex ($\Gamma_h>\Gamma_l$) and those that are concave ($\Gamma_h<\Gamma_l$).  The former are consistent with the expectation from single-zone inverse-Compton models of VHEGR sources \citep{Ghisellini:1998,Fermi_Blazafr_SED,2FHL}.  The latter suggest the need for an additional spectral component, either due to an additional comptonizing population or alternative emission source \citep{2013ApJ...768...54B, 2015MNRAS.448..910C, 2016A&A...588A.110Z}.

Uncertainties on the fit parameters were obtained via a Monte Carlo analysis.  Trial fluctuations in the fit parameters were constructed from normal distributions with standard deviations given by the Fisher matrix error estimates.  An estimate of the allowed range was obtained by taking the collection of parameter values for which the log-likelihood, i.e., $\chi^2$, increased by unity, shown by the gray bands in Figure~\ref{fig:SEDex}.  This was especially important for sources with only upper limits at high energies, and thus for which the uncertainty in the high-energy SEDs were highly asymmetric.

Generally, the normalization, $E_p$, and $\Gamma_l$ were tightly constrained; this is in part a selection effect as each object is a well-characterized \Fermi source.  Thus in our set of bright, nearby gamma-ray bright AGN we fix these to their observed (normalization) or fitted ($E_p$ and $\Gamma_l$) values.
In contrast, $\Gamma_h$ is considerably more uncertain, often as a result of a high $E_p$ and larger uncertainties or upper limits on the intrinsic high-energy flux estimates.  Therefore, for the purpose of generating ICC halo realizations we stochastically choose $\Gamma_h$ over the permitted range.  Because this is typically asymmetric, permitting either much smaller or larger values of $\Gamma_h$, we assume that the probability of a given $\Gamma_h$ is well approximated by two one-sided normal distributions centered at the best fit value with standard deviations set by the range obtained by the Monte Carlo procedure above and below.

Key intrinsic target parameters that enter the generation of mock realizations of the sample are the number of observed source photons, $z$, and the source intrinsic SED fit parameters.  These are listed in Table~\ref{tab:sample} for the \Fermi targets that are used in this paper (see also Section \ref{sec:opt}).

\begin{figure*}
  \begin{center}
    \includegraphics[width=0.32\textwidth]{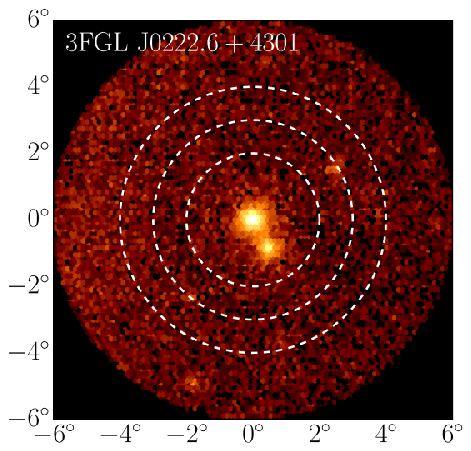}
    \includegraphics[width=0.32\textwidth]{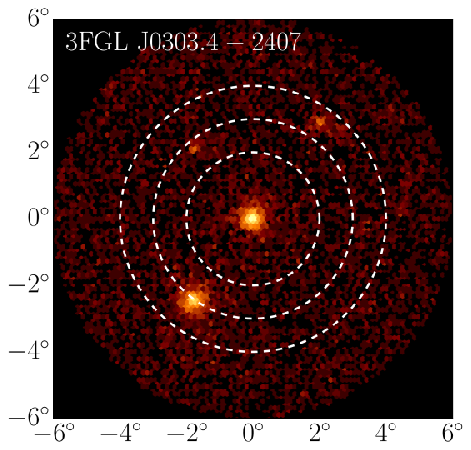}
    \includegraphics[width=0.32\textwidth]{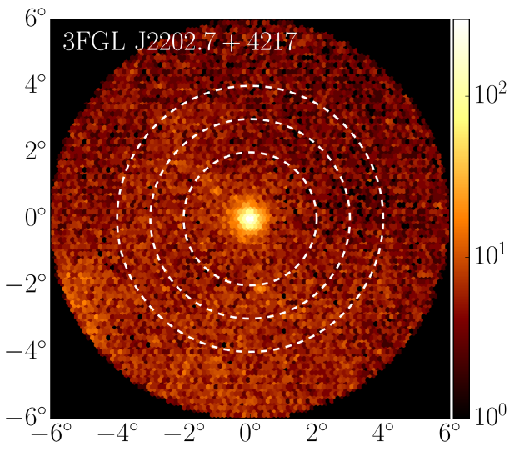}
  \end{center}
  \caption{Fluence maps of example excluded (due to neighboring sources, left), restricted (center), and ideal (right) \Fermi sources.  Color indicates the number of photons within a given pixel. Dashed circles have radii of $2^\circ$, $3^\circ$, and $4^\circ$. }\label{fig:imgex}
\end{figure*}

\subsection{Local Gamma-ray Neighborhood}
In addition to the intrinsic source requirements, the diffuse, large-scale nature of the ICC halos places constraints on the neighborhood of targets.  While these are essentially limits on potential contaminating features, for the purpose of target selection this reduces neighboring sources and large-scale background gradients.

Even weak neighboring sources can produce a large bias in the angular power spectra.  While we defer a characterization of this signal to Section \ref{sec:contam}, an initial target-list cut was made to remove all sources with bright neighbors within $2^\circ$.  Beyond $2^\circ$ neighbors were permitted, though the area over which the power spectrum analysis was performed was restricted to prevent contamination.  This is illustrated in Figure~\ref{fig:imgex}, which shows examples of excluded, restricted, and ideal sources.

The gamma-ray background varies substantially from source to source as a result of the different sky location.  This is dominated by the diffuse Galactic component, and becomes noticeably worse at low Galactic latitudes, where it imparts substantial gradients in the gamma-ray counts.  Rather than attempting to model this component we exclude sources with strong background gradients visible over scales of $10^\circ$ (typically corresponding to $|l|<18^\circ$) and make an estimate of the background photon density for each target source individually.  In practice, the presence of a background gradient appears dominantly in the dipolar power, and thus is distinguishable from the bipolar signals of interest (see Section \ref{sec:contam}).

Therefore, the above comprise the extrinsic source selection criteria: no neighbors within $2^\circ$, restricting from $4^\circ$ as necessary, and no large-scale gradients in the background flux. Key extrinsic target parameters that enter the generation of mock realizations of the sample are the maximum non-contaminated angular radius, $R_{\rm max}$, and the total number of front- and back-converted photons within the permitted region due to the source and background.  There are 27 objects that are sufficiently isolated and satisfy the intrinsic source parameter selection criteria, which are shown in Tables \ref{tab:sample2} and \ref{tab:fits}. Final selection of the \textit{Fermi} targets listed in Table~\ref{tab:sample} will be described in section~\ref{sec:opt}.

\newpage
\subsection{Jet Opening and Viewing Angles} \label{sec:angles}
\begin{figure}
  \begin{center}
    \includegraphics[width=\columnwidth]{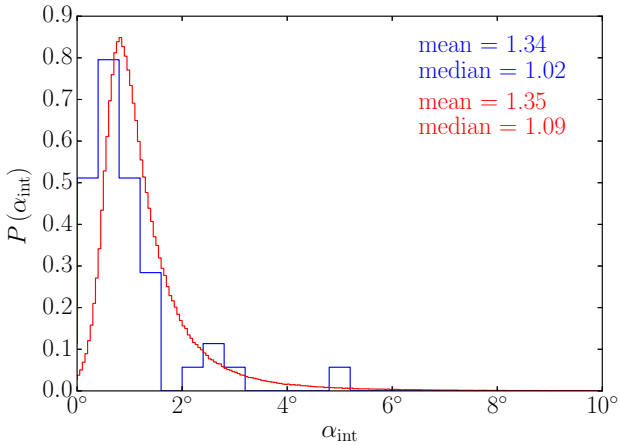}
  \end{center}
  \caption{Intrinsic parsec-scale radio opening angle (FWHM) distribution for \Fermi sources in the MOJAVE data base \citep[adapted from][]{Push_etal:09}.  Shown by the red line is the fitted generalized $\Lambda$ distribution from which we draw intrinsic parsec-scale radio opening angles.} \label{fig:aint}
\end{figure}

\begin{figure}
  \begin{center}
    \includegraphics[width=\columnwidth]{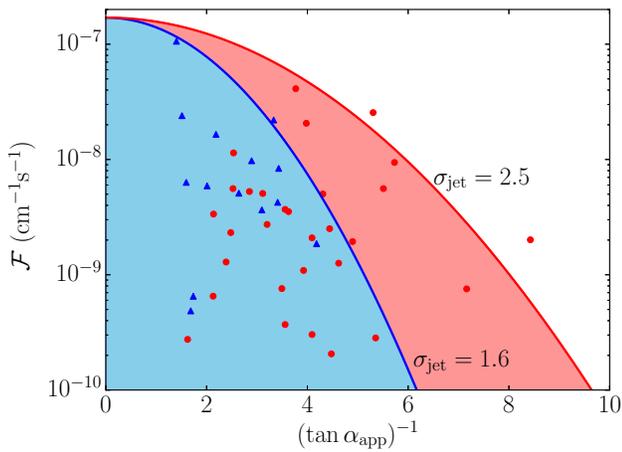}
  \end{center}
  \caption{Flux vs. viewing angle in units of radio jet width, approximated by $(\tan\alpha_{\rm app})^{-1}$ where $\alpha_{\rm app}$ is the apparent parsec-scale radio opening angle for hard ($\Gamma_l<2.1$, blue) and soft ($\Gamma_l>2.1$, red) \Fermi sources in the MOJAVE data base.  Fits to the upper-envelope assuming a Gaussian jet are shown by the light-blue filled and light-red filled regions.} \label{fig:Faapp}
\end{figure}

For most \Fermi blazars the viewing and intrinsic opening angles are poorly constrained.  However, the subset of objects that also appear in the sample monitored by the MOJAVE group provides some guidance on the parsec-scale {\em radio} opening and viewing angles \citep{Push_etal:09}.  Generally, \Fermi blazars exhibit radio jets that are intrinsically more narrow than the gamma-ray dim population.  Figure~\ref{fig:aint}, adapted from \citet{Push_etal:09}, shows the distribution of intrinsic radio opening angles (full-width half-max, FWHM), which peaks near the median of $\circp{1}{0}$ and falls rapidly thereafter.  This is well-fit by a generalized $\lambda$ distribution with parameters given in Appendix \ref{app:Lambda}, shown in the figure by the red line.

Interpreting the implications for the gamma-ray jet opening angles is complicated for a number of reasons.  First, obtaining the radio opening angles is itself challenging.  While measuring the apparent (projected) opening angle is straightforward, determining the intrinsic opening angle typically requires kinematic information obtained from multiple widely-separated epochs of imaging observations.  Second, the radio and gamma-ray opening angles need not be the same, and are generally different.  Moreover, the current uncertainty in the gamma-ray emission mechanism precludes using spectral information to relate the two directly.

To assess the relationship between the two we begin with the following assumptions:
\begin{enumerate}
\item The flux distribution of \Fermi sources is limited from above, i.e., they exhibit a maximum intrinsic luminosity.
\item The intrinsic opening angle of the gamma-ray and radio jets are proportional, i.e., broad radio jets are broad gamma-ray jets and vice versa.
\item The gamma-ray jet is structured as a Gaussian with a source-dependent standard deviation, $\theta_{\rm jet}$.  That is, the gamma-ray flux within the jet observed at an angle $\Theta$ is given by
  \begin{equation}
    \mathcal{F}=\mathcal{F}_0 e^{-\Theta^2/2\theta_{\rm jet}^2}\,.
  \end{equation}
\end{enumerate}
The first is a good approximation in practice since \Fermi only sees the bright end of the hard gamma-ray blazar luminosity function.  The second is natural given that the overwhelming majority of gamma-ray bright AGN are blazars \citep{2LAC,3LAC}.  The quantitative consequence is that the gamma-ray jet scale, $\theta_{\rm jet}$, is related to the intrinsic radio jet FWHM, $\alpha_{\rm int}$, by some proportionality constant:
\begin{equation}
  \theta_{\rm jet} = \sigma_{\rm jet} \alpha_{\rm int}\,.
\end{equation}
The third is already made in the gamma-ray halo models described in \citet{BowTiesI} and employed here.  Importantly, note that the assumed structure is effectively a condition on the {\em apparent} jet opening angle; while generally $\tan \alpha_{\rm app} = \tan \alpha_{\rm int} / \sin\Theta$, for small radio opening and viewing angles this is approximately $\tan\alpha_{\rm app}\approx\alpha_{\rm int}/\Theta$, and hence the condition imposed by the third can be cast in terms of the {\em apparent} opening angle alone:
\begin{equation}
  \mathcal{F}=\mathcal{F}_0 e^{-1/2\sigma_{\rm jet}^2 \tan^2\alpha_{\rm app}}\,.
\end{equation}
This has two consequences.  First, the number of \Fermi blazars for which $\alpha_{\rm app}$ is known is comparatively large.  Second, because $\alpha_{\rm app}$ is directly measured during a single epoch of radio observations it is much better known than $\alpha_{\rm int}$ or $\Theta$.

While the value of $\mathcal{F}_0$ is unknown for any given source, the presence of an upper limit implies that \Fermi sources should populate the region in the $\mathcal{F}$-$\alpha_{\rm app}^{-1}$ plane defined by some $\mathcal{F}_{0, \rm max}$.   Show in Figure~\ref{fig:Faapp}, this is the clearly the case -- sources with large $\alpha_{\rm app}^{-1}$ have systematically lower fluxes, falling in a manner consistent with the Gaussian dependence posited.  This presents a direct way in which to measure $\sigma_{\rm jet}$ by fitting the envelope of points in the $\mathcal{F}$-$\alpha_{\rm app}^{1}$ plane.

Interestingly, the value obtained depends strongly on spectral hardness.  Softer sources ($\Gamma_l>2.1$) have systematically broader gamma-jets, with $\sigma_{\rm jet}\approx 2.5$; the hard sources of interest here ($\Gamma_h<2.1$) have 45\% narrower jets, with $\sigma_{\rm jet}\approx 1.6$.  It is this latter relationship we adopt.

Note that this implies that the gamma-ray emission is beamed over a substantially larger angle than subtended by the parsec-scale radio jet.  The ratio of the FWHMs of the gamma-ray and parsec-scale radio jets is $(8\ln2)^{1/2}\sigma_{\rm jet}=3.77$, diffusing the emission over a solid-angle nearly 14 times larger.
This suggests that the beaming of the gamma-ray emission is similar, qualitatively, to the presumed jet structure near its base, i.e., within the collimating region.

This does not mean, however, that we anticipate large beaming corrections to the apparent gamma-ray flux or that there should be a large population of non-blazar gamma-ray sources observed by \Fermi.  Generally, parsec-scale jets are more tightly collimated in comparison to the typical beaming angle of the radio {\em emission}, with
\begin{equation}
  \alpha_{\rm int}\approx\frac{0.3}{\Gamma_{\rm jet}}\,,
\end{equation}
where $\Gamma_{\rm jet}$ is the jet Lorentz factor (not to be confused with a photon spectral index); this continues to hold for the \Fermi subset of MOJAVE sources \citep{Push_etal:09}.  That is, the typical angular scale over which the radio emission is beamed, $\Gamma_{\rm jet}^{-1}$, is itself three times the parsec-scale jet FHWM.  Comparing the gamma-ray and radio beaming angle gives $\theta_{\rm jet}\Gamma_{\rm jet}\approx0.48$ for the hard gamma-ray blazars.  That is, for these sources the gamma-ray emission is moderately more beamed than the radio emission.  Despite having a $\theta_{\rm jet}\Gamma_{\rm jet}$ that is roughly 40\% larger, this remains true for the soft gamma-ray blazars -- the gamma-ray emission is more beamed that the radio emission.

Because we do not have an a priori estimate of either $\theta_{\rm jet}$ or $\Theta$ for any particular source, for the purpose of creating ICC halo realizations we randomly generate values using the following procedure:
\begin{enumerate}
\item We select a value of $\alpha_{\rm int}$ from the distribution shown in Figure~\ref{fig:aint} and described in Appendix \ref{app:Lambda}.
\item Based on $\alpha_{\rm int}$ we estimate the relativistic beaming angle to be $\Gamma_{\rm jet}^{-1}\approx\alpha_{\rm int}/0.3$.  To appear as a blazar we must be viewing the source within this angle, and thus choose $\Theta$ from an isotropic distribution within $\Gamma_{\rm jet}^{-1}$.
\item We generate $\theta_{\rm jet}=\sigma_{\rm jet}\alpha_{\rm int}$ with $\sigma_{\rm jet}=1.6$.
\end{enumerate}

\subsection{IGMF Strength and Orientation} \label{sec:BSO}
In the presence of a large-scale IGMF the ICC halo images depend upon the assumed strength and orientation of the IGMF.  Without any prior knowledge of either the strength or orientation of the IGMF we consider variations in both.  In the case of the former we set the strength of the IGMF today, $B_0$, and assign the local strength at the source to be the appropriately redshifted value, i.e., the local magnetic field strength is
\begin{equation}
  B = B_0 (1+z)^2\,.
  \label{eq:fixedB0}
\end{equation}
For a given analysis we assume that all sources have the same value of $B_0$, assessing the detectability of the ICC halos as a function of field strength.  Because we restrict our attention to $z<0.5$ this has at most a factor of two impact on the assumed $B$.

Realizations for the orientation of the field is generated from an isotropic distribution for each image independently.  That is, while in this limit we assert that the field is coherent over scales large in comparison with the gamma-ray jet widths, we permit large variations over the distances between \Fermi AGN in our sample.  This is consistent with a picture in which the ICC halos are produced in cosmological voids in which the IGMF has been imprinted from early times.  We note that this is rather pessimistic -- correlations in the orientations of ICC halos from neighboring VHEGR sources would permit coherent stacking, which we ignore here.

\section{Asymmetric Signatures of Halos} \label{sec:signatures}

As introduced in Section \ref{sec:cartoon} we exploit the anticipated structure in the ICC halos by constructing a statistical measure of the anisotropy.  To do this we focus on the angular power spectra defined in terms of an event-specific polar angle, $\theta_j$, about the source center, given in Equation \ref{eq:PS}.  This is a natural statistic for the reasons discussed in Section \ref{sec:cartoon}: it is sensitive in particular to anisotropic structure yet independent of an absolute rotation of the source orientation.  Moreover, it is only weakly contaminated by the known systematics in the \Fermi LAT instrument responses and the background source population, and in a way that is easily distinguishable from the ICC halo signal of interest.  There are, however, a number of practical steps, which we discuss here.  These include removing coordinate aberration from the \Fermi images about a source (Section \ref{sec:FF}), identifying the source center robustly (Section \ref{sec:MLCF}), and masking source contamination (Section \ref{sec:Mask}). We then stack the power spectra from multiple sources (Section \ref{sec:stack}) and determine our final optimized source list (Section \ref{sec:opt}).

\subsection{Converting to Locally Euclidean Coordinates} \label{sec:FF}
At large latitudes coordinate aberration induces angular structure in the images of even cylindrically symmetric sources.  Therefore, we must first approximately flatten the gamma-ray images from \Fermi, i.e., transform to a set of flat coordinates.  Because we are interested only in the angular structure of the gamma-ray distribution about the central source $(\alpha_s,\delta_s)$, i.e., we are not concerned with its radial structure, fully flattening the images is unnecessary.  Rather it is sufficient to remove the angular distortion.

To do this we begin with the gamma-ray positions in equatorial coordinates $(\alpha_j,\delta_j)$ and perform a rotation to align the reported source position along the polar axis.  That is on the unit sphere we set the gamma-ray positions to
\begin{equation}
  \begin{aligned}
    x_j &= \sin\delta_s \cos\delta_j \cos(\alpha_j-\alpha_s) - \cos\delta_s \sin\delta_j\\
    y_j &= \cos\delta_j \sin(\alpha_j-\alpha_s)\\
    z_j &= \cos\delta_s \cos\delta_j \cos(\alpha_j-\alpha_s) + \sin\delta_s \sin\delta_j\,,\\
  \end{aligned}
\end{equation}
from which we obtain the angular positions by projecting along $\hat{z}$:
\begin{equation}
  \alpha'_j = \frac{x_j \cos^{-1}(z_j)}{\sqrt{x_j^2+y_j^2}}
  \quad\text{and}\quad
  \delta'_j = \frac{y_j \cos^{-1}(z_j)}{\sqrt{x_j^2+y_j^2}} \,.
\end{equation}

\subsection{Maximum-Likelihood Center Finding} \label{sec:MLCF}
The source positions reported by \Fermi provide an excellent initial estimate of the source locations.  However, these are obtained from a different set of event reconstructions (Pass 7R\_V15 SOURCE) from those employed here (Pass 8R2\_V6 ULTRACLEANVETO).  Moreover, even potentially small offsets provide an obvious systematic uncertainty that will generate spurious power at $m=1$.  While we discuss how such an error may be naturally identified in the structure of the angular power spectra directly in Section \ref{sec:contam}, we also make an effort to mitigate this directly via an improved estimate of the source location. This also has the effect of treating the mock images in the same fashion as the real data -- both will have small offsets in the source location based on the particular photon realization.

To identify the source location we make a maximum-likelihood estimate of the source location, $(\alpha'_s,\delta'_s)$.  The likelihood was taken to be composed of a Gaussian source of known size on top of a uniform background within a specified angular size on the sky.  The result is an estimate for the source location and ratio of the source-background fluences.  Additional details of the method can be found in Appendix \ref{app:center}.

\begin{figure}
  \centering
  \includegraphics[scale=0.4]{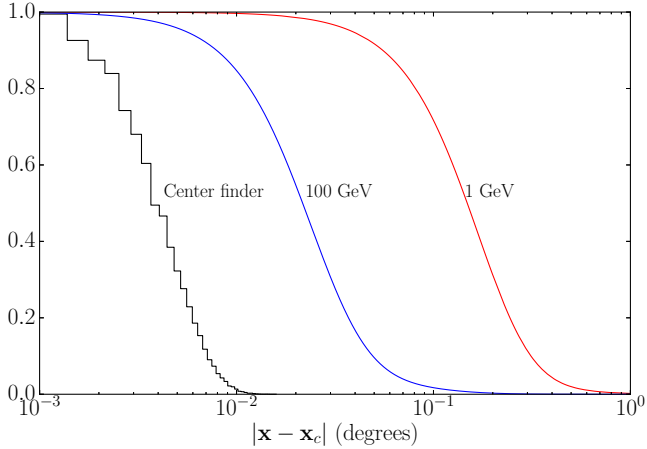}
  \caption{Distribution of displacements between the true and reconstructed centers for a number of mock sources with typical fluences in the 1~GeV-100~GeV band.  For reference the Pass 8R2\_V6 ULTRACLEANVETO PSFs for front-converted events at 1~GeV (red) and 100~GeV (blue) are shown.  In all cases the reconstructed center locations are at least an order of magnitude better than the typical PSF width.}\label{fig:center}
\end{figure}

While the \Fermi PSF is neither Gaussian nor independent of energy, and thus our likelihood does not formally describe the \Fermi response for a point source, both have proven to be adequate approximations for our purpose, producing highly accurate source location estimates.  We verified this by generating mock point source images following the algorithm described in Section \ref{sec:mocks}, which utilizes the fully energy-dependent \Fermi Pass 8R2\_V6 PSF, and generating source location estimates.  The distribution of offsets is compared to the \Fermi Pass 8R2\_V6 PSFs at low and high energies in Figure~\ref{fig:center}.  In all cases our estimate is roughly an order of magnitude better than the characteristic width of the \Fermi Pass 8R2\_V6, a reflection of the large number of photons in the gamma-ray maps of the bright sources of interest.

We generate a final set of source-centered positions
\begin{equation}
  \alpha_j'' = \alpha_j'-\alpha'_s
  \quad\text{and}\quad
  \delta_j'' = \delta_j'-\delta'_s\,,
\end{equation}
which we convert into polar coordinates:
\begin{equation}
  r_j \equiv \sqrt{{\alpha_j''}^2+{\delta_j''}^2}
  \quad\text{and}\quad
  \theta_j = \tan^{-1}(\delta_j''/\alpha_j'')\,.
\end{equation}
It is these $\theta_j$ that enter into Equation (\ref{eq:PS}) to construct the object-specific angular power spectra.

\subsection{Source Masking and Contamination Mitigation} \label{sec:Mask}
The primary sources of noise in the power spectra estimates are the photons from the source and in the background.  Beyond careful source selection, e.g., removing objects with obvious contaminating neighbors or strong background gradients, there is little that can be done regarding the background.  However, this is not true for the source itself.

Any structure on scales comparable to the PSF width will be erased, eliminating the value of photons near the source.  At the same time, the direct photons from a point source contribute dominantly within this region.  Therefore, prior to computing the angular power spectrum for each object we apply an energy dependent mask, excising the region inside the 68\% containment radius of the Pass 8R2\_V6 ULTRACLEANVETO PSF at each energy, independently for the front and back detectors.  This eliminates at once both a region without a significant anisotropy signal and a substantial source of noise.  We assess the impact of variations in the size of the excluded region in Section \ref{sec:6895}, generally finding that it is negligible.

For most objects the ICC halos extend over angular scales that are large in comparison to the mask, rendering the mask moot.  However, for present-day IGMF strengths less than $3\times10^{-17} (1+z)^2~\G$, corresponding to either weak fields or high-$z$, the ICC halo can lie completely within the masked region.\footnote{This limit arises from combining the angular scale implied by a deflection in Equation (\ref{eq:alphadef}) and the cosmological evolution of a fixed strength field, given in Equation (\ref{eq:fixedB0}) assuming that the source is sufficiently close that the angular diameter distance is similar to $\Dpp$.  Thus, we set $\alpha_{\rm def}/2 \approx \circp{0}{2}$, comparable to the typical mask size near 10~GeV, from which we obtain $B_0 = B (1+z)^{-2} = 3\times10^{-17} (1+z)^2$.  Note that the {\em observed} IC gamma-ray energy in Equation (\ref{eq:alphadef}), $E_{\rm IC}$, does not redshift.}  In effect this simply extends the constraint on ICC halo size already imposed by the finite resolution of the LAT marginally.

\subsection{Example Single-Source Power Spectra}

\begin{figure}
  \begin{center}
    \includegraphics[width=\columnwidth]{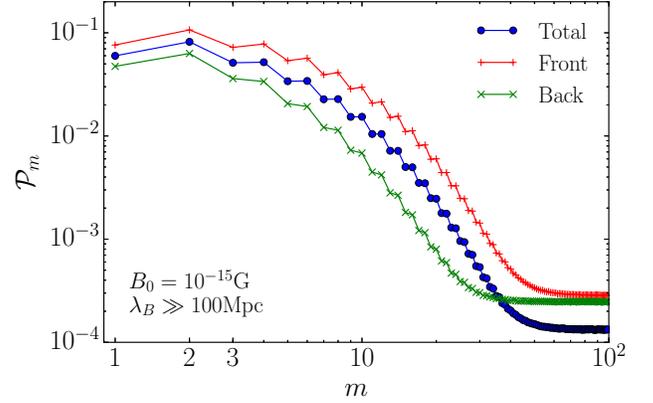}
  \end{center}
  \caption{Median mock power spectra for a bright \Fermi source containing a large-scale, uniform IGMF constructed from front-converted events (red +'s), back-converted events (green x's), and all events (blue circles).  The assumed IGMF strength is $10^{-15}~\G$ and $\Gamma_l = 1.7$, $z=0.3$, $\Theta=5^\circ$, $\theta_{\rm jet}=3^\circ$, $\Gamma_h=2.5$, $E_p=1~\TeV$, with on-axis fluence of $5000~\ph$, a background photon density of $50~\ph/\deg^2$, and a magnetic field oriented $60^\circ$ to the line of sight and $170^\circ$ from the horizontal axis.  At large $m$ all cases asymptote to the Poisson limit, $\P_m\approx N^{-1}$. } \label{fig:FBcomp}
\end{figure}

\begin{figure}
  \begin{center}
    \includegraphics[width=\columnwidth]{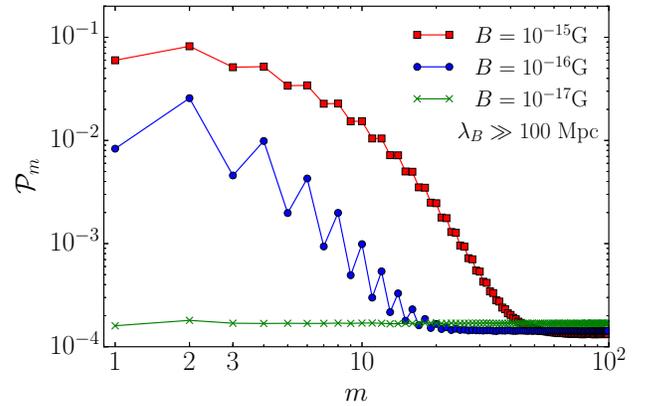}
  \end{center}
  \caption{Median mock power spectra for a bright \Fermi source assuming different IGMF structures.  Power spectra for a large-scale, uniform IGMF with strength $10^{-15}~\G$ (red squares), $10^{-16}~\G$ (blue circles), $10^{-17}~\G$ (green crosses) are shown; decreasing field strength results in a smaller low-$m$ power excess. In all cases $\Gamma_l = 1.7$, $z=0.3$, $\Theta=5^\circ$, $\theta_{\rm jet}=3^\circ$, $\Gamma_h=2.5$, $E_p=1~\TeV$, with on-axis fluence $5000~\ph$, a background photon density of $50~\ph/\deg^2$, and a magnetic field oriented $60^\circ$ to the line of sight and $170^\circ$ from the horizontal axis.  } \label{fig:ICCcomp_hom}
\end{figure}

\begin{figure}
  \begin{center}
    \includegraphics[width=\columnwidth]{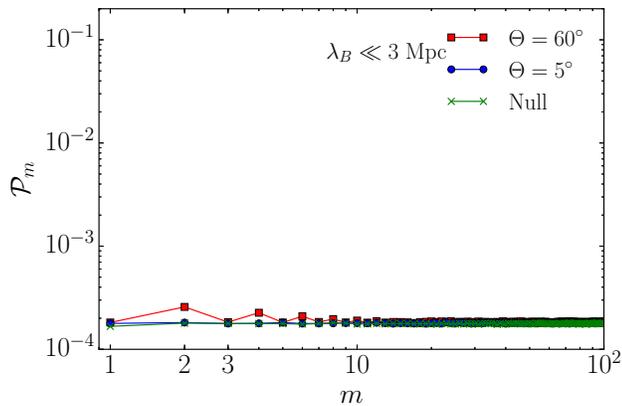}
  \end{center}
  \caption{Median mock power spectra for a bright \Fermi source assuming a small-scale, tangled IGMF with $B_0\gtrsim10^{-14}~\G$ for objects seen at large ($\Theta=60^\circ$, red squares) and small ($\Theta=5^\circ$, blue circles) viewing angles.  The power spectrum in the absence of an ICC halo is shown form comparison in green crosses. In all cases $\Gamma_l = 1.7$, $z=0.3$, $\theta_{\rm jet}=3^\circ$, $\Gamma_h=2.5$, $E_p=1~\TeV$, with on-axis fluence $5000~\ph$ and a background photon density of $50~\ph/\deg^2$.} \label{fig:ICCcomp_iso}
\end{figure}

\begin{figure*}
  \begin{center}
    \includegraphics[width=\columnwidth]{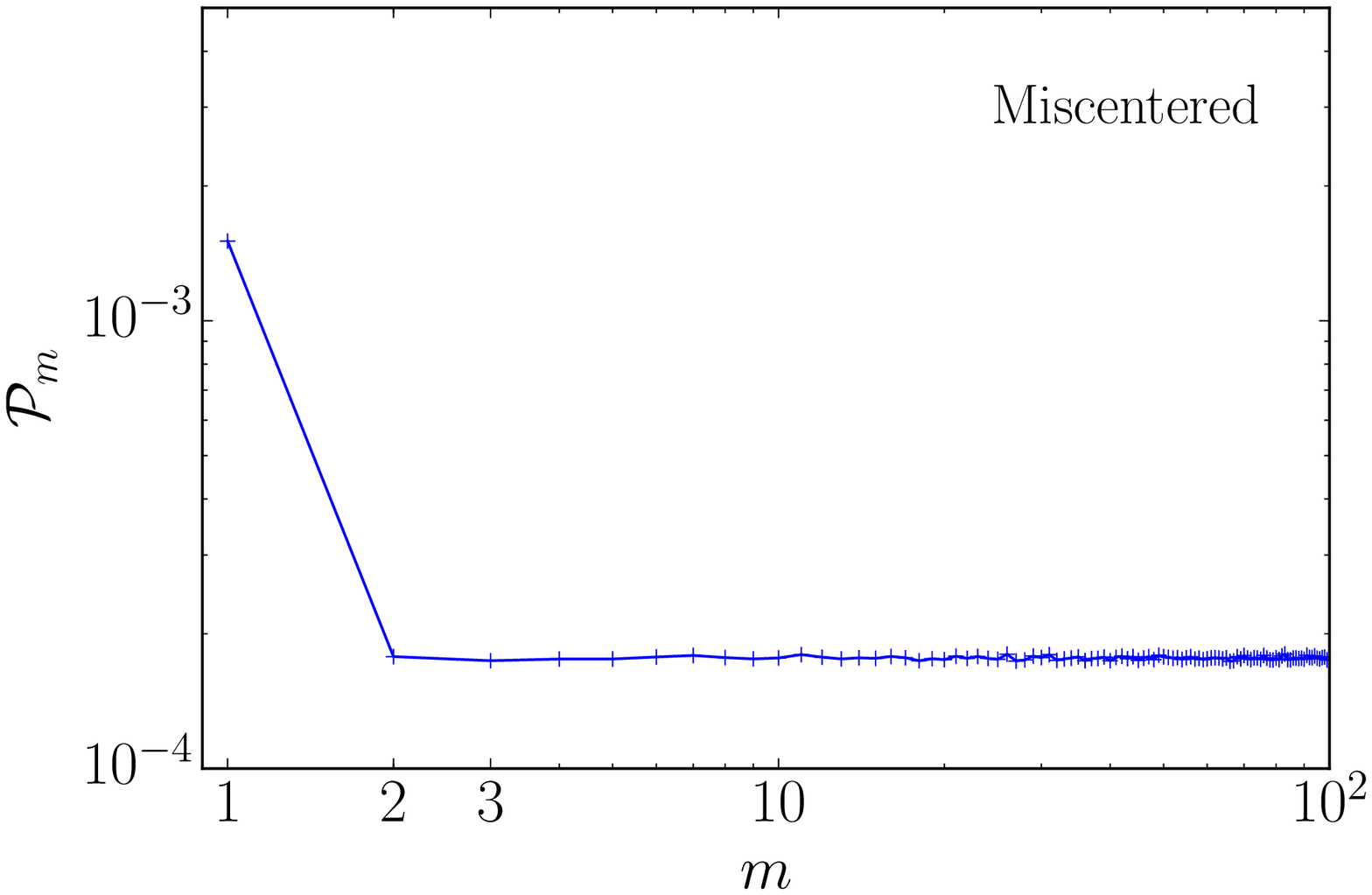}
    \includegraphics[width=\columnwidth]{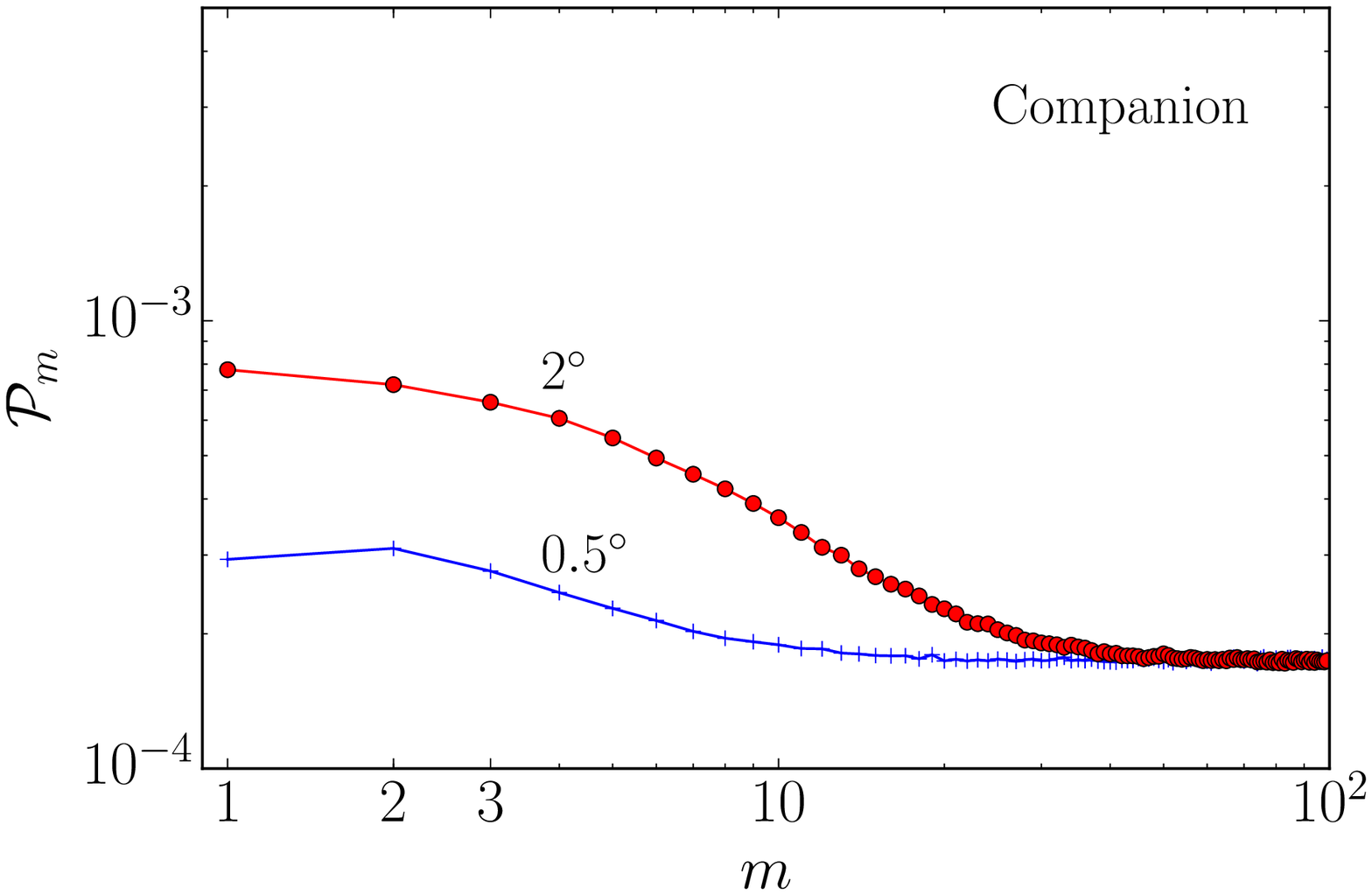}  
    \includegraphics[width=\columnwidth]{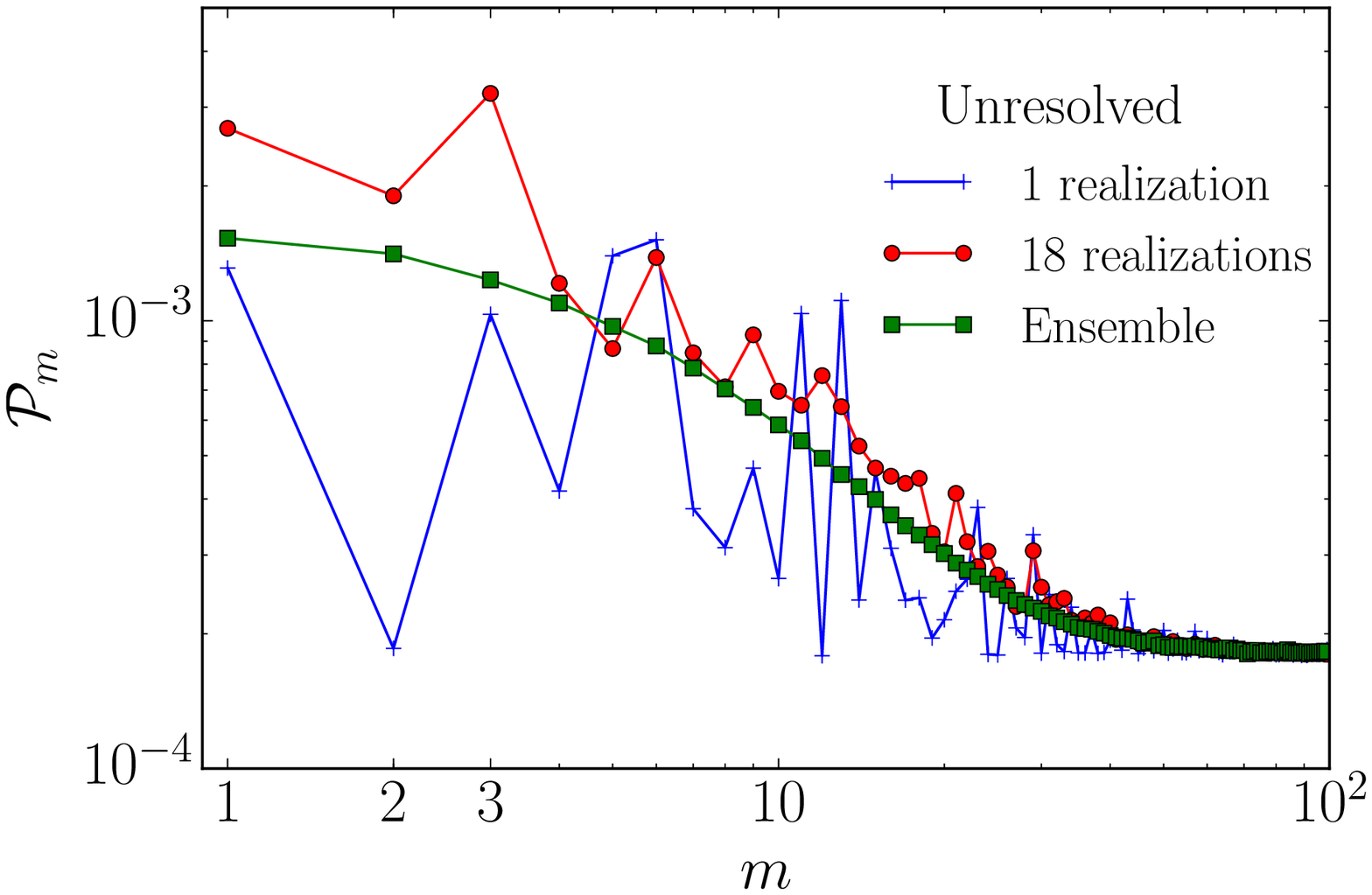}
    \includegraphics[width=\columnwidth]{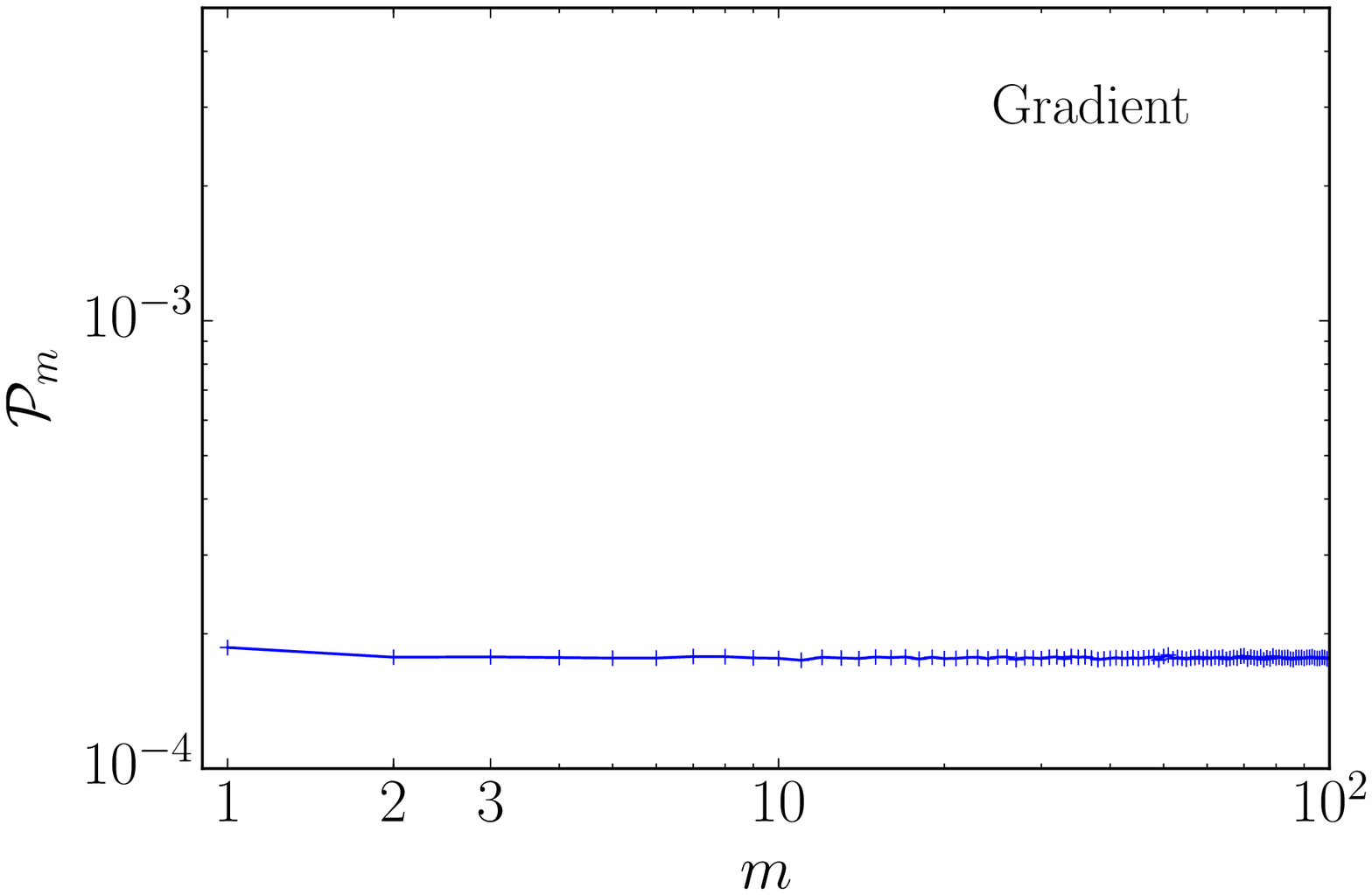}
  \end{center}
  \caption{Median mock power spectra for a bright \Fermi source without an ICC halo but various contaminating systematics.  Top left: a miscentered image, shifted by $\circp{0}{05}$, corresponding to roughly 10\% of the typical width of the 1~GeV Pass 8R2\_V6 ULTRACLEANVETO-front PSF.  Top right: a faint companion (150~ph) that lies well above the \Fermi detection threshold (roughly 5~ph) but below the visual inspection threshold, located at $\circp{0}{5}$ (blue) and $2^\circ$ (red).  Bottom left: a background comprised of many unresolved gamma-ray point sources consistent with the flux distribution of \Fermi point sources for a single realization (blue), 18 realizations (red), and an ensemble of realizations (green).  Bottom right: a weak gradient in the background comparable to that of the visual inspection threshold.  In all cases $\Gamma_l = 1.7$, $z=0.3$, $\Theta=5^\circ$, $\theta_{\rm jet}=3^\circ$, $\Gamma_h=2.5$, $E_p=1~\TeV$, with on-axis fluence $5000~\ph$ and a background photon density of $50~\ph/\deg^2$.} \label{fig:badcomp}
\end{figure*}

Example average power spectra for a single, bright source are shown in Figures \ref{fig:FBcomp}--\ref{fig:badcomp}.  These include both the power spectra anticipated from the various ICC halo models under consideration and those from a variety of potential contaminants.  Analytical computations for approximate cases of potential relevance are also collected in Appendix \ref{app:ExPS}.

\subsubsection{ICC Halos}
Power spectra may be made for events converted separately in the front, back, or entirety of the LAT.  While front-converted events have a smaller PSF, and therefore maintain small-scale power more effectively, the improvement in event statistics arising from the near-doubling of the gamma-ray number when back-converted events are included produces an overall improvement in the ability to identify ICC halos.  This is clearly evident in the comparison between the power at small and large $m$ in Figure~\ref{fig:FBcomp}, which shows the each power spectrum class individually.

In all cases at large $m$ the $\P_m$ reach the Poisson noise limit, producing a characteristic flattening at $\P_m\approx N^{-1}$, indicating the effective number of image photons used.  The key discriminant that provides evidence for ICC halos is the disparity between $\P_2$ and this floor, for which the power spectrum constructed from all events is largest.  Thus, henceforth, we show only the power spectra for the entire event list, i.e., including both front- and back-converted events.

A comparison of the ICC halo signal for different assumptions regarding the IGMF and orientation are shown in Figures \ref{fig:ICCcomp_hom} and \ref{fig:ICCcomp_iso}.  In contrast to the null case, the clear signal for an ICC halo is the large quadrupolar power, i.e., $\P_2$, in comparison to the Poisson limit.  Moreover, the clear oscillatory nature is a signature for the anticipated near bipolar symmetry.  Importantly, this ``sawtooth'' structure is a key systematic diagnostic, differentiating a true ICC halo signal from potential power spectrum contaminants (see the following subsection).

Nevertheless, the magnitude and structure of the power spectrum is strongly sensitive to the assumed IGMF geometry and source viewing angle.  For acute viewing angles, i.e., $\Theta\approx\theta_{\rm jet}$, the large-scale, uniform IGMF models are most significantly distinct from the null case.  In contrast, in this case the small-scale, tangled IGMF is difficult to distinguish as a result of the extreme foreshortening and dilution.  However, for oblique viewing angles, i.e., $\Theta\gg\theta_{\rm jet}$, the small-scale, tangled IGMF models do exhibit noticeable structure in the power spectrum; we will address possibilities for exploiting gamma-ray observations of oblique jets elsewhere.

For the large-scale uniform IGMF models the power excess extends beyond the even multipoles as a consequence of the breaking of the bipolar symmetry by the intrinsic structure of the jet.  When viewed along the jet axis a near-perfect symmetry exists as the positrons and electrons gyrate in opposite directions, generating halo emission on opposite sides of the source.  However, when viewed at angles comparable to the jet width one component is suppressed by the comparative deficit in pair production due to the reduced VHEGR flux.  This is insufficient to remove the characteristic ``sawtooth'' pattern in any case.

Weaker IGMFs produce smaller ICC halo signals, vanishing below $10^{-17}~\G$ as a result of the source mask.  IGMFs that are much stronger than $10^{-14}~\G$ also produce weaker ICC halo signatures in the power spectra as a result of the dilution of the gamma-ray flux due to the multiple gyrations.
Thus, in principle, apart from simple detection, which is the focus of this work, it should be possible to characterize a large-scale IGMF given a measurement of the gamma-ray angular power spectra.

Also visible in Figure~\ref{fig:ICCcomp_hom} is a two-zone analog of a gamma-ray excess at large angular radii from the central source, i.e., the signal described in Figure~\ref{fig:chbin}.  In the stacked angular power spectrum this takes the form of a systematic drop in $\P^{\rm tot}_m$ at large $m$ that systematically grows with increasing $B_0$.  This arises from the ICC halos moving photons from within the central masked region to outside, where they are included in the angular power spectrum estimate, decreasing the Poisson noise limit.  As discussed earlier, exploiting this signal requires knowing the gamma ray background and radial structure of the PSF to high accuracy a priori.

\subsubsection{Contaminants} \label{sec:contam}
Low-multipole angular power can also be produced by features that are independent of ICC halos.  These include systematic errors in the generation of the angular power spectra, angular structure in the \Fermi Pass 8R2\_V6 ULTRACLEANVETO PSF, and unresolved features in the background.  Here we quantitatively consider each of these.  Importantly, none produce the quadrupole-dominated, sawtooth structure in the angular power spectrum characteristic of the bipolar structures associated with the ICC halos.  

As has already been described in Section \ref{sec:MLCF}, the location of the central source may be identified with an accuracy that significantly exceeds the typical PSF width of \Fermi.  Nevertheless, centering errors combined with the large radial gradients in the gamma-ray flux away from the source center produce a natural source of error in the angular power spectrum (see, e.g., Appendix \ref{app:MCGPS}).  To assess the magnitude of this error, we intentionally offset a bright gamma-ray source typical of the \Fermi sample by $\circp{0}{05}$, corresponding to roughly 10\% of the typical width of the 1~GeV Pass 8R2\_V6 ULTRACLEANVETO-front PSF (top).  As seen in the top panel of Figure~\ref{fig:badcomp} this leads to spurious power at low multipoles, dominated by the dipole, falling rapidly with $m$, and joining the typical Poisson noise by $m=2$.  The large dipole-quadrupole power ratio makes this easily distinguishable from a signal attributable to ICC halos.

Unresolved sources within the background surrounding the primary \Fermi AGN also provide a natural source for dipole angular power.  This arises in two instances, the first of which is a single neighboring object just below our exclusion threshold.  Given that the \Fermi detection threshold is 5~ph \citep{1FHL}, this is unlikely to produce a substantial contribution for images comprised of many thousands of events.  Adding a source that is obviously visible (150~ph, intermediate to the companions in the left and center panels of Figure~\ref{fig:imgex}) produces a notable excess of power at low multipoles (top-right panel of Figure~\ref{fig:badcomp}).  The degree of this excess depends on the location of the contaminating neighbor, becoming larger for more distant unresolved sources.  In practice, this is many times over the \Fermi detection threshold and would already be excluded.  Much weaker companions would produce a correspondingly weaker contribution to the power spectrum.  In all cases, as before it is dominated by the dipole component and fails to exhibit the sawtooth morphology.

The second way in which unresolved sources enter is through the origin of the background itself.  At high energies ($>50$~GeV) nearly the entirety of the extragalactic gamma-ray background has been resolved into point sources \citep{50GevBkgnd}.  Extending this to lower energies results in a clustering of photons about the brightest background objects, and therefore low-multiple power arising from a handful of unresolved neighbors\footnote{This is extremely pessimistic.  Even at high latitudes the smooth, Galactic contribution to the gamma-ray background is substantial.  Nevertheless, this gives an extreme estimate of the impact of a highly structured background.}.  In practice, the extended fields about individual targets in Table~\ref{tab:sample} do include known 3FGL sources with up to 40~ph, well above the \Fermi detection threshold.  Thus, shown in the bottom-left panel of Figure~\ref{fig:badcomp} is the angular power spectrum arising from a background comprised of a population of sources below the 40~ph threshold and distributed according to the 2FGL flux distribution \citep{2FGL,PaperVa}.  Because the latter is formally divergent, we filled in the background from the high-flux end, beginning just below the \Fermi threshold and stopping when the required number of background photons were obtained.  For all of the background sources we adopted a photon spectral index of $2.4$, consistent with the 1~GeV--100~GeV background \citep{Fermi_EGRB2010,Fermi_EGRB2015}.

Here it is useful to distinguish between random realizations of the background sources, i.e., the positions and fluxes of the unresolved sources that comprise the background, and random realizations of photons drawn from the sources.  Even after averaging over a full ensemble of photon realizations a single background source realization produces a stochastic angular power spectrum exhibiting large variations at low multipoles, shown by the blue line in the bottom-left panel of Figure~\ref{fig:badcomp}.  This is a result of the structure imposed on the background via the locations and strengths of the background sources and cannot be overcome by collecting additional observations.  Averaging over realizations of the background sources and photons results in a smooth angular power spectrum, similar to that arising from a nearby, unresolved source (green line in the bottom-left panel of Figure~\ref{fig:badcomp}).

In principle, we may directly measure this background angular power spectrum using nearby empty fields.  However, our ability to remove it is fundamentally limited by the variance due to moderate number of background realizations presented by the source list in Table~\ref{tab:sample}.  Nonetheless, even after averaging over 18 sources (the number of sources in Table~\ref{tab:sample}) the fluctuations are already substantially reduced (red line in bottom-left panel of Figure~\ref{fig:badcomp}).  Regardless, the stochastic structure lacks the telltale sawtooth morphology of the ICC halos.

Azimuthal structure of the \Fermi Pass 8R2\_V6 ULTRACLEANVETO PSF will also generate angular power within the source photons.  Note that this is not true for the distribution of the uniform gamma-ray background -- even a highly anisotropic PSF cannot impart structure on a uniform field of photons.
The angular structure of the \Fermi Pass 8R2\_V6 ULTRACLEANVETO PSF arises from the square geometry of the LAT (see also Section \ref{sec:PSF}).  The instantaneous PSF exhibits only 5\% variations in the PSF at the energies of interest \citep{LAT_perf}.  Over timescales short in comparison to years this is substantially reduced by the rotation of the \Fermi during solar tracking and the eight-fold symmetry of the LAT.

In Appendix \ref{app:PSFPS} we make an estimate of the residual PSF-induced angular power, assuming that cumulative gamma-ray image is comprised of many epochs during which the roll angle of \Fermi is highly correlated.  The duration of these epochs correspond to roughly the time for \Fermi to rotate by $45^\circ$, after which it effective rotates through the entirety of the PSF as a result of the LAT's square geometry.  Presuming that the fluence is relatively evenly distributed over the past 8~yr, even for optimistic assumptions regarding the structure of the PSF, the estimated residual angular power is less than 1\% of the anticipated Poisson noise.  Moreover, it vanishes identically for $m=2$ as a result of the LAT's symmetry, exhibiting power only for $m=4$ and its harmonics.

Finally, we considered a linear gradient in the background photon density.  Because we explicitly select sources for which there is no apparent large-scale gradients in the background, we again set the value to our effective detection threshold, corresponding to a variation in the photon density of 20\% across the $4^\circ$ image.  As shown in figure \ref{fig:badcomp}, the impact on the angular power spectrum is very small, weakly modifying the dipole power primarily.

It is important to note in summary two key results regarding all of the potential the systematic uncertainties arising from contaminants:  First, their {\em shape} is qualitatively different from the distinct signatures of the bimodal ICC halos.  Second, their {\em magnitude} is far smaller than that expected from the ICC halos arising from a large-scale, uniform IGMF.  As a result they should be readily distinguishable.

\subsection{Combining Multiple Sources} \label{sec:stack}

Finally, to increase the fidelity of the angular power spectrum we stack the estimates from multiple sources.  Unlike stacking the images directly, this preserves the anisotropic signal; a rotation of any image, corresponding to setting $\theta_j\rightarrow\theta_j+\varphi$, leaves the $\P_m$ unchanged as may be verified by inspection of Equation (\ref{eq:PS}).  It does, however, reduce the intrinsic scatter in the power spectrum estimate.

In principle, this may be optimized via the weighting assigned to individual sources -- images with higher numbers of intrinsic photons will produce better intrinsic power spectrum estimates and thus may be given additional weight in the stacking processes.  The natural way to do this is the variance-weighting, giving the smallest variance in $\P_m$ at each $m$.  In practice, we found that the dominance of the source counts by a handful of objects (e.g., Mkn 421) led to an associated dominance of the power spectra estimate, eliminating much of the power of the stacking process.

It is also possible to exploit the spectra of the ICC halos and/or the redshift-dependence of the ICC halo extent to provide more optimal weightings.  We found that no such effort made a substantial impact on the ability to distinguish the power in the quadrupole and the neighboring odd multipoles, the key observable for our bimodal ICC halos.  The reason is simply that while the ICC halos are typically harder than the gamma-ray background they are only marginally so, limiting the ability to spectrally separate the two components.

Therefore, we define our stacked power spectrum by the arithmetic average of the individual source power spectra:
\begin{equation}
  \P_m^{\rm tot} = \frac{1}{N_{\rm src}} \sum_{\rm src} \P_m^{\rm src}\,,
\end{equation}
where $\P_m^{\rm src}$ is the single-source power spectrum defined by Equation (\ref{eq:PS}).

\subsection{Source List Optimization} \label{sec:opt}

The ability to generate simulated realizations of the \Fermi sky enables us to theoretically optimize the list of \Fermi sources that are ultimately stacked.  That is, apart from gross properties of the sources (e.g., SEDs and redshifts), we can select the group most likely to collectively produce the apparent signatures of ICC halos in the stacked angular power spectrum {\em without} looking at the actual structures of these images.

The SEDs described in Section \ref{sec:SEDs} provide guidance on which sources are likely to be bright above a TeV.  However, while this is a necessary condition, it is not sufficient to produce bright ICC halos.  The halo itself is impacted by the source distance (among other parameters that are marginalized over).  The ability to detect the halo is impacted by the local background.  Thus, armed with the ability to simulate halos from the 27 sources that are sufficiently isolated, appear in the 2FHL and 3LAC and have known redshifts, we applied a final optimization step designed to maximize the ability to detect ICC halos.

We begin by defining a halo-model specific detection likelihood statistic:
\begin{equation}
  P_{\rm det} = \int dp P^{\rm h}_{>\P_2}(p) P^{\rm n}_{\P_2}(p) \,,
\end{equation}
where $P^{\rm n}_{\P_2}(p)$ is the probability of finding $\P^{\rm tot}_2=p$ in the null case, i.e., without a halo, and $P^{\rm h}_{>\P_2}(p)$ is the cumulative probability associated with $\P^{\rm tot}_2\ge p$ when a halo is present.  This is the probability that the $m=2$ power from the given halo model exceeds that from the null case, marginalized over the probabilities of both.  For the null case, $P_{\rm det}=1/2$, i.e., in the absence of an ICC halo the probability that $\P_2$ will exceed that from the null case is simply 50\%.  This translates directly into the probability that a given halo model will produce excess power in the quadrupole, i.e., the probability that an ICC halo is detectable.  From a collection of $N_{\rm real}$ realizations of the full sample of isolated, hard \Fermi AGN we estimate $P_{\rm det}$ via
\begin{equation}
  P_{\rm det} \approx \frac{1}{N_{\rm real}} \sum_{\rm real} \frac{1}{51} \sum_{m=50}^{100} \Theta(\P^{\rm tot}_2 - \P^{\rm tot}_m)\,,
\end{equation}
where $\Theta(x)$ is the Heaviside function and we have used the high-$m$ power spectra as a proxy for the null case.  This is necessarily a function of the particular set of sources included, changing as a result of all of the systematic inputs into the construction of the ICC halo models and the stacked power spectra.  Thus, for a given halo model we can optimize the list of input sources by maximizing the simulated values of $P_{\rm det}$.

This maximization is achieved in a restricted sense in practice: we order the list of sources by intrinsic TeV brightness and then construct a sequence of source lists, each comprised of the preceding list and the next-brightest TeV source.  For example, the first list includes Mkn~421, the second, Mkn~421 and 3FGL~J2347.0+5142, etc.  The value of each source is then evaluated when it is first included; if it increases $P_{\rm det}$ it is kept, if it decreases $P_{\rm det}$ it is removed.  This procedure results in a set of optimal sources for each halo model. 

This list varies between halo models due to the differences in the ICC halos produced.  As a result in Table~\ref{tab:sample} we present four separate samples corresponding to four different choices of $B_0$ for the uniform IGMF halo models: $B_0=10^{-17}~\G$, $10^{-16}~\G$, $10^{-15}~\G$, and $10^{-14}~\G$.  In practice they are nested: the first sample is comprised of the brightest four sources, the second sample includes the first and the next six sources, etc.  This is a result of the smaller apparent sizes of the halos associated with weaker IGMFs, and thus preferring a brighter and nearer AGN sample.  It is this collection of source samples that we will use henceforth.

We performed a similar procedure to optimize the source list for the tangled magnetic fields.  This was less successful due to the low intrinsic signal in the power spectra.  We found little difference among the various samples chosen and therefore simply adopted the set from the uniform IGMF model with $B_0=10^{-15}~\G$.

It is important to note that this optimization procedure was performed {\em entirely} using the {\em simulated} images.  That is, apart from the source SEDs and measured redshifts the {\em actual} source structure played no role.  Hence, we have in no way begged the question by having done so.  Rather, this is precisely the chief advantage of simulating the data -- it permits identifying the key signatures of the ICC halos and optimize the procedure to the detect them.

\section{Monte Carlo Confidence Level Estimates} \label{sec:MCCLs}

The median angular power spectra presented in the previous section while indicative of the impact of different image features are poor representations of the angular power spectra associated with a single source realization.  The fractional uncertainty in the power at a given multiple for a single image is order unity (see Appendix \ref{app:ExPSw}), implying large deviations are typical.  This is ameliorated by stacking the power spectra from multiple images as described in Section \ref{sec:stack}, which formally decreases the scatter by roughly $N_{\rm src}^{-1/2}$.  However, variations in the underlying intrinsic source and IGMF properties act to increase the fluctuations in $\P_m$.

Here we report the resulting 95\% confidence-level regions for the various ICC halo models, and thereby IGMF models.  The meaning of these is similar to that of a likelihood; they represent the probability of finding a given value of $\P^{\rm tot}_m$ at a specified $m$ for a given IGMF.  As such, they present a natural way to assess the single realization afforded by the actual \Fermi data.

Given the approximately bimodal nature of the ICC halos we have focused on two primary observables: the quadrupolar power $\P^{\rm tot}_2$ and the sawtooth morphology.  The latter is primarily in service to separating contributions to the angular power spectrum from ICC halos from other sources, including the potential sources of systematic error discussed in Section \ref{sec:contam}.

There are a number of additional potential observables.  Typically, strong correlations between nearby multipoles limits the ability to leverage deviations for many $m$ to improve statistical weight.  Nevertheless, key systematics may be assessed using the distribution of $\P^{\rm tot}_m$ at large $m$, set by the Poisson noise and therefore the effective number of gamma rays used to construct the power spectrum estimate.

Thus, the essential experiment is to compare the low-$m$ $\P^{\rm tot}_m$ to the predictions of the various IGMF/halo models.  Confidence in the simulation of the \Fermi sources and their subsequent stacked angular power spectrum is obtained by comparing their large-$m$ characteristics.  The confidence with which any IGMF may be detected or excluded is then set by the degree to which the measured $\P^{\rm tot}_m$ is inconsistent with the predicted value at, for example, $m=2$.

The key theoretical input is then the anticipated {\em distribution} of $\P^{\rm tot}_m$ for a given IGMF model.  Here we describe how we estimate the ranges over which the angular power spectra can vary for a given halo model, and what these ranges are for the particular IGMF models described in Section \ref{sec:mocks}.

\subsection{Generating Mock \Fermi Samples}
The confidence with which a given IGMF can be detected or excluded based upon the gamma-ray image angular power spectra depends critically on a quantitative estimate of the range of, distribution of, and correlations among, the $\P_m$ for the stacked power spectra of the \Fermi sample.  To assess this we perform a Monte Carlo sampling of the \Fermi sample and its associated $\P^{\rm tot}_m$.  This synthesizes the steps described in Sections \ref{sec:mocks}--\ref{sec:signatures} for a given assumption regarding the geometry and strength of the IGMF.

In summary, for each realization of the \Fermi sample we generate a set of realizations of each source in Table~\ref{tab:sample} consistent with their known intrinsic properties (e.g., $N_{\rm ph}$ and $\mathcal{B}$) and our estimates for the distributions of their unknown intrinsic properties (e.g., $\Gamma_h$, $\theta_{\rm jet}$, $\Theta$, etc.).  Within each image we choose the number of source and background photons from a Poisson distribution, taking $N_{\rm ph}$ and $\pi R_{\rm img}^2 \mathcal{B}$ as means of the distribution, respectively (see Table~\ref{tab:sample}).  The source photons are distributed among the direct component and the ICC halo according to the particular halo model appropriate for the IGMF under consideration as described in Section \ref{sec:confusion}.  For large-scale, uniform IGMF models we further permit its local strength and orientation to vary as described in Section \ref{sec:BSO}.

The resulting list of realized $\P^{\rm tot}_m$ provide a proxy for the distribution of stacked angular power spectra associated with a given IGMF model.  To quantify this distribution we quote one-sided 95\% confidence-level regions, i.e., the values of $\P^{\rm tot}_m$ that 95\% of the realizations lie above, and independently the value that 95\% lie below; because they are one-sided these are subtly different from the normal $2\sigma$ errors.  This procedure of using one-sided confidence levels enables rejecting a model hypothesis at 95\% confidence if it lies outside the one-sided half-interval.  We also use the sample to explore the correlations between stacked angular power at different multipoles; when ICC halos contribute substantially to the $\P^{\rm tot}_m$ the multipoles are generally strongly correlated, maintaining the sawtooth structure.

\subsection{Anticipated Inverse Compton Halo Power Spectra}
We now turn to the probability distributions of $\P^{\rm tot}_m$ for various halo/IGMF models of interest.  Note that in many cases the power at nearby multipoles are strongly correlated.

\subsubsection{Null Case}
We begin with the null hypothesis -- the absence of any ICC halo component.  In this case the image consists solely of the central point source and uniform background for each object in the \Fermi sample.  This represents the baseline against which all subsequent models that include ICC halos are compared.  Here we use the $10^{-15}~\G$, large-scale, uniform IGMF optimized source list.

\begin{figure}
  \begin{center}
    \includegraphics[width=\columnwidth]{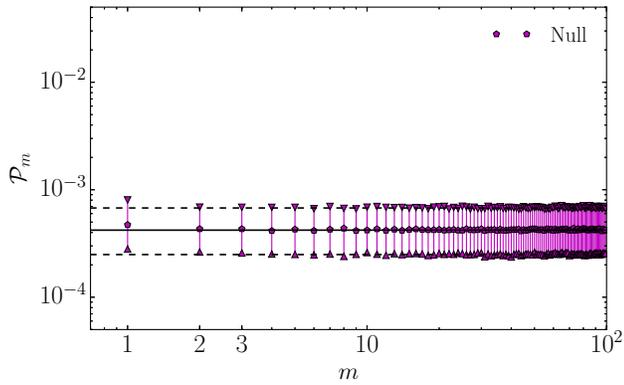}
  \end{center}
  \caption{Mock angular power spectra distributions when no ICC halos are present.  The median value is shown by the filled pentagon, the triangles denote the one-sided 95\% confidence interval.  In comparison the black solid and dashed lines show the Poisson noise limit and its one-sided 95\% confidence intervals, respectively, defined collectively by combining the distributions of all multipoles.  This is produced with the $10^{-15}~\G$, large-scale, uniform IGMF optimized source list.}\label{fig:PSnull}
\end{figure}

\begin{figure}
  \begin{center}
    \includegraphics[width=\columnwidth]{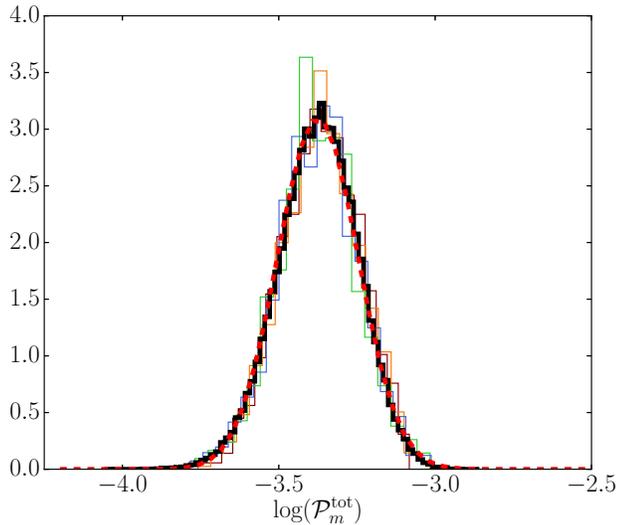}
  \end{center}
  \caption{Distribution of mock $\P^{\rm tot}_m$ about the median for the first 4 multipoles ($m=2$, 3, 4, 5 from red to blue) and the first 100 multiples combined (black).  For comparison a log-normal fit is shown (red dashed line) with mean $-3.38$ and standard deviation $0.129$.  This is produced with the $10^{-15}~\G$, large-scale, uniform IGMF optimized source list. }\label{fig:PSnoiseln}
\end{figure}

The median angular power spectrum, including Front- and Back-converted events, and its 95\% one-sided confidence limits are shown in Figure~\ref{fig:PSnull}.  As anticipated there is no structure in this case -- the power spectrum is well fit by a constant corresponding to the effective Poisson noise limit.  The distribution about the median values of $\P^{\rm tot}_m$ are uncorrelated between multipoles and well fit by a log-normal distribution, shown in Figure~\ref{fig:PSnoiseln}.

We would expect any contaminating systematic (e.g., centering errors or dim neighbors) to contribute in a fashion similar to that seen for the single source power spectra presented in Section \ref{sec:contam}.

\subsubsection{Small-scale, Tangled IGMF}

\begin{figure}
  \begin{center}
    \includegraphics[width=\columnwidth]{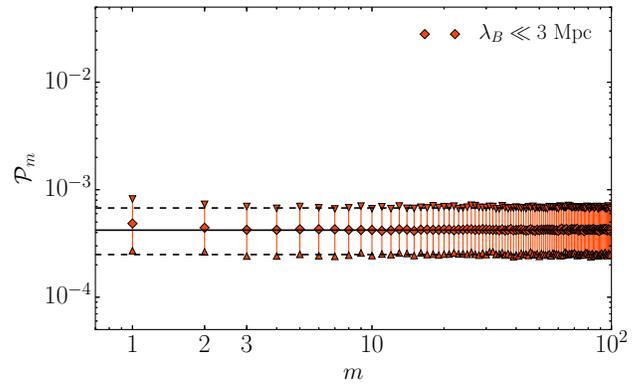}
  \end{center}
  \caption{Mock angular power spectra distributions when ICC halos associated with a small-scale, tangled IGMF are present.  The median value is shown by the filled diamonds, the triangles denote the one-sided 95\% confidence interval.  In comparison the black solid and dashed lines show the Poisson noise limit and its one-sided 95\% confidence intervals, respectively, defined collectively by combining the distributions of multipoles with $m\ge50$.  This is produced with the $10^{-15}~\G$, large-scale, uniform IGMF optimized source list. }\label{fig:PSiso}
\end{figure}

\begin{figure}
  \begin{center}
    \includegraphics[width=1.1\columnwidth]{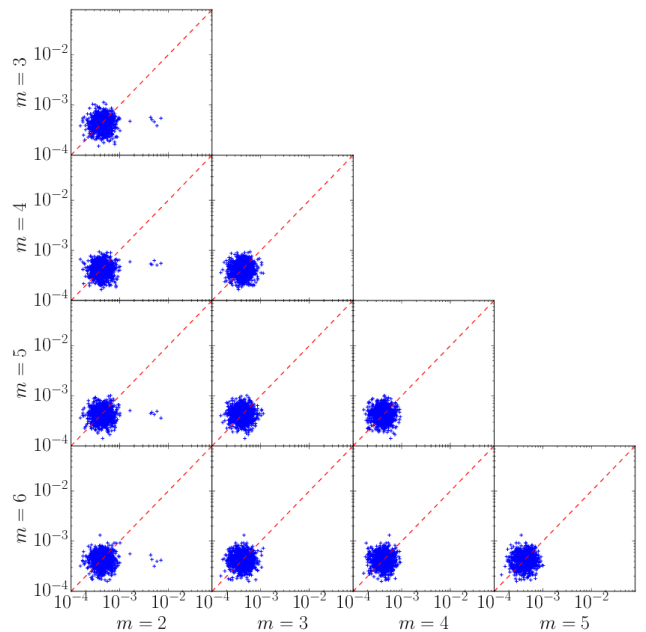}
  \end{center}
  \caption{Pairwise correlations between the angular power in $m=2$, 3, 4, and 5 multipoles for each mock realization of the \Fermi sample when ICC halos associated with a small-scale, tangled IGMF.  For reference the one-to-one line is shown by the red-dashed line.  This is produced with the same set of realizations used to construct Figure~\ref{fig:PSiso}, and utilizes the $10^{-15}~\G$, large-scale, uniform IGMF optimized source list.}\label{fig:Corriso}
\end{figure}

\begin{figure*}
  \begin{center}
    \includegraphics[width=\columnwidth]{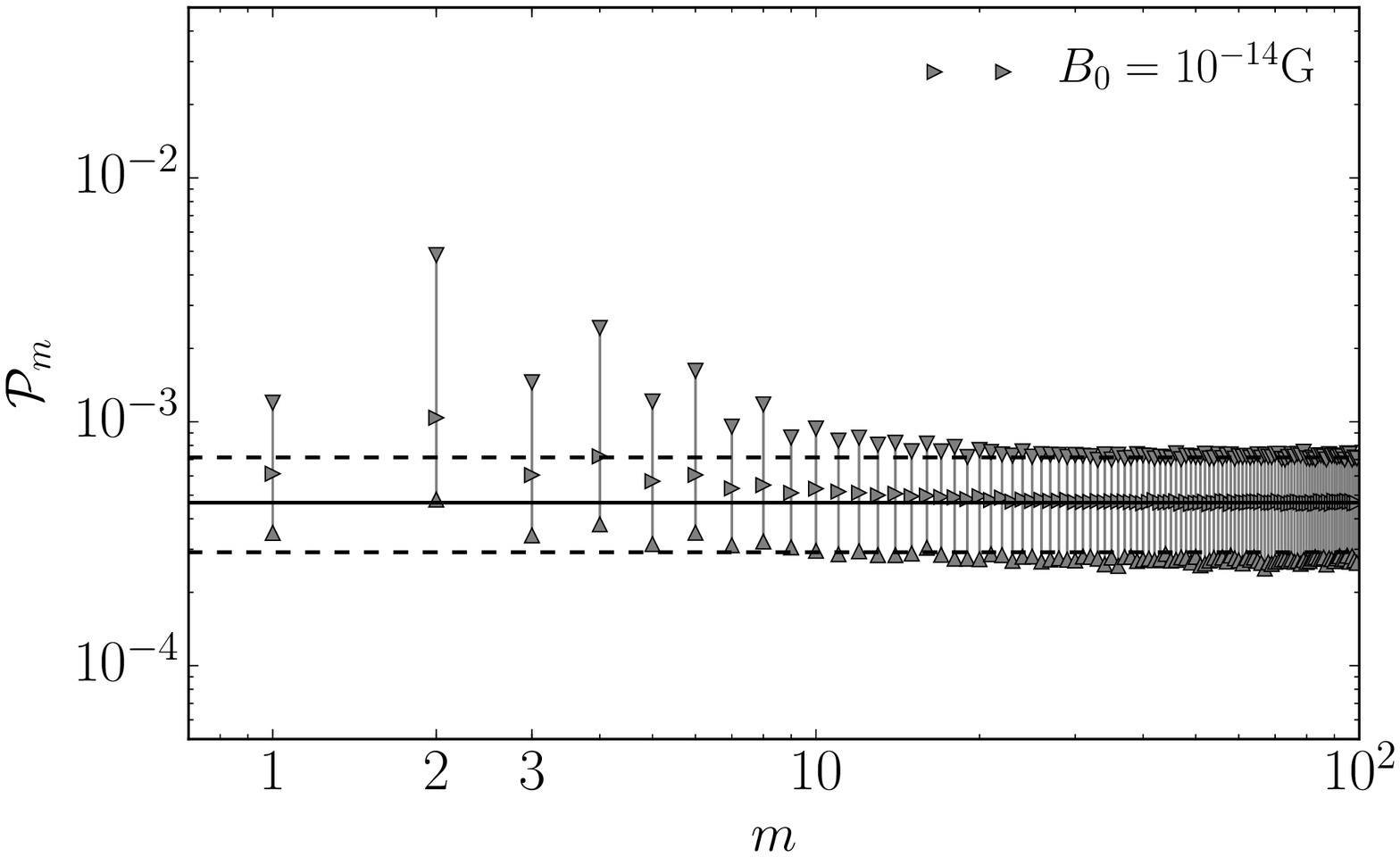}
    \includegraphics[width=\columnwidth]{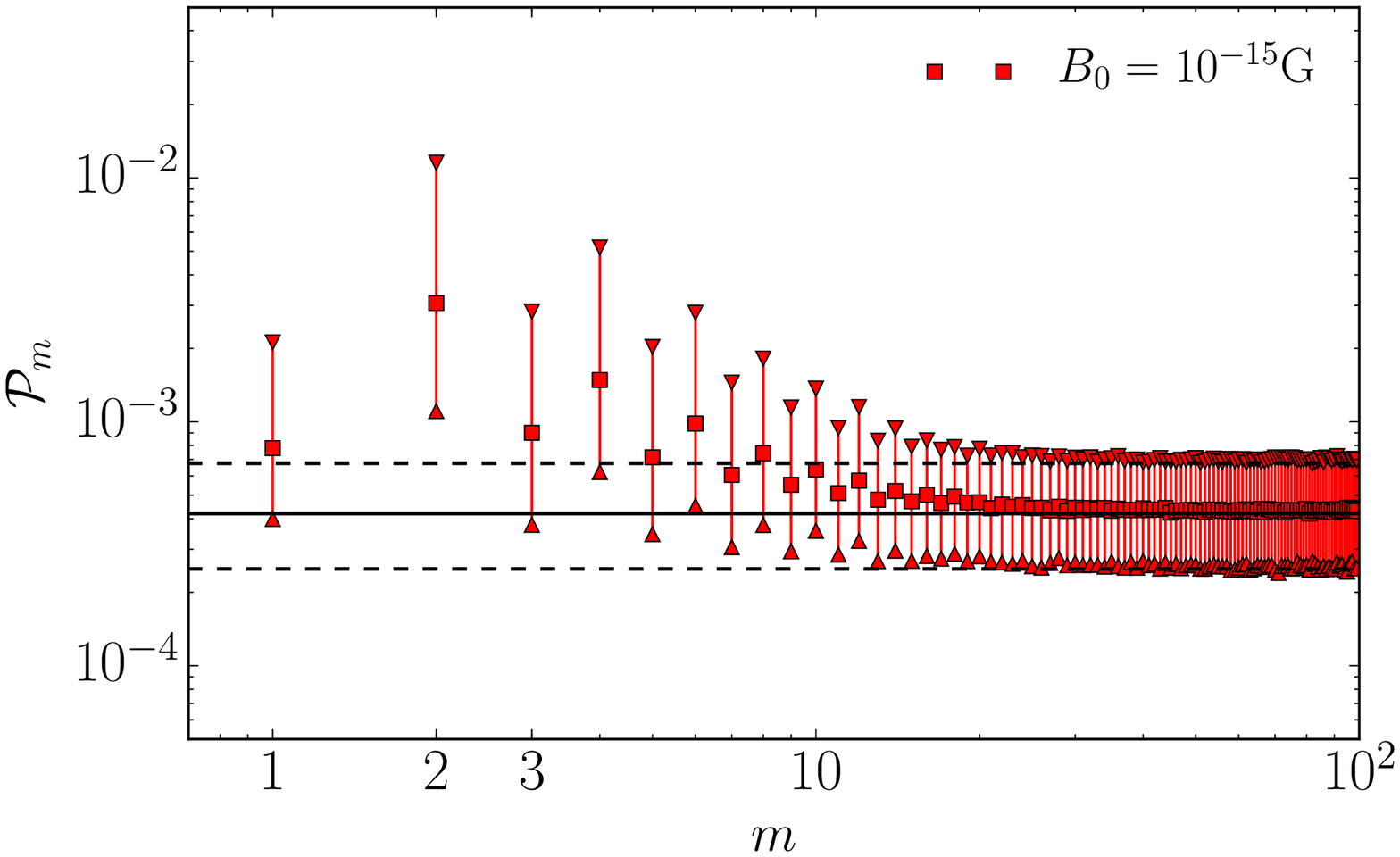}
    \includegraphics[width=\columnwidth]{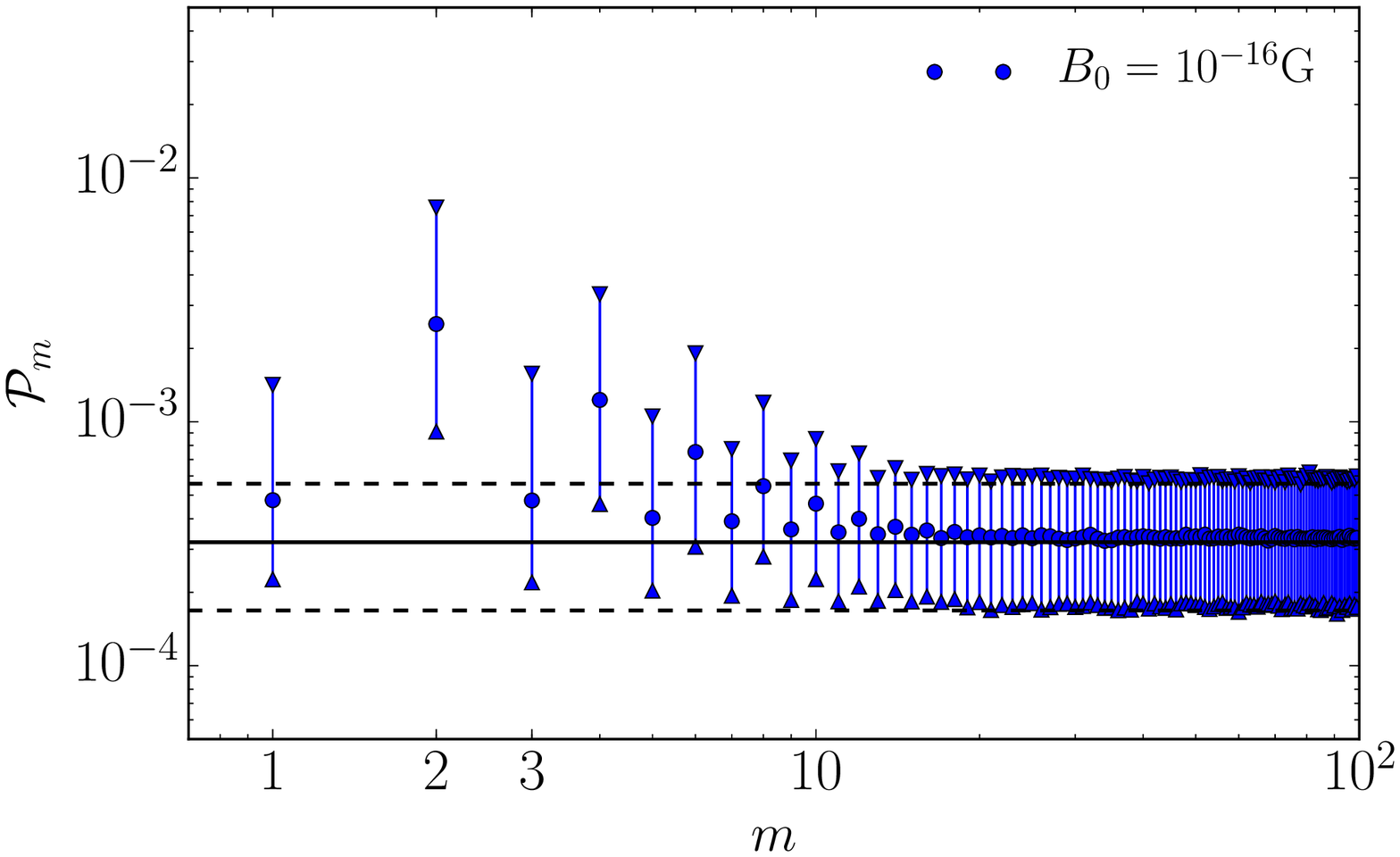}
    \includegraphics[width=\columnwidth]{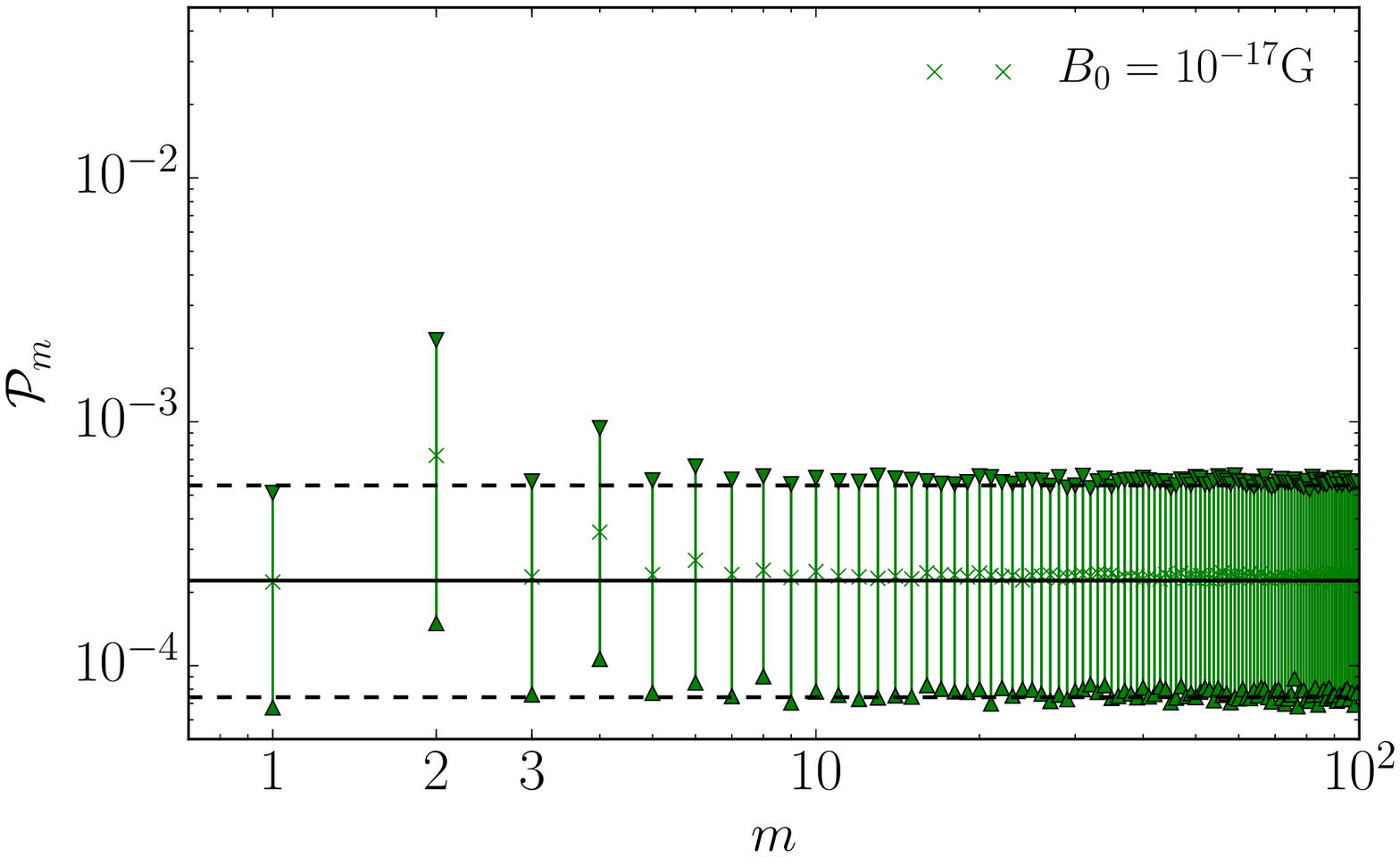}
  \end{center}
  \caption{Mock angular power spectra distributions when ICC halos associated with a large-scale, uniform IGMF are present for $B_0=10^{-15}~\G$ (grey triangle, upper left), $B_0=10^{-15}~\G$ (red squares, upper right), $10^{-16}~\G$ (blue circles, lower left), and $10^{-17}~\G$ (green crosses, lower right).  The median value is shown by the filled points, the triangles denote the one-sided 95\% confidence interval.  In comparison the black solid and dashed lines show the Poisson noise limit and its one-sided 95\% confidence intervals, respectively, defined collectively by combining the distributions of multipoles with $m\ge50$.  Each case was produced using the associated optimized source list (see Table~\ref{tab:sample}). }\label{fig:PSweak}
\end{figure*}

\begin{figure}
  \begin{center}
    \includegraphics[width=1.1\columnwidth]{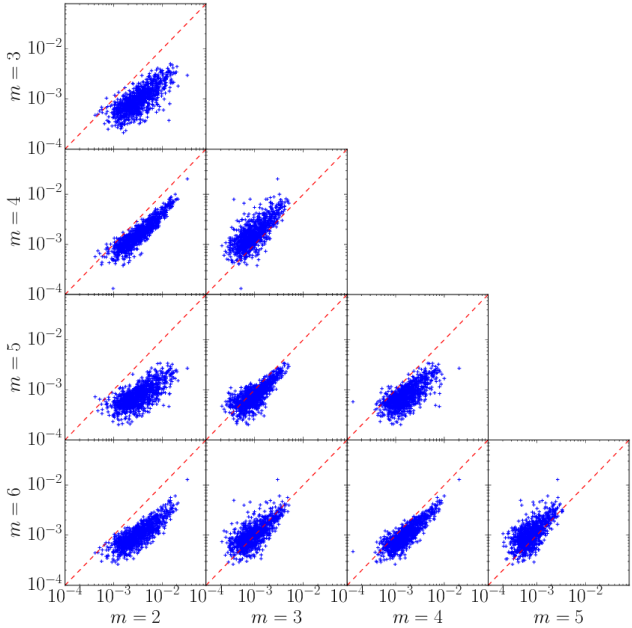}
  \end{center}
  \caption{Pairwise correlations between the angular power in $m=2$, 3, 4, and 5 multipoles for each mock realization of the \Fermi sample when ICC halos associated with a large-scale, uniform IGMF with $B_0=10^{-15}~\G$.  For reference the one-to-one line is shown by the red-dashed line.  This is produced with the same set of realizations used to construct the associated panel of Figure~\ref{fig:PSweak}.}\label{fig:Corrweak}
\end{figure}

\begin{figure}[!t]
  \begin{center}
    \includegraphics[width=1.1\columnwidth]{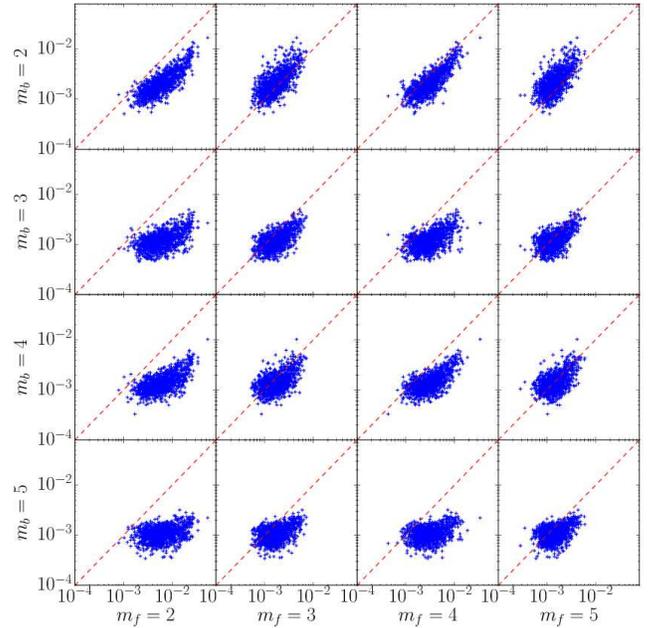}
  \end{center}
  \caption{Pairwise cross-correlations between the angular power in $m=2$, 3, 4, and 5 multipoles obtained from Front- ($m_f$) and Back-converted ($m_b$) events for each mock realization of the \Fermi sample when ICC halos associated with a large-scale, uniform IGMF with $B_0=10^{-15}~\G$.  For reference the one-to-one line is shown by the red-dashed line.  This is produced with the same set of realizations used to construct the associated panel of Figure~\ref{fig:PSweak}.}\label{fig:CrossCorrweak}
\end{figure}

The median and 95\% confidence regions for the stacked angular power spectra for images containing ICC halos associated with a small-scale, tangled IGMF are shown in Figure~\ref{fig:PSiso}.  As anticipated by Figure~\ref{fig:ICCcomp_iso} in comparison to Figure~\ref{fig:PSnull}, and less directly by Figure~\ref{fig:haloreals}, there is no discernable contribution from an ICC halo.  This is a direct result of the combination of a low number of halo photons and the extreme foreshortening that results from the small viewing angles typical of the \Fermi sources.

The cross-correlation of different multipoles, shown in Figure~\ref{fig:Corriso} reveals that for a small fraction of realizations ($<5\%$) there are large deviations in the quadrupolar power.  This suggests that alternative samples, composed of sources viewed at more oblique angles, may provide more promise for assessing the existence of ICC halos in the presence of a small-scale, tangled IGMF.  We leave such an analysis for future work.

\begin{figure}[!t]
  \begin{center}
    \includegraphics[width=\columnwidth]{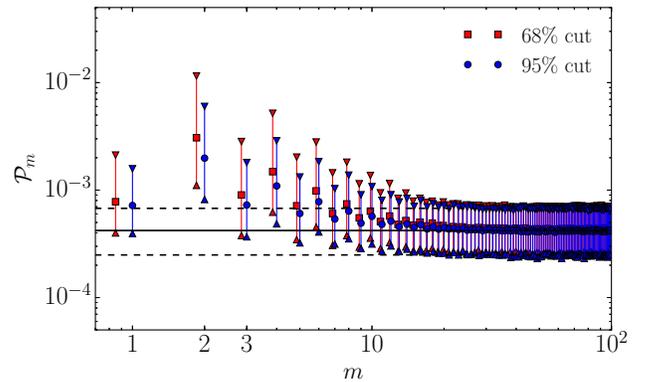}
  \end{center}
  \caption{Mock angular power spectrum assuming a large-scale, uniform IGMF with $B_0=10^{-15}~\G$ when the ``source'' region is 68\% and 95\% of the Pass 8R2\_V6 ULTRACLEANVETO PSF, respectively.  The median value is shown by the filled points, the triangles denote the one-sided 95\% confidence intervals.  In comparison the black solid and dashed lines show the Poisson noise limit and its one-sided 95\% confidence intervals, respectively, defined collectively by combining the distributions of multipoles with $m\ge50$.  This is produced with the $10^{-15}~\G$, large-scale, uniform IGMF optimized source list.}\label{fig:PS95}
\end{figure}

\subsubsection{Large-scale, Uniform IGMF}
Unlike the other models considered, the large-scale, uniform IGMF model has a free parameter in the form of the current IGMF strength.  Thus, we show the distributions of stacked angular power for a variety of magnetic field strengths in Figure~\ref{fig:PSweak}.

These all exhibit excess power at low-$m$, becoming less significant as the field strength nears $10^{-17}~\G$, as anticipated from the single-image power spectra in Section \ref{sec:signatures}.  For $B_0>10^{-17}~\G$ the deviations from the Poisson noise are substantial, rising with increasing field strength; for $B_0=10^{-15}~\G$ the excess power extends to $m\approx30$.  Importantly, note that for $B_0=10^{-15}~\G$ and $10^{-16}~\G$ the quadrupolar power is {\em inconsistent} with that for the null hypothesis, implying that the angular power spectrum will immediately distinguish between these two IGMF/halo models.

The breaking of the bipolar symmetry is clearly apparent in the stacked angular power, corresponding to the large power at odd multipoles.  This is a natural consequence of the larger probability of viewing angles near the maximum permitted (see Section \ref{sec:mocks}).  Nevertheless, the sawtooth structure is directly evident in the medians of the $\P^{\rm tot}_m$ for all cases.

Less obvious is that this remains true in individual realizations despite the large apparent variations implied by the extent of the 95\% confidence regions; as seen in Figure~\ref{fig:Corrweak}, the power in all multiples are strongly correlated, moving the sawtooth pattern up and down together.  This is particularly true for the quadrupole, which overwhelmingly dominates all of the odd multipoles in more than 99\% of the realizations, failing to do so only in the most pessimistic cases.  This remains true for the Front- and Back-converted events independently, as shown in Figure~\ref{fig:CrossCorrweak}, which shows the cross-correlation of the two event populations.  {\em Thus, it is always possible to distinguish the large-scale, uniform IGMF model from the many potential sources of contamination in the angular power spectrum that show a smoothly falling excess of power at low-$m$ generally.}  The strong correlation among multipoles also means, however, that it remains possible that the power excess is low at all $m$, providing a fundamental limit on the minimum $B_0$ that can be reliably differentiated from the null case.

\subsection{Sensitivity to Source-Halo Confusion} \label{sec:6895}

How the potential confusion between source, halo and background photons is dealt with modifies the number of halo photons at large angular separations, and thus potentially the significance with which ICC halo contribution to the power spectrum may be identified.  We therefore consider modifications to the scheme described in Section \ref{sec:confusion}, extending the ``source'' region, i.e., the region within which photons will be assumed to contribute to the number of source photons to the 95\% containment radius of the Pass 8R2\_V6 ULTRACLEANVETO PSF.  As discussed in Section \ref{sec:confusion} this extends to large angular radii, typically well beyond the point at which an extended component would be clearly visible.  Nevertheless, as shown in Figure~\ref{fig:PS95}, this has little impact on the ability to identify ICC halos in the angular power spectrum.

\subsection{Future Prospects with \Fermi}
\begin{figure*}
  \begin{center}
    \includegraphics[width=\columnwidth]{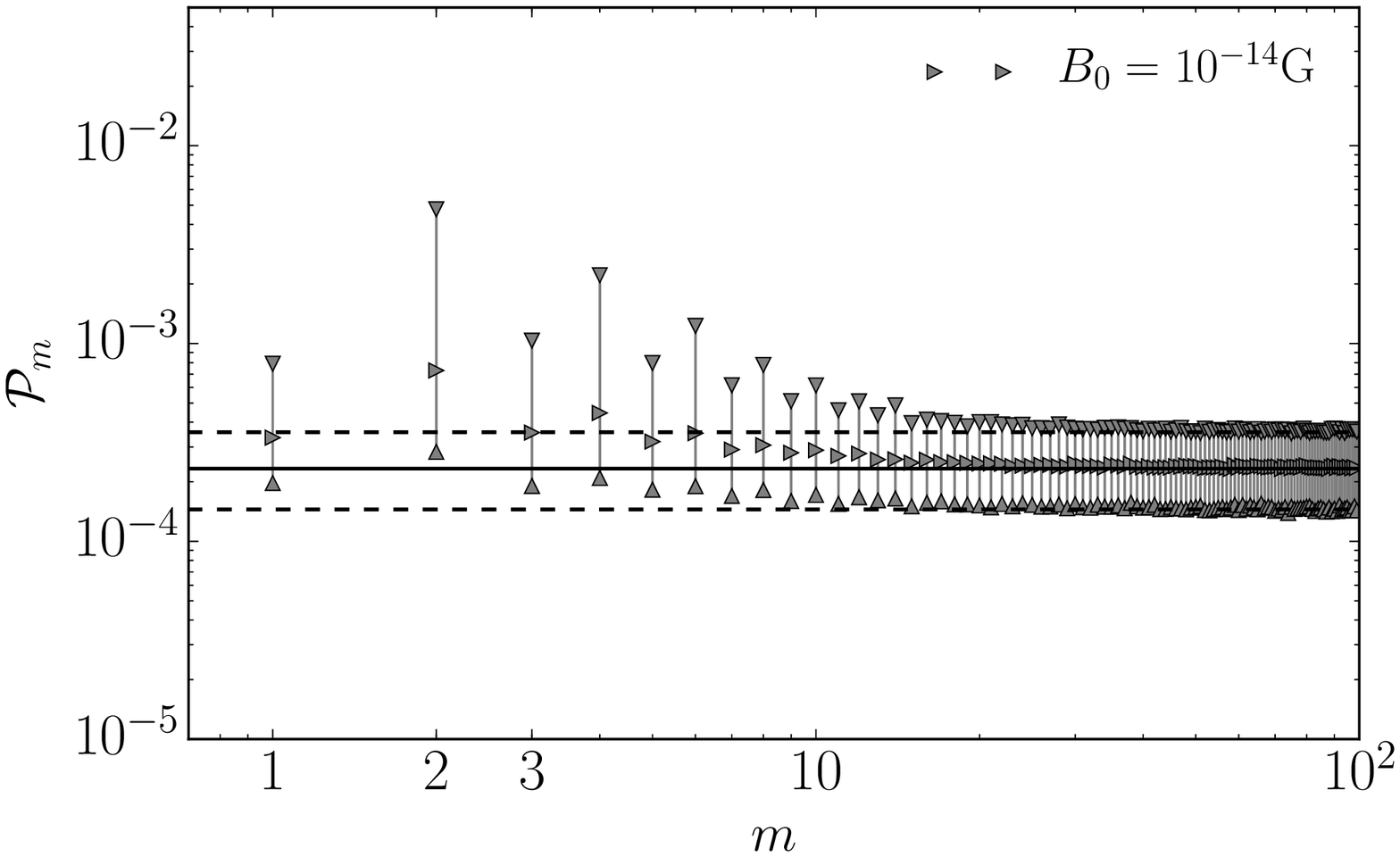}
    \includegraphics[width=\columnwidth]{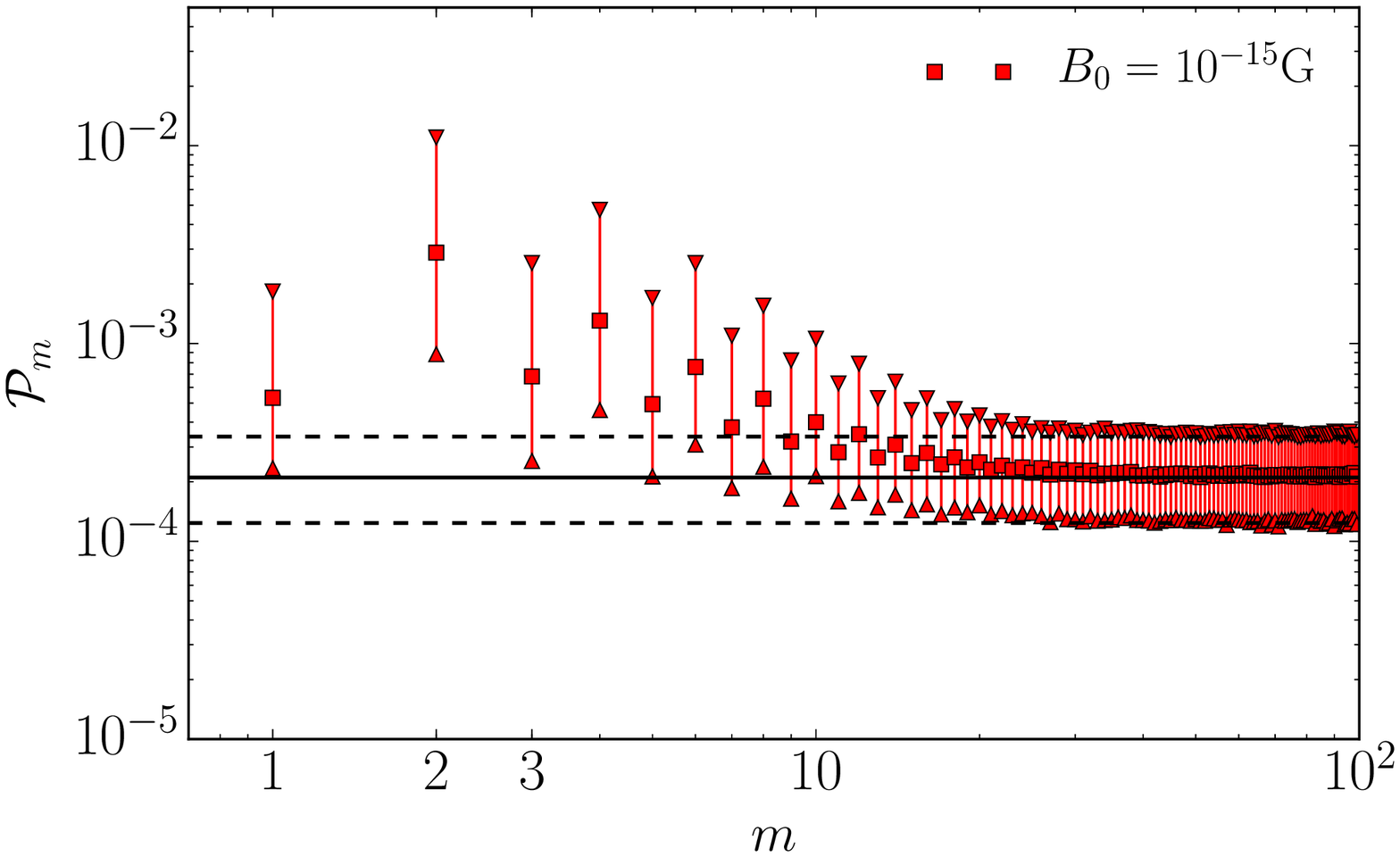}
    \includegraphics[width=\columnwidth]{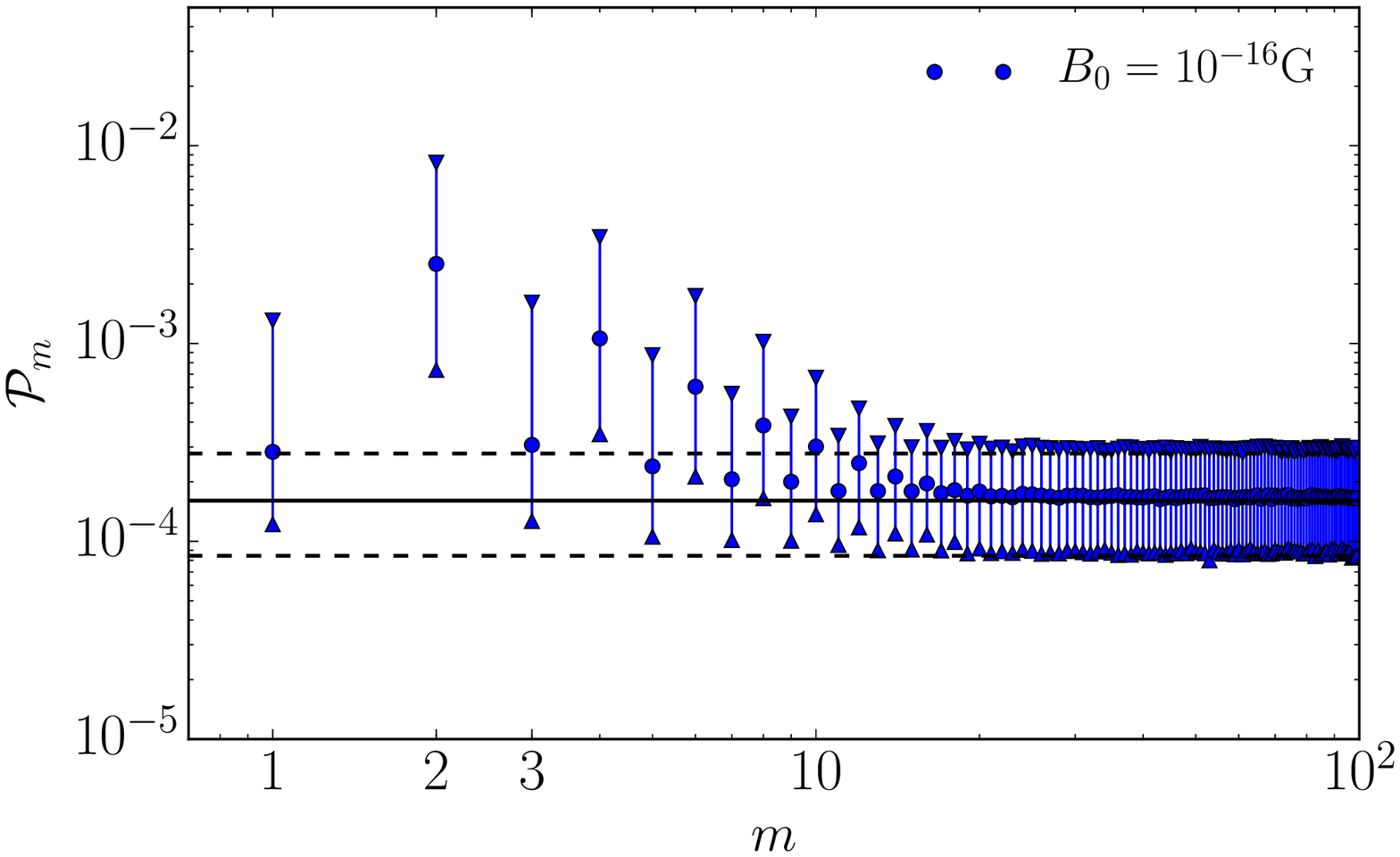}
    \includegraphics[width=\columnwidth]{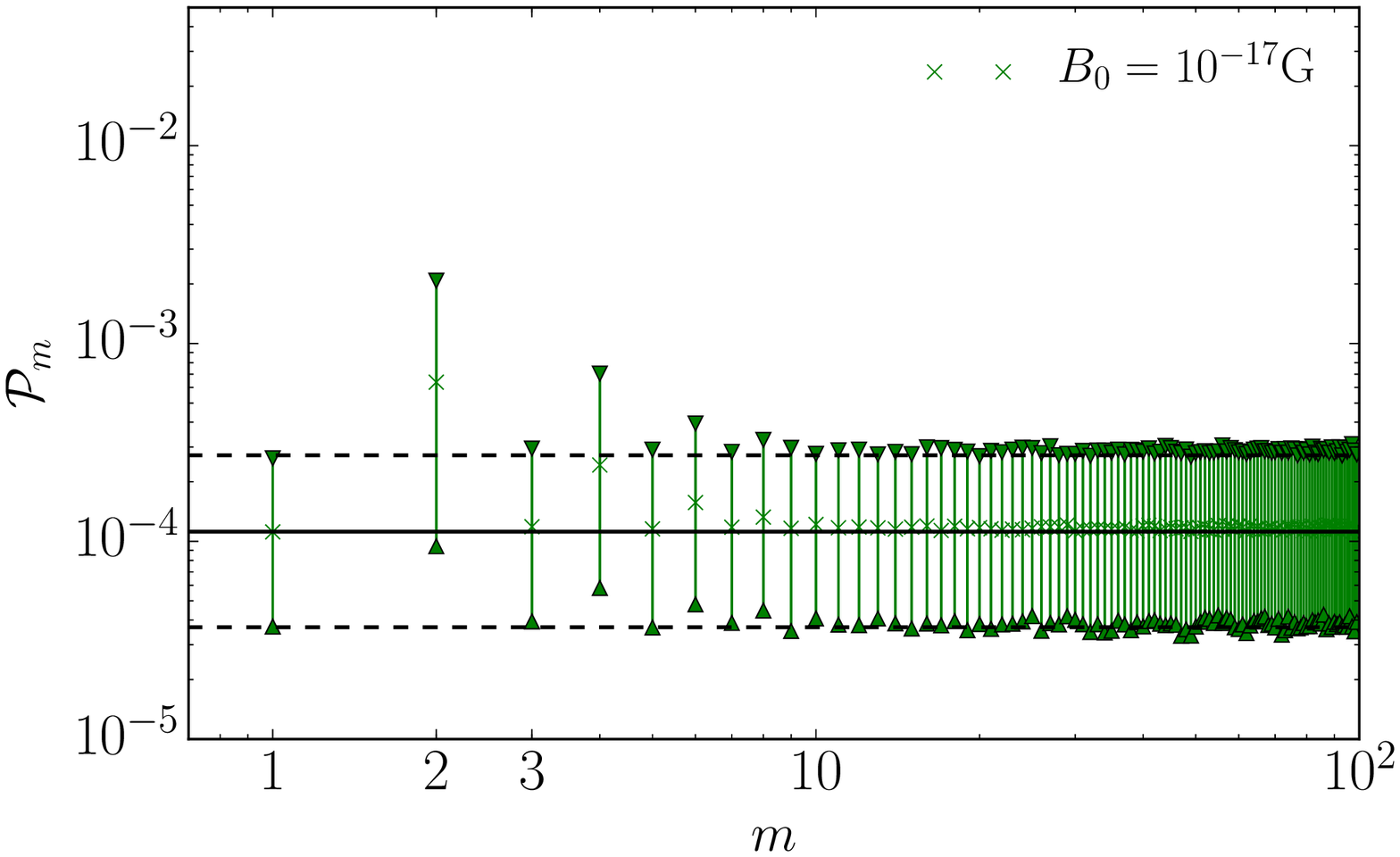}
  \end{center}
  \caption{Mock angular power spectra distributions with double the number of photons(i.e., a projected total of 16 years of \Fermi observations), when ICC halos associated with a large-scale, uniform IGMF are present for $B_0=10^{-15}~\G$ (grey triangle, upper left), $B_0=10^{-15}~\G$ (red squares, upper right), $10^{-16}~\G$ (blue circles, lower left), and $10^{-17}~\G$ (green crosses, lower right).  The median value is shown by the filled points, the triangles denote the one-sided 95\% confidence interval.  In comparison the black solid and dashed lines show the Poisson noise limit and its one-sided 95\% confidence intervals, respectively, defined collectively by combining the distributions of multipoles with $m\ge50$.  Each case was produced using the associated optimized source list (see Table~\ref{tab:sample}).}\label{fig:PSweak2x}
\end{figure*}

The significance estimates shown in Figures \ref{fig:PSnull}, \ref{fig:PSiso}, and \ref{fig:PSweak} are all for the current \Fermi data set.  However, the operational lifetime of \Fermi may be considerably longer.  Therefore, we also consider a gamma-ray data set that is doubled in size, i.e., the results of 16 years of \Fermi observations.  To facilitate a direct comparison and in the interest of simplicity we keep the source list identical.

As shown in Figure~\ref{fig:PSweak2x}, the significance with which a large-scale, uniform IGMF can be detected or excluded is significantly improved.  Of $10^3$ realizations that assume an IGMF strength of $B_0=10^{-15}~\G$, none were found that had a $P_2$ below the 95\% confidence upper limit of the null model.  Thus, it would be possible to detect or exclude ICC halos associated with moderate IGMF strengths with significances well above $3\sigma$.

More importantly, the multipoles for which this may be done extends beyond the quadrupole and includes the $m=4$ mode.  At $B_0=10^{-15}~\G$ this may be similarly excluded or detected with significance $\gtrsim3\sigma$; of $10^3$ realizations 4 lay below the 95\% confidence upper limit of the null model.  As a result, should any of the systematic contaminants described in Section \ref{sec:contam} be present (e.g., weak neighbors, unresolved backgrounds, etc.) the additional statistical power afforded by the increased fluence may prove critical to verifying any putative halo signal.  However, the range of IGMF strengths that may be detected or excluded is not improved substantially.

For the null and small-scale, tangled IGMF halo models the result is primarily a reduced range of variation about a reduced Poisson noise limit; the ability to detect the latter is not substantially improved.

Note that this provides only a very pessimistic view of the value of an extended \Fermi mission, even within the narrow confines of the detection of anisotropic ICC halos.  A doubling of gamma-ray fluence not only increases the number of photons available in the sources listed in Table~\ref{tab:sample}, but improves the significance with which structure can be identified in sources not listed in Table~\ref{tab:sample}.  As a result, the collection of \Fermi sources that contribute positively to $P_{\rm det}$ will grow beyond those in Table~\ref{tab:sample} at all values of $B_0$.  As a result, the ability to detect ICC halos will benefit from an increased collection of sources, which we have not addressed, as well as the improved fluence.

\section{Conclusions}

In \citet{BowTiesI} we presented a semi-analytical computation of ICC halos and the generation of realizations around \Fermi sources.  There we found that the putative ICC halos that surround bright VHEGR sources are generally anisotropic.  The origin of the anisotropy depends primarily on the structure of the IGMF in the AGNs vicinity, and otherwise weakly on its remaining intrinsic and observational properties.  For small-scale, tangled IGMFs the halo structure is driven by the beamed nature of the VHEGR emission; for large-scale, uniform IGMFs the halo structure is a result of the opposite gyration of electrons and positrons coupled with the geometry of inverse Compton scattering in the high-energy limit.

In all cases, however, the structure of the gamma-ray map is bimodal.  Here we suggest an explicit experiment to detect ICC halos using this bimodal structure that is insensitive to the many systematic uncertainties in the LAT instrumental response.  This exploits the angular structure imposed by the bimodal structure instead of the radial structure often used.  We propose to do this by generating one-dimensional angular power spectra, $\P_m$, about each VHEGR source, following the application of an energy-dependent mask to remove the contribution of the direct emission from the central source.  These have the considerable advantage that, unlike the images, they retain evidence of angular structure in the image when stacked regardless of absolute orientation.

The presence of bimodal structures in the underlying gamma-ray count maps generally produces a quadrupolar power excess (i.e., at $m=2$) and oscillating thereafter.  The magnitude of the low-power excess and the depth of the oscillations depend on the intrinsic source parameters and the structure of the IGMF.  Nevertheless, the sawtooth nature is generally present, both among different IGMF models and intrinsic parameter distributions as well as individual realizations of the images.  Importantly, this is qualitatively distinct from the smooth power excesses induced by many potential observational and astronomical contaminants, including undetected neighbors, an unresolved point-source background, large background gradients, failures to center sources properly, and structure in the LAT instrument response.

Via Monte Carlo simulation of the images from VHEGR \Fermi sources we have produced an optimized experiment for detecting ICC halos.  That is, the ability to generate a large ensemble of realizations of images for the full sample enables the {\em theoretical} prediction of the anticipated {\em distribution} of the power at each $m$.  We have therefore optimized both the procedure and sample of \Fermi AGN.  While the procedure is independent of the details underlying the ICC halo generation, the optimal list of \Fermi sources is not; we report the optimized source samples for detecting ICC halos associated with large-scale, uniform IGMFs in Table \ref{tab:sample}.  It is important to note that apart from the source SEDs and measured redshifts, this was performed {\em entirely} using {\em simulated} images, and thus is not colored by actual structures within the gamma-ray images.

Based upon this optimized experiment we are able to define confidence levels at which a given halo model can be detected {\em or} excluded using current and future \Fermi data.  We find that this depends on the origin of the ICC halo structure.  ICC halos generated in the presence of a small-scale, tangled IGMF are unlikely to be detected using the \Fermi AGN sample.  This is a direct consequence of the extreme foreshortening associated with the small viewing angles implied for most gamma-ray bright AGN and the intrinsic reduction in the ICC halo surface brightness associated with the isotropization of the generating pair population.

However, the ICC halos generated in the presence of a large-scale, uniform IGMF are detectable at high confidence for IGMF strengths between $10^{-17}~\G$ and $10^{-14}~\G$, well matched to the range implied by gamma-ray observations of bright, nearby \Fermi AGN.  At weaker IGMF strengths the putative halos are sufficiently spatially concentrated that they are confused with the central source.  At stronger IGMF strengths they are sufficiently spatially diffused that they are confused with neighboring sources.  Were \Fermi to continue operations for roughly eight more years the significance with which ICC halos could be detected or excluded would increase.  

In a forthcoming companion publication we apply this to the existing sample of suitable \Fermi blazars, placing constraints on the geometry and strength of the IGMF.

\acknowledgments A.E.B.~and M.S.~receive financial support from the Perimeter Institute for Theoretical Physics and the Natural Sciences and Engineering Research Council of Canada through a Discovery Grant.  Research at Perimeter Institute is supported by the Government of Canada through Industry Canada and by the Province of Ontario through the Ministry of Research and Innovation.  C.P.~acknowledges support by the European Research Council under ERC-CoG grant CRAGSMAN-646955 and by the Klaus Tschira Foundation. P.C.~gratefully acknowledges support from the NASA ATP program through NASA grant NNX13AH43G, and NSF grant AST-1255469. A.L.~receives financial support from an Alfred P. Sloan Research Fellowship, NASA ATP Grant NNX14AH35G, and NSF Collaborative Research Grant 411920 and CAREER grant 1455342. E.P. acknowledges support by the Kavli Foundation.

\begin{appendix}

\section{Generalized Lambda Distribution Fit of Intrinsic Opening Angles} \label{app:Lambda}
The generalized lambda distribution (G$\lambda$D) provides a convenient parameterization of probability distributions \citep{Ramberg:1974:AMG:360827.360840}.  A key feature of the G$\lambda$D is that it parameterizes the {\em inverse} of the {\em cumulative} probability distribution instead of the probability density directly.  This is particularly useful for generating random variates from the parameterized probability distribution.

The G$\lambda$D, with the Ramberg and Schmeiser (RS) parameterization, is defined by four constants, $\lambda_{1,2,3,4}$, in terms of which
\begin{equation}
  F^{-1}(u)
  =
  \lambda_1 + \frac{u^{\lambda_3}-(1-u)^{\lambda_4}}{\lambda_2}\,,
\end{equation}
where $u\in[0,1]$ is the cumulative probability and $F^{-1}(u)$ is a map to the dependent variable.  Drawing a random variate from the desired distribution is then reduced to choosing a uniform variate on $[0,1]$ and evaluating $F^{-1}(u)$.

We performed a non-parametric, fit of the cumulative distribution of intrinsic radio opening angles, $\alpha_{\rm int}$ from \citep{Push_etal:09}, using the numerical statistics environment R \citep{team2014r} and the G$\lambda$D RS fitting package by GLDEX \citep{GLDEX,SuGLDEX} and performed a maximum liklihood fit. A KS test was performed to analyze the quality of the fit and the result was a p-value $0.9704$ i.e. consistent with the assumption that they are drawn from the same distribution.  The four G$\lambda$D parameters are $\lambda_1=0.8261001$, $\lambda_2=-0.63510851$, $\lambda_3=-0.07892053$, $\lambda_4=-0.2946635$, and the resulting probability distribution is shown in Figure~\ref{fig:aint}.

\section{Maximum-Likelihood Center Finding} \label{app:center}
Here we describe how we produce refined high-accuracy estimates of the source location.  We begin with a parameterized model for the probability distribution of photons in an image that describes a Gaussian source on a uniform background.  This has three parameters: the source position, $\bmu$, the ratio of photons in the background to those in the source, $A$, and the width of the source distribution, $\sigma$.  We will assume that the last is fixed by the characteristic size of the \Fermi PSF.  Finally, and critically, we will only perform the center-fitting within a known circular window with angular radius $R$ about some fixed initial position.  Thus, the probability of a given photon being located at position $\bx$ given parameter $\bp=(\bmu,A)$ is
\begin{equation}
  P(\bx|\bp)
  =
  \Theta(R-|\bx|) N
  \left[
    \frac{1}{2\pi\sigma^2} e^{-(\bx-\bmu)^2/2\sigma^2} + \frac{A}{\pi R^2}
    \right]\,,
\end{equation}
These are not all independent, related by the normalization condition upon $P(\bx|\bp)$:
\begin{equation}
  \begin{aligned}
    1 &= \int d^2\!x\,P(\bx|\bp)
    = \int_0^R r dr \int_0^{2\pi} d\theta 
    N \left[
      \frac{1}{2\pi\sigma^2} e^{-(r^2-2r\mu\cos\theta + \mu^2)/2\sigma^2} + \frac{A}{\pi R^2}
      \right]
    =
    NA + N e^{-\mu^2/2\sigma^2}
    \int_0^{R/\sigma} dx \, x \, e^{-x^2/2} I_0(xm)\,,
  \end{aligned}
\end{equation}
where $x\equiv r/\sigma$, $m\equiv\mu/\sigma$, and $I_0(x)$ is the modified Bessel function of the first kind.  This final integral is difficult to perform in practice due to the presence of $I_0$.  It maybe be simplified via various asymptotic expansions, though in practice, it is most easily evaluated numerically.

However it is computed, the resulting integral is a function of $R^2$, $\sigma^2$, and $\mu^2$ only, and thus we define:
\begin{equation}
  I(R^2,\sigma^2,\mu^2) 
  \equiv
  e^{-\mu^2/2\sigma^2}
  \int_0^{R/\sigma} dx \, x \, e^{-x^2/2} I_0(xm)\,.
\end{equation}
Regardless how $I(R^2,\sigma^2,\mu^2)$ is computed, the normalization condition becomes
\begin{equation}
N = \frac{1}{A + I(R^2,\sigma^2,\mu^2)}\,,
\end{equation}
and thus, our probability distribution is
\begin{equation}
  P(\bx|\bp)
  =
  \Theta(R-|\bx|)
  \frac{
    (\pi R^2/2\pi\sigma^2)
    e^{-(\bx-\bmu)^2/2\sigma^2} 
    + 
    A
  }{
    \pi R^2 \left[ A + I(R^2,\sigma^2,\mu^2) \right]
  }
  \,.
\end{equation}
The corresponding likelihood for a set of $\{\bx_j\}$ is then
\begin{equation}
L(\{\bx_j\}|\bp)
=
-2 \sum_j
\log
\left\{
\frac{
  (\pi R^2/2\pi\sigma^2)
  e^{-(\bx_j-\bmu)^2/2\sigma^2} 
  + 
  A
}{
  \pi R^2 \left[ A + I(R^2,\sigma^2,\mu^2) \right]
}
\right\}
=
-2\sum_j\log\left(\frac{\Delta_j + A}{I + A}\right) + \text{const.}\,,
\end{equation}
where we ignore the constant $\pi R^2$ term and define
\begin{equation}
\Delta_j = \frac{\pi R^2}{2\pi\sigma^2}
e^{-(\bx_j-\bmu)^2/2\sigma^2}\,.
\end{equation}
Finally, to avoid issues with $A$ dropping below zero, we replace it with $e^\alpha$, which is positive definite.  Thus,
\begin{equation}
L = -2\sum_j\log\left(\frac{\Delta_j + e^\alpha}{I + e^\alpha}\right) + \text{const.}\,.
\end{equation}
We supplement these with the derivatives:
\begin{equation}
\frac{\partial L}{\partial \mu}
=
-2 \sum_j
\left[
\frac{\Delta_j}{\Delta_j + e^\alpha}
\frac{(\bx_j-\bmu)}{\sigma^2} 
-
\frac{2\bmu}{I + e^\alpha}
\frac{\partial I}{\partial \mu^2}
\right]
\quad\text{and}\quad
\frac{\partial L}{\partial\alpha}
=
-2 \sum_j
e^\alpha
\left[
\frac{1}{\Delta_j+e^\alpha}
-
\frac{1}{I+e^\alpha}
\right]\,.
\end{equation}
The values of $\bmu$ and $\alpha$ are then obtained via numerically minimizing $L$.

\section{Example Angular Power Spectra} \label{app:ExPS}
Here we present a number of angular power spectra for toy models.  These are primarily for illustrative and contextual purposes -- power spectra in the text are obtained for physically motivated mock images in various circumstances.  In all cases we make use of the following general definitions.
The angular power associated with a given image is
\begin{equation}
\P_m = \frac{1}{N^2} \left| \sum_j e^{i m \theta_j} \right|^2
=
\frac{1}{N^2} \sum_{j,k} e^{i m (\theta_j-\theta_k)}
=
\frac{1}{N}
+
\frac{1}{N^2} \sum _{j\ne k} e^{i m (\theta_j-\theta_k)}\,.
\end{equation}
Note that generally, for $m=0$ we have $\P_0 = N^{-2} \sum_{j,k} 1 = 1$ independent of the details of the distribution of $\theta_j$.

To compute the statistical properties of $\P_m$ for $m>0$ we must specify the probability distribution of the $\theta_{N,m}$, which we describe via the probability density of finding a photon at $\theta$, $\wp(\theta)$.  (The independence of $m=0$ is a consequence of the normalization condition on $\wp(\theta)$.)  Hence, assuming each photon is independent, the mean is given by
\begin{equation}
\begin{aligned}
\left< \P_m \right>
=
\int d\theta_1 \dots d\theta_N \, 
\wp(\theta_1)\dots\wp(\theta_N)
\left[
\frac{1}{N}
+
\frac{1}{N^2} \sum _{j\ne k}
e^{i m (\theta_j-\theta_k)}
\right]
=
\frac{1}{N}
+
\frac{N-1}{N}
\int d\theta_1 d\theta_2 \, 
\wp(\theta_1)\wp(\theta_2)
e^{i m (\theta_1-\theta_2)}
=
\frac{1}{N} + \frac{N-1}{N} \left|\varphi_m\right|^2
\end{aligned}
\end{equation}
where $\varphi_m$ is the characteristic function of the photon distribution:
\begin{equation}
\varphi_m \equiv \int d\theta \wp(\theta) e^{im\theta}\,.
\end{equation}
The variance may be computed in a similar way, though some care must be taken with counting the degeneracies of the various terms, ultimately yielding for $m>0$
\begin{equation}
\begin{aligned}
\sigma_{\P_m}^2
&=
\left< \P_m^2 \right>  - \left<\P_m\right>^2\\
&=
\frac{1}{N^4}
\int d\theta_1 \dots d\theta_N \, 
\wp(\theta_1)\dots\wp(\theta_N)
\left[
  N^2
  + 2 N \sum_{j\ne k} e^{im(\theta_j-\theta_k)}
  + \sum_{j\ne k, a\ne b} e^{im(\theta_j-\theta_k+\theta_a-\theta_b)}
\right]
- \left[ N + N(N-1) \left|\phi_m\right|^2 \right]^2\\
&=
\frac{1}{N^4}
\int d\theta_1 \dots d\theta_N \, 
\wp(\theta_1)\dots\wp(\theta_N)
\left[
  N^2
  + 2 N \sum_{j\ne k} e^{im(\theta_j-\theta_k)}
  + \sum_{j=b\ne k=a} 1
  + \sum_{j=b\ne k\ne a} e^{im(\theta_a-\theta_k)}
  + \sum_{j\ne b\ne k= a} e^{im(\theta_j-\theta_b)}
\right.\\
&\qquad\qquad
\left.
  + \sum_{j=a\ne k=b} e^{im(2 \theta_j-2\theta_k)}
  + \sum_{j=a\ne k\ne b} e^{im(2 \theta_j-\theta_k-\theta_b)}
  + \sum_{j\ne a\ne k=b} e^{im(\theta_j-2 \theta_k+\theta_a)}
  + \sum_{j\ne k\ne a\ne b} e^{im(\theta_j-\theta_k+\theta_a-\theta_b)}
\right]
- \left[ N + N(N-1) \left|\phi_m\right|^2 \right]^2\\
&=
\frac{1}{N^4}
\bigg[
  N^2 + 2 N^2(N-1) \left|\varphi_m\right|^2 
+ N(N-1) + 2N(N-1)(N-2)\left|\varphi_m\right|^2
+ N(N-1) \left|\varphi_{2m}\right|^2\\
&\qquad\quad
+ N(N-1)(N-2)\left( \varphi_{2m}^* \varphi_m^2 + \varphi_{2m} {\varphi_m^*}^2 \right)
+ N(N-1)(N-2)(N-3) \left|\varphi_m\right|^4
- N^2 - 2N^2 (N-1)\left|\varphi_m\right|^2  - N^2(N-1)^2 \left|\varphi_m\right|^4
\bigg]\\
&=
\frac{N-1}{N^3}\left[
1
+
(2N-4)\left|\varphi_m\right|^2 
-
( 4N-6) \left|\varphi_m\right|^4
+
\left|\varphi_{2m}\right|^2 
+
(N-2) \left( \varphi_{2m}^* \varphi_m^2 + \varphi_{2m} {\varphi_m^*}^2 \right)
\right]\,.
\end{aligned}
\end{equation}

Identifying $\left<\P_m\right>$ and $\sigma_{\P_m}$ for specific models now requires the specification of $\varphi_m$.  We consider a handful of specific cases below.

\subsection{Isotropic Background}
We begin with an isotropic background, i.e.,
\begin{equation}
  \wp(\theta) = \frac{1}{2\pi}
  \quad\Rightarrow\quad
  \varphi_m = \delta_{m0}\,,
\end{equation}
and thus for $m>0$
\begin{equation}
  \left< \P_m \right>
  =
  \frac{1}{N}
  \quad\text{and}\quad
  \sigma^2_{\P_m}
  =
  \frac{N-1}{N^3}\,,
\end{equation}
Note that this does not imply a uniform background, only one that is isotropic about the chosen image center.  This is simply the Poisson noise limit, and as is typical in power spectrum estimation, based on a single source the variance in the estimate of $\P_m$ is comparable to its mean.

\subsection{Offset Gaussian} \label{app:MCGPS}
We now consider a peaked source offset from the origin, as may occur if the source location estimate is in error.  Here we make simplifying assumption that the source Gaussian, i.e., set the 2-dimensional photon probability density to be 
\begin{equation}
\wp(\alpha,\delta) 
= 
\frac{1}{2\pi \sigma^2} e^{-[(\alpha-\eta\sigma)^2+\delta^2]/2\sigma^2}
=
\frac{1}{2\pi \sigma^2} e^{-(r^2+\eta^2\sigma^2)/2\sigma^2} 
\left( 
1 
+ 
\frac{\eta r\cos\theta}{\sigma}
+ 
\frac{\eta^2 r^2\cos^2\theta}{2 \sigma^2}
+ 
\frac{\eta^3 r^3\cos^3\theta}{6 \sigma^3}
 + \dots \right)\,.
\end{equation}
Integrating over radius, this gives the needed angular probability density:
\begin{equation}
\wp(\theta) = \int r dr \wp(\alpha,\delta)
=
\frac{1}{2\pi}
\left[
1
+
\sqrt{\frac{\pi}{2}} \eta \cos\theta
+
\eta^2 \cos^2\theta
+
\sqrt{\frac{\pi}{8}} \eta^3 \cos^3\theta
+
\dots
\right]\,.
\end{equation}
Therefore, the characteristic function is
\begin{equation}
\varphi_m
=
\delta_{m0}
+
\frac{1}{2} \sqrt{\frac{\pi}{2}} \eta (\delta_{m1}+\delta_{m-1})
+
\frac{1}{4} \eta^2
( \delta_{m2} + 2\delta_{m0} + 2\delta_{m-2} )
+
\frac{1}{8} \sqrt{\frac{\pi}{8}} \eta^3 
( \delta_{m3} + 3 \delta_{m1} + 3 \delta_{m-1} + \delta_{m-3} )
+
\dots
\end{equation}
To lowest order in $\eta$ at each $m$, this gives
\begin{equation}
\left< \P_m \right>
=
N + N(N-1)
\left[
\delta_{m0}
+
\frac{\pi}{8}\eta^2 \delta_{m1}
+
\frac{1}{16}\eta^4 \delta_{m2}
+
\frac{\pi}{512}\eta^6 \delta_{m3}
+
\dots
\right]\,.
\end{equation}
Note that the contribution to the high-$m$ multipoles falls off as $\eta^{2m}$.  Not surprisingly, the variance also falls off quite rapidly in $\eta$:
\begin{equation}
\sigma_{\P_m}^2
\simeq
\sum_j 
2 N^3 
\left[
\delta_{m0}
+
\frac{\pi}{8}\eta^2 \delta_{m1}
+
\frac{1}{16}\eta^4 \delta_{m2}
+
\frac{\pi}{512}\eta^6 \delta_{m3}
+
\dots
\right]\,.
\end{equation}

We can construct a general formula, using the integral
\begin{equation}
\int_0^\infty dr r^{n+1} e^{-r^2/2} = 2^{n/2}\Gamma\left(\frac{n}{2}+1\right)\,,
\end{equation}
with which we have
\begin{equation}
\wp(\theta) 
= 
\frac{1}{2\pi} \sum_{j=0}^\infty \frac{2^{j/2}}{j!}\Gamma\left(\frac{j}{2}+1\right) \eta^j \cos^j\theta
\quad\Rightarrow\quad
\varphi_m = \sum_{j=0}^\infty 
\frac{2^{-j/2}}{j!}\Gamma\left(\frac{j}{2}+1\right) \eta^j
\sum_{k=-j}^j \frac{j!}{k! (j-k)!} \delta_{mk}\,,
\end{equation}
in terms of which we have the lowest-order terms in $\eta$ for each
$m$ in the power spectrum
\begin{equation}
\left< \P_m \right>
\approx
\left|\varphi_m\right|^2
\approx
2^{-m}\left[\frac{1}{m!}\Gamma\left(\frac{m}{2}+1\right)\right]^2 \eta^{2m}
<
2^{-3m/2} \eta^{2m}
\quad\text{and}\quad
\sigma_{\P_m}^2
\approx
2 \left|\phi_m\right|^2
<
2^{1-3m/2} \eta^{2m}\,.
\end{equation}

\subsection{Weak Image Components} \label{app:ExPSw}
There are many instances where the image is comprised of a dominant, isotropic component and sub-dominant structured component.  In this case we generally have
\begin{equation}
  \wp(\theta) = \frac{1-\eta}{2\pi} + \eta h(\theta)
  \quad\Rightarrow\quad
  \varphi_m = \delta_{m0} + \eta \psi_m\,,
\end{equation}
where $\psi_m$ is the structure function associated with the unit-normalized $h(\theta)$ perturbing distribution, with strength described by order parameter $\eta\ll1$ that is equal to the ratio of the photons associated with the perturbation to those in the isotropic background.  Then, for $m>0$
\begin{equation}
  \left< \P_m \right>
  \approx
  \frac{1}{N} + \eta^2 \left|\psi_m\right|^2
  \quad\text{and}\quad
  \sigma_{\P_m}^2
  \approx
  \frac{1}{N^2}
  +
  \frac{2}{N} \eta^2\left|\psi_m\right|^2\,.
\end{equation}
Note that again $\sigma_{\P_m}\approx \P_m$, which motivates combining multiple estimates of $\P_m$ to obtain improved statistical significance of any detection.

\subsubsection{Gaussian Neighbor}
Distant peaked sources, arising from unidentified neighbors, will also induce dipolar power.  Again we make the simplifying assumption that the neighboring source is Gaussian, which if it is sufficiently far is approximately described by 
\begin{equation}
h(\theta) = \frac{e^{-(\theta-\theta_0)^2/2 w^2}}{\sqrt{2\pi} w} 
\quad\Rightarrow\quad
\psi_m = e^{i m\theta_0 - w^2 m^2/2}\,,
\end{equation}
where $\theta_0$ and $w =\sigma/\Delta r$ are the location and width of the companion in polar angle measured about the central source.  The associated mean power spectrum is
\begin{equation}
\left< \P_m \right>
\approx
\frac{1}{N}
+
\eta^2 e^{-w^2 m^2}
\quad\text{and}\quad
\sigma_{\P_m}^2
\approx
\frac{1}{N^2}
+
\frac{2}{N}
\eta^2e^{-w^2 m^2}\,.
\end{equation}
As anticipated this generates substantial dipolar power, smoothly falling off on a scale inverse set by the width.

\subsubsection{Bimodal Gaussian}
As an approximation to what the angular power spectrum from a bi-modal halo may look like we consider a bimodal Gaussian, consisting of two sources located $180^\circ$ apart.  That is,
\begin{equation}
h(\theta) = \frac{1}{\sqrt{8\pi}w}\left( e^{-(\theta-\theta_0)^2/2w^2}+e^{-(\theta-\theta_0+\pi)^2/2w^2} \right)
\quad\Rightarrow\quad
\psi_m
=
\Delta_m e^{im\theta_0 - w^2 m^2/2}\,,
\end{equation}
where $\Delta_m$ is unity for even $m$ and zero for odd $m$.  This is similar to the Gaussian neighbor, with the exception of the factor of $\Delta_m$ that arises from the beating between the two sources.  Hence, the mean power spectrum is,
\begin{equation}
\left< \P_m \right>
\approx
\frac{1}{N}
+
\eta^2
\Delta_m e^{- w^2 m^2}
\quad\text{and}\quad
\sigma_{\P_m}^2
\approx
\frac{1}{N^2} + \frac{2}{N} \eta^2 \Delta_m e^{- w^2 m^2}\,.
\end{equation}
Note that this exhibits the clear sawtooth indicative of the bimodal structures generally.

\section{Estimate of PSF-Induced Angular Structure} \label{app:PSFPS}
We make a quantitative estimate of the impact of the intrinsic anisotropy of the \Fermi LAT PSF.  To do this we follow the description in \S5.2.3 of \citet{LAT_perf} and adopt an azimuthally dependence for the PSF of the form
\begin{equation}
  f(\theta) = 1 + q_0 \xi^{q_1}
  \quad\text{where}\quad
  \xi = \frac{4}{\pi} \left| \left(\theta\mod\frac{\pi}{2}\right) -\frac{\pi}{4} \right|\,.
\end{equation}
where $q_0$, $q_1$, and $\xi$ have the same meaning as in Equation (15) of \citet{LAT_perf}.  The impact of this is to modify the distribution of photons in the image, inducing a small $m=4$ perturbation to the angular probability distribution.

The gamma rays in a typical image from a bright \Fermi AGN are accumulated over many years, and therefore the LAT presents different orientations.  Were the orientation completely independent for each photon, the net PSF would become isotropic and therefore any remaining structure would be absent.  However, we may imagine that subsequent photons have LAT orientations that are closely correlated.  The roll angle of the LAT changes primarily as a result of the fixed telescope orientation relative to the Sun.  The square geometry of the LAT implies that the angular impact on PSF is effectively averaged by the time it has rotated by $45^\circ$, corresponding to roughly 45 days.  Over this timescale a non-varying \Fermi source would accumulate less than 2\% of its total fluence or roughly 100 photons.

Thus, we estimate the residual power from $n\approx50$ epochs of $N\approx100$ events, assuming perfect correlation in the LAT orientation for the latter.  Within each subgroup there is a fixed roll angle, $\vartheta_\mu$, and hence relative to the orientation of the LAT the polar angle of the photons are $\theta_{j,\mu} = \theta_j - \vartheta_\mu$.  Between subgroups the roll angle is effectively uncorrelated.  Thus, for $m>0$,
\begin{equation}
  \left< \P_m \right>
  =
  \frac{1}{(nN)^2}
  \sum_{\mu,\nu} \sum_{j,k}
  \left<
    e^{im(\theta_j-\theta_k)} e^{-im(\vartheta_j+\vartheta_k)}
  \right>
  =
  \frac{1}{n N^2}
  \sum_{j,k}
  \left<
    e^{im(\theta_j-\theta_k)}
  \right>\,,
\end{equation}
where the independence of different $\vartheta_\mu$ was employed.

Following the prefactor of $n^{-1}$ is simply the angular power spectrum for the single epoch, for which the photon distribution is perturbed from isotropy.  This is precisely the case considered in Appendix \ref{app:ExPSw}.  Here, the perturbation is $h(\theta) = \xi/\pi$, where we have set $q_1=1$ for simplicity and implicitly subsumed $q_0$ into $\eta=2q_0$.  The associated characteristic function is
\begin{equation}
  \psi_m = \int d\theta \frac{\xi}{\pi} e^{im\theta}
  =
  \left(1+e^{im\pi/2}+e^{im\pi}+e^{3im\pi/2}\right)
  \int_0^{\pi/2} d\theta \frac{\xi}{\pi} e^{im\theta}
  =
  -
  \frac{i}{2} 
  \Sigma_m
  e^{im\pi/4}
  j_1\left(\frac{m\pi}{4}\right)\,,
\end{equation}
where $\Sigma_m$ is unity for $m$ that are divisible by $4$ and zero otherwise, and $j_1(x)$ is the spherical Bessel function of the first kind of order 1.  Therefore, the corresponding mean angular power spectrum is
\begin{equation}
  \left< \P_m \right>
  =
  \frac{1}{n}
  \left[
    \frac{1}{N} + \eta^2 \Sigma_m \frac{1}{4} j_1^2\left(\frac{m\pi}{4}\right)
  \right]
  \lesssim
    \frac{1}{nN} +  \frac{\eta^2}{n} \frac{\Sigma_m}{4\pi^2 m^2}\,.
\end{equation}
The excess is non-vanishing only for $m=4$ and its harmonics; in particular there is no power at $m=2$.  For a typical value of $\eta\approx0.2$, corresponding to $q_0\approx0.1$, the excess at $m=4$ is $\lesssim 10^{-6}$ which is much smaller than the $(nN)^{-1}\gtrsim10^{-4}$.

In practice, the angular power associated with structure in the PSF is further reduced by the fraction of the total photons those from the source compose.  This is impacted both by the surface brightness of the background, typically contributing 25\% for the brightest sources, and the source mask, typically cutting the source contribution by half.  This makes only an order unity correction.

\section{Source Sample} \label{app:SEDs}
Here we collect the images and fits for the 18 sources listed in Table \ref{tab:sample} that comprise the full set of sources within the optimized source lists.  Details of the fitting process are described in Section \ref{sec:SEDs}; we only summarize these here.

There were 84 sources selected on the basis of their 2FHL 50~GeV--2~TeV flux, as well as having a redshift below 0.5 and a 3LAC counterpart.  Gamma-ray SEDs were produced by compiling the 0.1-0.3~GeV, 0.3-1~GeV, 1-3~GeV, 3-10~GeV, and 10-100~GeV flux measurements reported in the 3LAC and 50-171~GeV, 171-585~GeV, and 585-2000~GeV reported in the 2FHL.  These were deabsorbed via Equation \ref{eq:tauobs} evaluated at the geometric center of the energy bin.  A maximum-likelihood fit of the broken-power-law SED model was performed to each candidate source, desribed by low- and high-energy spectral indexes, $\Gamma_l$, $\Gamma_h$, a pivot energy $E_p$, and a normalization.

Sources for which the source-frame TeV flux, i.e., $E^2 dN/dE$ at $(1+z)$~TeV, was above $10^{-9}~{\GeV\,\cm^{-2}\s^{-1}}$, were then visually inspected for neighboring sources within $2^\circ$ or large background gradients.  Sufficiently isolated, TeV-bright sources comprise the source sample employed here, consisting of the 27 objects listed in Tables \ref{tab:sample2} and \ref{tab:fits}.  Note that the sources that comprise the optimized source lists are chosen from these 27 after simulating the ability to detect ICC halos (see Section \ref{sec:opt}).

The observed and deabsorbed SEDs and our fits for the 18 sources used here are shown in Figures \ref{fig:compnew1} and \ref{fig:compnew2}.

\begin{figure}
  \begin{center}
    \includegraphics[width=0.5\textwidth]{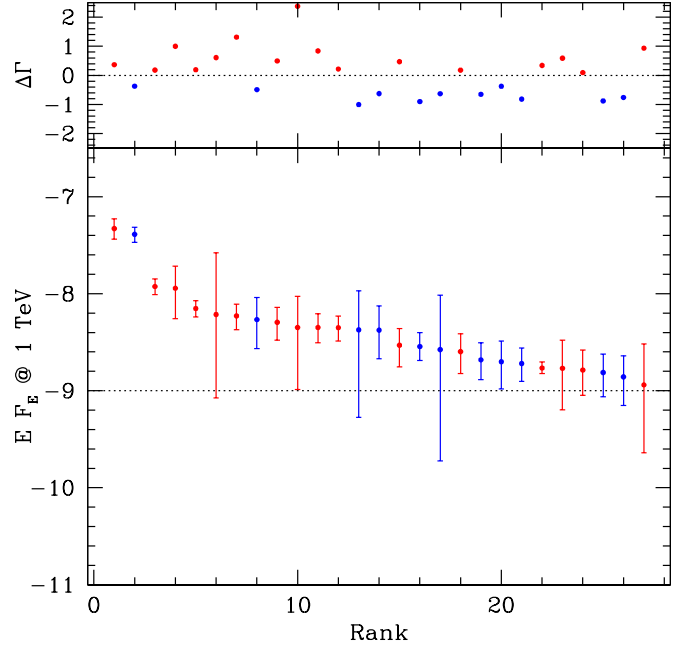}
  \end{center}
  \caption{The rank-ordered list of TeV fluxes.  The nominal cutoff is shown by the dotted line.  Sources are ranked by the 1$\sigma$ lower limit on the TeV fluxes.  Sources are color coded by SED type: convex ($\Gamma_h>\Gamma_l$) and concave ($\Gamma_h<\Gamma_l$) SEDs are shown in red and blue, respectively.} \label{fig:TeVFluxes}
\end{figure}

\begin{deluxetable*}{cccccccc}
  \tablecaption{List of Sources\label{tab:sample2}}   
  \tablehead{
    \multicolumn{4}{c}{Names} &
    &
    &
    \multicolumn{2}{c}{Fluxes}
    \\
    3FGL &
    2FHL &
    TeV &
    Other &
    Class\tablenotemark{a} &
    $z$\tablenotemark{b} &
    $F_{\rm 3FGL}$\tablenotemark{c} &
    $F_{\rm 2FHL}$\tablenotemark{d}
    \\
    & & & & 
    &
    &
    (${\rm ph\,cm^{-2}s^{-1}}$) & (${\rm ph\,cm^{-2}s^{-1}}$) 
  }
  \startdata

J1104.4+3812 & J1104.4+3812 & J1104+3811 & Mkn 421 & bll & 0.031 & 3.03\e{-8} & 1.24\e{-9} \\
J1653.9+3945 & J1653.9+3945 & J1653+3945 & Mkn 501 & bll & 0.0337 & 9.74\e{-9} & 4.78\e{-10} \\
J0449.4-4350 & J0449.4-4349 & J0449-4350 & PKS 0447-439 & bll & 0.205 & 1.03\e{-8} & 1.65\e{-10} \\
J0508.0+6736 & J0507.9+6737 & J0507+6737 & 1ES 0502+675 & bll & 0.34 & 2.12\e{-9} & 1.63\e{-10} \\
J1015.0+4925 & J1015.0+4926 & J1015+4926 & 1ES 1011+496 & bll & 0.212 & 7.35\e{-9} & 1.62\e{-10} \\
J2000.0+6509 & J2000.1+6508 & J1959+6508 & 1ES 1959+650 & bll & 0.047 & 5.84\e{-9} & 1.57\e{-10} \\
J0650.7+2503 & J0650.7+2502 & J0650+2503 & 1ES 0647+250 & bll & 0.203 & 2.16\e{-9} & 1.26\e{-10} \\
J2009.3-4849 & J2009.4-4849 & J2009-4849 & PKS 2005-489 & bll & 0.071 & 3.55\e{-9} & 9.70\e{-11} \\
J0303.4-2407 & J0303.3-2407 & J0303-2407 & PKS 0301-243 & bll & 0.26 & 5.74\e{-9} & 8.62\e{-11} \\
J1444.0-3907 & J1443.9-3909 & --- & PKS 1440-389 & bll & 0.0654 & 3.23\e{-9} & 8.22\e{-11} \\
J2347.0+5142 & J2347.1+5142 & J2346+5142 & 1ES 2344+514 & bll & 0.044 & 2.42\e{-9} & 7.48\e{-11} \\
J0648.8+1516 & J0648.6+1516 & J0648+1516 & RX J0648.7+1516 & bll & 0.179 & 1.45\e{-9} & 5.62\e{-11} \\
J0543.9-5531 & J0543.9-5533 & --- & 1RXS J054357.3-55320 & bll & 0.273 & 1.61\e{-9} & 5.27\e{-11} \\
J1512.8-0906 & J1512.7-0906 & J1512-0906 & PKS 1510-08 & fsrq & 0.36 & 4.11\e{-8} & 4.59\e{-11} \\
J1120.8+4212 & J1120.8+4212 & --- & RBS 0970 & bll & 0.124 & 1.16\e{-9} & 4.18\e{-11} \\
J1117.0+2014 & J1116.9+2014 & --- & RBS 0958 & bll & 0.138 & 1.74\e{-9} & 3.85\e{-11} \\
J2250.1+3825 & J2249.9+3826 & J2250+3824 & B3 2247+381 & bll & 0.119 & 1.10\e{-9} & 3.38\e{-11} \\
J0627.0-3529 & J0626.9-3528 & --- & PKS 0625-35 & rdg & 0.05494 & 1.43\e{-9} & 2.92\e{-11} \\
J0622.4-2606 & J0622.4-2604 & --- & PMN J0622-2605 & bll & 0.41449 & 1.14\e{-9} & 2.58\e{-11} \\
J2131.5-0915 & J2131.4-0914 & --- & RBS 1752 & bll & 0.449 & 8.63\e{-10} & 2.49\e{-11} \\
J2016.4-0905 & J2016.5-0904 & --- & PMN J2016-0903 & bll & 0.367 & 1.40\e{-9} & 1.86\e{-11} \\
J0115.8+2519 & J0115.8+2519 & --- & RX J0115.7+2519 & bll & 0.358 & 1.15\e{-9} & 1.83\e{-11} \\
J0757.0+0956 & J0756.8+0955 & --- & PKS 0754+100 & bll & 0.266 & 1.86\e{-9} & 1.73\e{-11} \\
J0958.6+6534 & J0958.3+6535 & --- & S4 0954+65 & bll & 0.367 & 1.38\e{-9} & 1.54\e{-11} \\
J2329.2+3754 & J2329.2+3754 & --- & NVSS J232914+375414 & bll & 0.264 & 8.40\e{-10} & 1.43\e{-11} \\
J1436.8+5639 & J1437.0+5639 & --- & RBS 1409 & bll & 0.15 & 4.71\e{-10} & 1.40\e{-11} \\
J1442.8+1200 & J1442.9+1159 & J1442+1200 & 1ES 1440+122 & bll & 0.16309 & 5.59\e{-10} & 1.39\e{-11}

\enddata
\tablenotetext{a}{Source class type: bll, fsrq, and rdg correspond to BL Lac, flat-spectrum radio quasars, and radio galaxy, respectively}
\tablenotetext{b}{3FGL value}
\tablenotetext{c}{1-100 GeV fluence from 3FGL}
\tablenotetext{d}{50-2000 GeV fluence from 3FGL}
\end{deluxetable*}

\begin{deluxetable}{cccccc}[ht]
  \tablecaption{Source SED Fit Parameters and TeV Flux\label{tab:fits}}   
  \tablehead{
    3FGL Name &
    $F_{\rm TeV}$\tablenotemark{a} &
    $f_0$ &
    $E_p$ &
    $\Gamma_l$ &
    $\Gamma_h$\tablenotemark{b}
    \\
    &
    (${\rm GeV\,cm^{-2}s^{-1}}$) &
    (${\rm ph\,cm^{-2}s^{-1}}$) &
    (GeV) &
    &
  }
  \startdata
\sidehead{Convex Spectrum Sources ($\Gamma_h>\Gamma_l$)}
J1104.4+3812 & ${ 4.69_{-1.04}^{+1.20}}\e{-8}$ & 6.53\e{-8} & 95.38 & 1.77 & $ 2.14^{+ 0.11}_{- 0.10}$ \\ %
J1653.9+3945 & ${ 1.14_{-0.58}^{+0.79}}\e{-8}$ & 3.21\e{-8} & 236.30 & 1.72 & $ 2.72^{+ 0.51}_{- 0.37}$ \\ %
J0449.4-4350 & ${ 5.07_{-1.75}^{+2.14}}\e{-9}$ & 1.58\e{-8} & 23.03 & 1.81 & $ 2.30^{+ 0.11}_{- 0.10}$ \\ %
J1015.0+4925 & ${ 1.19_{-0.21}^{+0.23}}\e{-8}$ & 7.89\e{-9} & 2.71 & 1.75 & $ 1.93^{+ 0.03}_{- 0.03}$ \\ %
J2000.0+6509 & ${ 4.49_{-3.45}^{+4.91}}\e{-9}$ & 1.15\e{-8} & 658.00 & 1.87 & $ 4.25^{+ 1.58}_{- 1.60}$ \\ %
J0650.7+2503 & ${ 6.10_{-5.25}^{+20.21}}\e{-9}$ & 8.75\e{-9} & 281.60 & 1.67 & $ 2.29^{+ 1.64}_{- 1.35}$ \\ %
J2009.3-4849 & ${ 1.15_{-0.92}^{+1.89}}\e{-9}$ & 7.08\e{-9} & 76.45 & 1.77 & $ 2.71^{+ 0.64}_{- 0.39}$ \\ %
J0303.4-2407 & ${ 7.03_{-1.28}^{+1.44}}\e{-9}$ & 5.68\e{-9} & 1.02 & 1.78 & $ 1.97^{+ 0.03}_{- 0.03}$ \\ %
J1444.0-3907 & ${ 2.94_{-1.18}^{+1.44}}\e{-9}$ & 5.43\e{-9} & 11.90 & 1.67 & $ 2.14^{+ 0.13}_{- 0.09}$ \\ %
J2347.0+5142 & ${ 4.47_{-1.21}^{+1.39}}\e{-9}$ & 2.43\e{-9} & 1.81 & 1.69 & $ 1.90^{+ 0.05}_{- 0.04}$ \\ %
J0648.8+1516 & ${ 4.49_{-1.36}^{+1.74}}\e{-9}$ & 1.21\e{-9} & 1.03 & 0.97 & $ 1.81^{+ 0.05}_{- 0.06}$ \\ %
J0543.9-5531 & ${ 5.90_{-1.64}^{+1.88}}\e{-9}$ & 1.17\e{-9} & 0.84 & 0.46 & $ 1.77^{+ 0.05}_{- 0.04}$ \\ %
J1512.8-0906 & ${ 1.71_{-0.21}^{+0.27}}\e{-9}$ & 7.24\e{-8} & 1.27 & 2.22 & $ 2.56^{+ 0.02}_{- 0.02}$ \\ %
J1120.8+4212 & ${ 1.70_{-1.07}^{+1.61}}\e{-9}$ & 2.92\e{-9} & 28.65 & 1.56 & $ 2.15^{+ 0.29}_{- 0.19}$ \\ %
J1117.0+2014 & ${ 2.53_{-1.03}^{+1.33}}\e{-9}$ & 1.89\e{-9} & 3.02 & 1.77 & $ 1.95^{+ 0.09}_{- 0.08}$ \\ %
J2016.4-0905 & ${ 1.63_{-0.74}^{+1.00}}\e{-9}$ & 1.50\e{-9} & 2.37 & 1.89 & $ 1.99^{+ 0.11}_{- 0.11}$ \\ %
\sidehead{Concave Spectrum Sources ($\Gamma_h<\Gamma_l$)}
J0508.0+6736 & ${ 4.09_{-0.70}^{+0.74}}\e{-8}$ & 1.32\e{-9} & 2.32 & 1.81 & $ 1.43^{+ 0.03}_{- 0.03}$ \\ %
J2250.1+3825 & ${ 1.91_{-0.66}^{+0.84}}\e{-9}$ & 8.83\e{-10} & 0.46 & 2.72 & $ 1.90^{+ 0.06}_{- 0.05}$ \\ %
J0627.0-3529 & ${ 2.85_{-0.80}^{+1.12}}\e{-9}$ & 1.01\e{-9} & 0.37 & 2.77 & $ 1.87^{+ 0.04}_{- 0.05}$ \\ %
J0622.4-2606 & ${ 5.41_{-2.69}^{+3.71}}\e{-9}$ & 8.75\e{-10} & 7.75 & 2.12 & $ 1.63^{+ 0.14}_{- 0.11}$ \\ %
J2131.5-0915 & ${ 4.21_{-2.08}^{+3.26}}\e{-9}$ & 6.00\e{-10} & 4.26 & 2.27 & $ 1.64^{+ 0.12}_{- 0.12}$ \\ %
J0115.8+2519 & ${ 2.08_{-0.78}^{+1.03}}\e{-9}$ & 9.38\e{-10} & 0.80 & 2.54 & $ 1.89^{+ 0.07}_{- 0.06}$ \\ %
J0757.0+0956 & ${ 2.65_{-2.46}^{+6.98}}\e{-9}$ & 8.99\e{-10} & 59.98 & 2.25 & $ 1.62^{+ 0.93}_{- 0.61}$ \\ %
J0958.6+6534 & ${ 4.25_{-3.71}^{+6.44}}\e{-9}$ & 5.73\e{-10} & 46.64 & 2.35 & $ 1.35^{+ 0.68}_{- 0.34}$ \\ %
J2329.2+3754 & ${ 1.99_{-0.95}^{+1.25}}\e{-9}$ & 6.34\e{-10} & 2.19 & 2.19 & $ 1.81^{+ 0.11}_{- 0.08}$ \\ %
J1436.8+5639 & ${ 1.39_{-0.68}^{+0.89}}\e{-9}$ & 3.18\e{-10} & 1.05 & 2.54 & $ 1.79^{+ 0.10}_{- 0.09}$ \\ %
J1442.8+1200 & ${ 1.54_{-0.67}^{+0.84}}\e{-9}$ & 3.57\e{-10} & 0.49 & 2.69 & $ 1.81^{+ 0.08}_{- 0.07}$ %
\enddata
\tablenotetext{a}{$E^2 dN/dE$ at $(1+z)$~TeV}
\tablenotetext{b}{Uncertainty in $\Gamma$ at $(1+z)$~TeV}
\end{deluxetable}

\begin{figure*}
  \begin{center}
    \includegraphics[width=\textwidth]{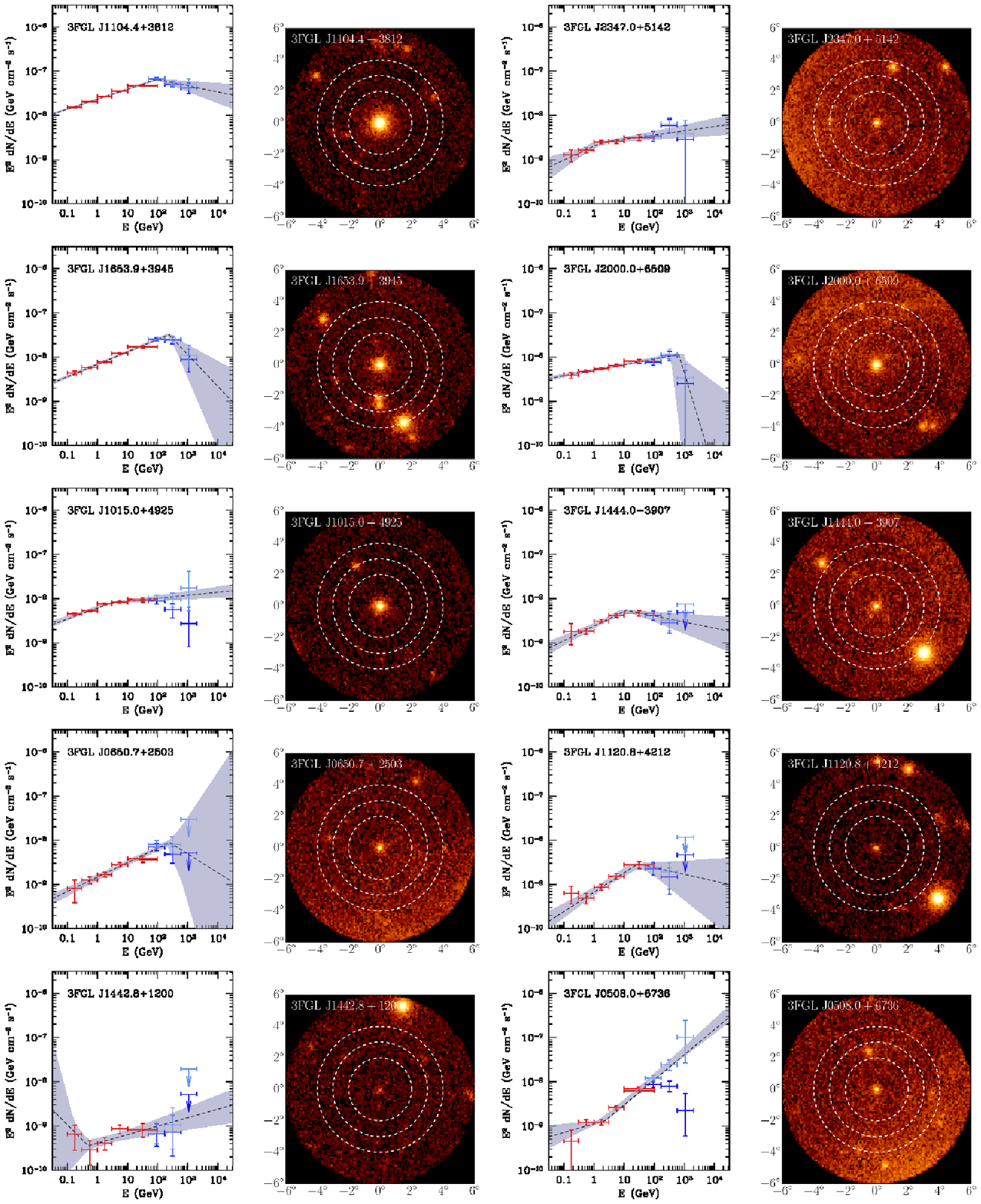}
  \end{center}
  \caption{Source Images and SEDs for objects in Table \ref{tab:sample} -- part 1. Left: Compilations of the 3FGL (red/orange) and 2FHL (blue/light blue) fluxes.  Also shown are the broken power-law fits (dashed lines) and associated uncertainty (grey region).  Right: 1--100~GeV count maps for the sources on the left.} \label{fig:compnew1}
\end{figure*}
\begin{figure*}
  \begin{center}
    \includegraphics[width=\textwidth]{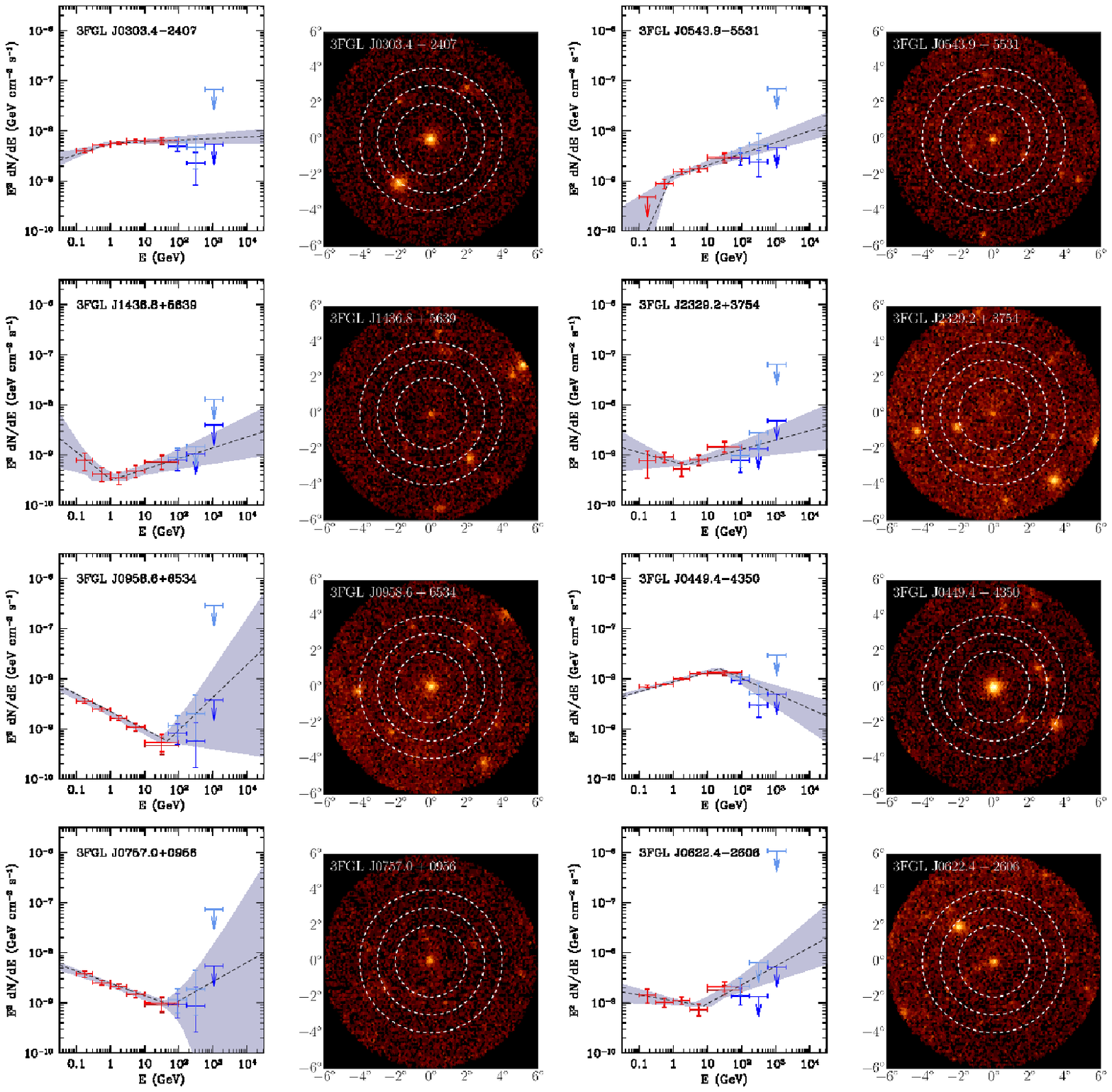}
  \end{center}
  \caption{Source Images and SEDs for objects in Table \ref{tab:sample} -- part 2. Left: Compilations of the 3FGL (red/orange) and 2FHL (blue/light blue) fluxes.  Also shown are the broken power-law fits (dashed lines) and associated uncertainty (grey region).  Right: 1--100~GeV count maps for the sources on the left.} \label{fig:compnew2}
\end{figure*}

\end{appendix}

\bibliography{bigmh.bib,bigmh_orig.bib}
\bibliographystyle{apj}

\end{document}